\documentclass[12pt,preprint]{aastex}
\usepackage{natbib}
\usepackage{multirow}
\usepackage{upgreek}
\usepackage[colorlinks=true, pdfstartview=FitV, linkcolor=blue,citecolor=blue, urlcolor=blue]{hyperref}
\usepackage{wasysym}
\usepackage{txfonts}
\usepackage{gensymb}
\usepackage{morefloats}
\usepackage{threeparttable}
\slugcomment{version \today}
\shorttitle{CHEERS results on MRK 573}
\shortauthors{Paggi et al.}

\begin{document}

\title{CHEERS RESULTS ON MRK 573: STUDY OF DEEP \textit{CHANDRA} OBSERVATIONS}

\author{Alessandro Paggi\altaffilmark{1}, Junfeng Wang\altaffilmark{1}, Giuseppina Fabbiano\altaffilmark{1}, Martin Elvis\altaffilmark{1} and Margarita Karovska\altaffilmark{1}}
\affil{\altaffilmark{1}Harvard-Smithsonian Center for Astrophysics, 60 Garden St, Cambridge, MA 02138, USA: \href{mailto:apaggi@cfa.harvard.edu}{apaggi@cfa.harvard.edu}}

\begin{abstract}
We present results on Mrk 573 obtained as part of the CHandra survey of Extended Emission-line Regions in nearby Seyfert galaxies (CHEERS). Previous studies showed that this source features {a} biconical emission in the soft X-ray {band} closely related with the Narrow Line Region as mapped by the [O \textsc{iii}] emission line and the radio emission, though on a smaller scale; we {investigate} the properties of soft X-ray emission from this source with new deep \textit{Chandra} observations. Making use of the subpixel resolution of the \textit{Chandra}/ACIS image and {PSF-}deconvolution, we resolve and study substructures in each ionizing cone. The two cone spectra are fitted with photoionization model, showing a mildly photoionized phase diffused over the bicone.
Thermal collisional gas at about \(\sim 1.1\mbox{ keV}\) and \(\sim 0.8\mbox{ keV}\) appears to be located between the nucleus and the ``knots" resolved in radio observations, and between the ``arcs" resolved in the optical images, respectively; this can be interpreted in terms of shock interaction with the host galactic plane.
The nucleus shows a significant flux decrease across the observations indicating variability of the AGN, with the nuclear region featuring higher ionization parameter with respect to the bicone region.
The long exposure allows us to find extended emission up to \(\sim 7 \mbox{ kpc}\) from the nucleus along the bicone axis. Significant emission is also detected in the direction perpendicular to the ionizing cones, disagreeing with the fully obscuring torus prescribed in the AGN unified model, and suggesting instead the presence of a clumpy structure.
\end{abstract}


\section{INTRODUCTION}\label{intro}

The interaction between active galactic nuclei (AGN) and their host galaxy is signaled by the narrow line region and extended narrow line region (NLR, ENLR). This [O \textsc{iii}] emitting region, extending on kpc scale and observed in many Seyfert galaxies, is widely interpreted as gas photoionized by the AGN \citep[e.g.,][]{2003ApJS..148..327S}.

A key tool to understand this interaction is provided by the study of soft X-ray emission from these sources. The overall structure of the X-ray emission appears to be morphologically correlated with the ENLR, suggesting a common physical origin \citep{2001ApJ...556....6Y,2003MNRAS.345..369I,2006A&A...448..499B,2011ApJ...742...23W}.

In the unified AGN model \citep{1993ARA&A..31..473A} the soft X-ray spectra from Seyfert 2 galaxies are  affected by emission from the circumnuclear medium, illuminated by the nuclear continuum. This is supported by high resolution spatial/spectral observations with both \textit{Chandra} and \textit{XMM-Newton} showing that soft X-ray emission arise from the circumnuclear medium photoionized by the nuclear continuum. Moreover, X-ray grating spectroscopy agrees with this picture, showing that the spectra can be interpreted as a blending of emission lines, with little contribution from collisionally ionized plasma \citep[see e.g.][]{2000ApJ...545L..81O,2002A&A...396..761B,2002ApJ...575..732K,2007MNRAS.374.1290G}.

Mrk 573 (\(z=0.017\), \citealt{2005AJ....129...73R}) is one of the brightest Seyfert galaxies observed by \textit{HST}, with a ENLR extending to \(\sim 9"\) corresponding to a projected size of about \(3\) kpc\footnote{In the following, we adopt the standard flat cosmology with \(\Omega_\Lambda = 0.73\) and \(H_0 = 70 \mbox{ km}\mbox{ s}^{-1}\mbox{ Mpc}^{-1}\) \citep{2011ApJS..192...18K}.}, where photoionzation by the central AGN appears to be the dominant process \citep{1999MNRAS.309....1F,2009ApJ...699..857S}. Mrk 573 is also also associated with a triple radio source, with a central core and two side knots \citep{1984ApJ...278..544U}.

The ENLR of Mrk 573 has been modeled by \citet{2010AJ....140..577F} making use of \textit{HST} STIS long-slit spectra; 
the kinematic model developed by these authors features an ionizing bicone with the NW cone tilted toward the observer line of sight, and with a half opening angle of \(53\degree\) which is close to the \(60\degree\) expected in the unifying structure for the inner quasar regions presented by \citet{2000ApJ...545...63E}.
The observed half opening angle of \(\sim 30\degree\) results from the intersection of the bicone with the galaxy plane.
The authors conclude that the circumnuclear emission is mainly due to the intersection of the ionizing bicone with the galactic disk, as supported by the observed emission arcs that coincide with outer dust lanes.

The X-ray emission from Mrk 573 has been previously studied by several authors. A study of EPIC/\textit{XMM-Newton} data by \citet{2005A&A...444..119G} showed for this source a strong Fe K\(\alpha\) line with an equivalent width larger than \(1\) keV, yielding to a classification as a Compton-thick object; high-resolution spectral analysis of RGS/\textit{XMM-Newton} data by \citet{2007MNRAS.374.1290G} led to an interpretation of the X-ray emission from this source in term of gas photoionized by the central AGN.
Recently, two papers concerning the study of soft X-ray emission from Mrk 573 with ACIS/\textit{Chandra} and RGS/\textit{XMM-Newton} data have been published \citep{2010MNRAS.405..553B,2010ApJ...723.1748G}; they both agree in interpreting this emission as dominated by two photoionized phases, with the contribution of a collisional phase.
{Making use of the archival \textit{Chandra} observation available in 2010, these authors found a clear correlations between the diffuse X-ray emission from Mrk 573, its ENLR as mapped by \textit{HST} data, and the triple radio source shown in the \textit{VLA} image, suggesting a physical interplay among these three components.}

In this paper we present a detailed imaging and spectroscopic study of new Mrk 573 ACIS/\textit{Chandra} data obtained as part of the CHandra survey of Extended Emission-line Regions in nearby Seyfert galaxies (CHEERS, \citealt{2010cxo..prop.3033W}). These new data, adding to the archival ones, reach an observation time of \(\sim 110\mbox{ ks}\). With this deep observation the unmatched \textit{Chandra} spatial resolution allows us to explore different regions of the Mrk 573 ENLR looking for a more comprehensive description of the underlying physical processes.

\section{OBSERVATIONS AND DATA REDUCTION}

\subsection{ACIS/\textit{Chandra} Data}

Mrk 573 was observed by \textit{Chandra} on 2006 November 18 for an exposure time of 40 ks (Obs. ID 07745, PI: Bianchi), and on 2010 September 16, 17 and 19, for an exposure time of 10, 53 and 17 ks respectively (Obs. IDs 12294, 13124, 13125); the last three observations were performed as part of the CHEERS survey. Level 2 event data were retrieved from the \textit{Chandra} Data Archive\footnote{\href{http://cda.harvard.edu/chaser}{http://cda.harvard.edu/chaser}} and reduced with the Chandra Interactive Analysis of Observations (CIAO; \citealt{2006SPIE.6270E..60F}) 4.3 software and the \textit{Chandra} Calibration Data Base (\textsc{caldb}) 4.4.3, adopting standard procedures. 

Time intervals of background flares exceeding \(3\sigma\) of the quiet level were excluded using the \textsc{lc\_sigma\_clip} task in source free regions of each observation; net exposure times are reported in Table \ref{obsprop}. The nucleus has no significant pile up, as measured by the CIAO \textsc{pileup\_map} tool.

We produced a merged image of the four observations, to take advantage of the longer exposure time and identify fainter signatures; to this end we used the \textsc{wavdetect} task to identify point sources in each observation, then we used the \textsc{reproject\_aspect} task to modify the aspect solution minimizing position differences between the sources found, and finally merged the images with the \textsc{merge\_all} script.

\subsection{Optical and Radio Data}

To compare the soft X-ray emission with optical and radio structures, we retrieved \textit{HST} and \textit{VLA} images of Mrk 573. In particular, we retrieved from the Hubble Legacy Archive\footnote{\href{http://hla.stsci.edu/hlaview.html}{http://hla.stsci.edu/hlaview.html}} the \textit{HST}/WFPC2 narrow band image with the FR533N filter tracking the [O \textsc{iii}]\(\uplambda 5007\) emission line, obtained as part of the GO program 6332 (PI: Wilson) on 1995 November 12
\citep{1998ApJ...502..199F,2003ApJS..148..327S}; this image was aligned to the X-ray images by matching the brightest pixel of the nuclear source in both images.
The [O \textsc{iii}] image (Figure \ref{deconv}, upper-right panel) shows the bright nucleus with two bright spots at \(\sim 1"\) and \(\sim 0.7"\) in the NW and SE directions, respectively; we also see two pairs of arcs in the two cones, the brighter inner arcs at \(\sim 2"\) and \(\sim 1.5"\) and the fainter outer arcs at \(\sim 4"\) and \(\sim 3"\) from the nucleus in the NW and SE directions, respectively (see also \citealt{1999MNRAS.309....1F,1999ApJ...525..685Q}).

We retrieved from the NASA/IPAC Extragalactic Database (NED\footnote{\href{http://ned.ipac.caltech.edu/}{http://ned.ipac.caltech.edu/}}) the VLA image of the \(6\mbox{ cm}\) (\(5 \mbox{ GHz}\)) observation performed in 1985 March 3 in A array configuration; this image clearly shows a central nucleus with a size \(\sim 1"\) and two side lobes of similar size aligned with the NLR (Figure \ref{deconv}, contours in lower-right panel, see also \citealt{1984ApJ...278..544U,1998ApJ...502..199F}).

\section{CHANDRA IMAGE ANALYSIS}

Imaging analysis was performed with the subpixel event repositioning (SER) procedure \citep{2003ApJ...590..586L} and without pixel randomization, to take advantage of the telescope dithering to allow subpixel binning of the images, using a pixel size smaller than the native one \(0.492"\) of the \textit{Chandra}/ACIS detector \citep[see, e.g.,][]{2004ApJ...615..161H,2007ApJ...657..145S,2010ApJ...708..171P,2011ApJ...729...75W}.

\subsection{Comparison with \textit{Chandra} PSF}

We performed Point Spread Function (PSF) simulations with Chandra Ray Tracer (ChaRT\footnote{\href{http://cxc.harvard.edu/chart/}{http://cxc.harvard.edu/chart/}}, \citealt{2003ASPC..295..477C}) taking into account the source spectrum, exposure time and off-axis angle; while the hard band images (\(2-10\mbox{ keV}\)) show a point-like structure consistent with the simulated PSF, the soft band images (\(0.3-2\mbox{ keV}\)) show extended morphology that closely resembles that of the NLR, as mapped by the [O \textsc{iii}] emission (see Figure \ref{deconv}).

\begin{figure}
\centering
\includegraphics[scale=0.40]{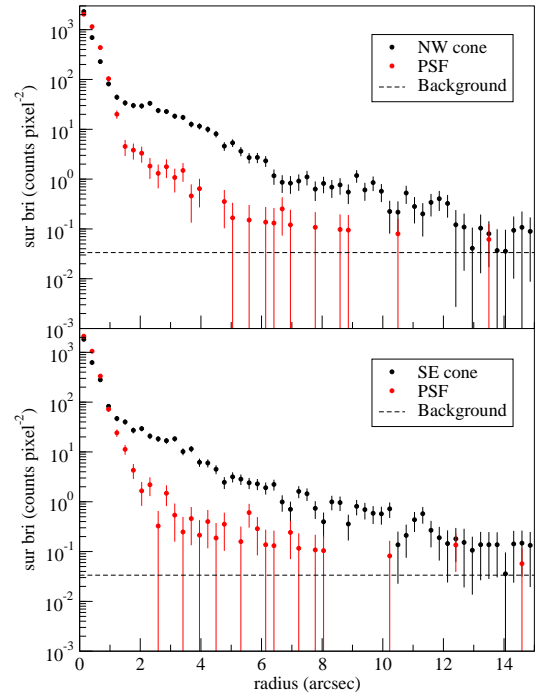}
\caption{Radial profiles of the ionizing cones (black dots) in comparison with the simulated PSF (red dots) in the \(0.3-2\mbox{ keV}\) range.}\label{coneprofiles}
\end{figure}

In Figure \ref{coneprofiles} we show radial profiles of the soft X-ray emission for the NW and SE direction  compared with the simulated PSF in the same band; while the nuclear emission in the inner \(\sim 1"\) is comparable with the PSF, the diffuse emission extends in both directions up to \(\sim 12"\).

\subsection{Image Deconvolution}\label{imagedeconv}

We applied to the merged images of the soft emission two different PSF-deconvolution algorithms, namely the Richardson-Lucy (R-L) \citep{1972JOSA...62...55R,1974AJ.....79..745L} and the Expectation through Markov Chain Monte Carlo (EMC2) \citep{2004ApJ...610.1213E,2005ApJ...623L.137K,2007ApJ...661.1048K}; while both methods show a similar extended morphology in the deconvolved images, the R-L algorithm yields a more grainy image reconstruction with respect to EMC2; the latter results are presented in Figure \ref{deconv}. In particular, in the upper panels we show a comparison between the reconstructed ACIS image with the subpixel binning (1/8 of the native pixel size, upper-left panel) and the [O \textsc{iii}] image (upper-right panel): the soft X-ray extended emission clearly shows structures in correspondence with the optical optical arcs 
\citep{1997ApJ...475..231H}; this correspondence is more clearly shown in the lower panels of the same figure, where [O \textsc{iii}] and radio contours are overlaid on the soft X-ray emission (lower-left and lower-right panel, respectively). 
While the radio emission appears more compact than the soft X-ray emission, the latter shows a  striking coincidence with the optical features and an interesting interplay between radio, optical and X-ray emissions, {as already suggested by \citet{2010MNRAS.405..553B} and \citet{2010ApJ...723.1748G} in their analysis of \textit{Chandra} obs. 07745.} In particular we see soft X-ray structures lying in coincidence or just in front of the inner [O \textsc{iii}] arcs, with the latter wrapping around the outer radio lobes.

\subsection{Diffuse emission}

\begin{figure}
\centering
\includegraphics[scale=0.45]{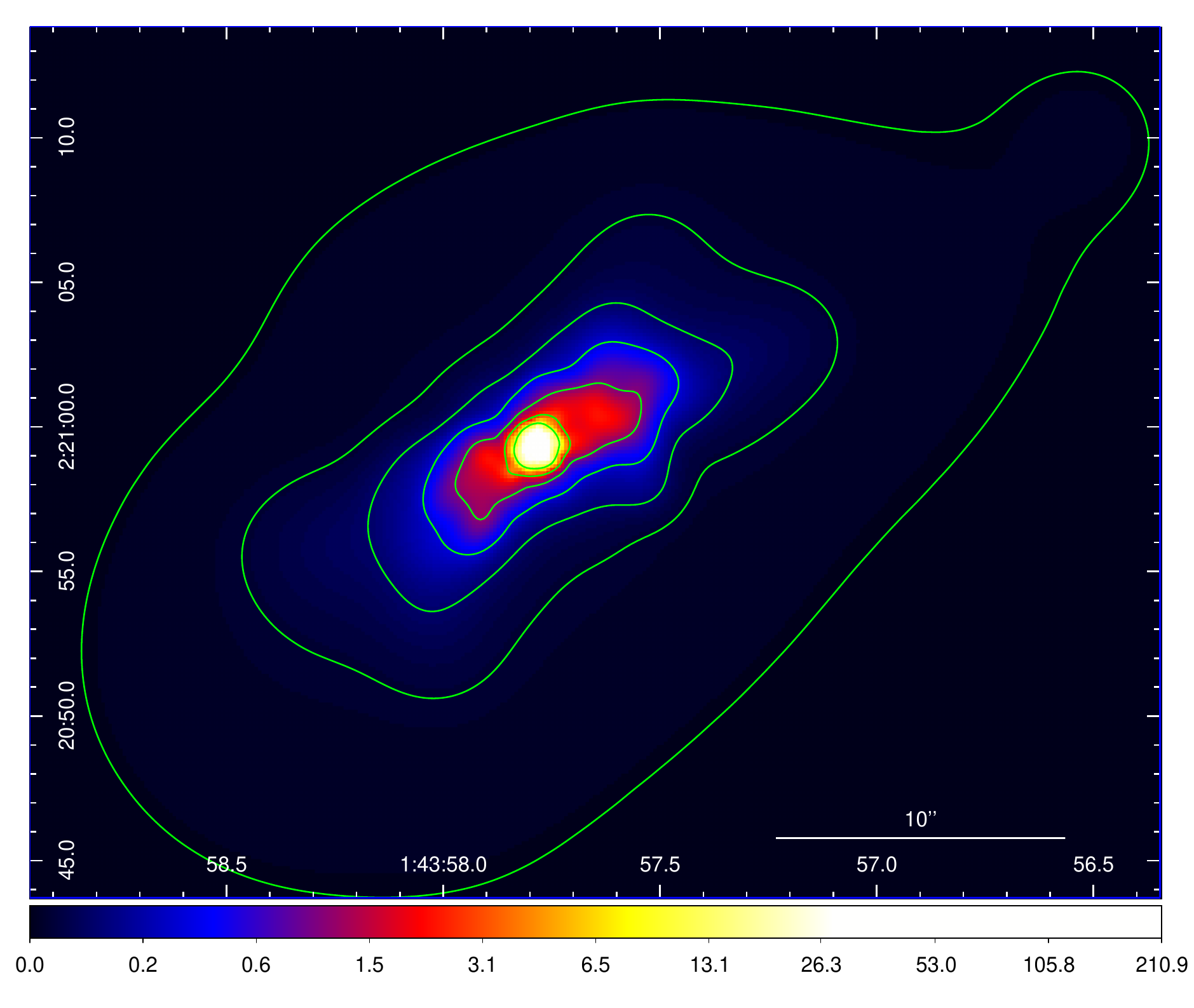}
\caption{Adaptive smoothing applied to subpixel binned ACIS-S image of the soft X-ray emission (0.3 - 2 keV); showing extended emission up to \(\sim 6\) kpc. Contours correspond to seven logarithmic intervals in the range of \(0.003-5\%\) with respect to the brightest pixel.}\label{csmooth}
\end{figure}

The deep exposure reached with CHEERS observations (\(\sim 110\mbox{ ks}\)) allows us to explore fainter features of the diffuse X-ray emission. To enhance these, we applied the adaptive smoothing procedure to the merged soft X-ray image using the \textsc{csmooth} tool \citep{2006MNRAS.368...65E}, with minimum and maximum significance S/N levels of \(2.5\) and \(3.5\), respectively; this smoothing procedure allows us to enhance fainter, extended features of the diffuse emission. The smoothed image is presented in Figure \ref{csmooth}; the bicone emission appears to extend to \(\sim 6\mbox{ kpc}\) in the NW direction and \(\sim 7\mbox{ kpc}\) in the SE direction, as also confirmed by the extended radial profiles presented in Figure \ref{extradprofile}.

\begin{figure}
\centering
\includegraphics[scale=0.45]{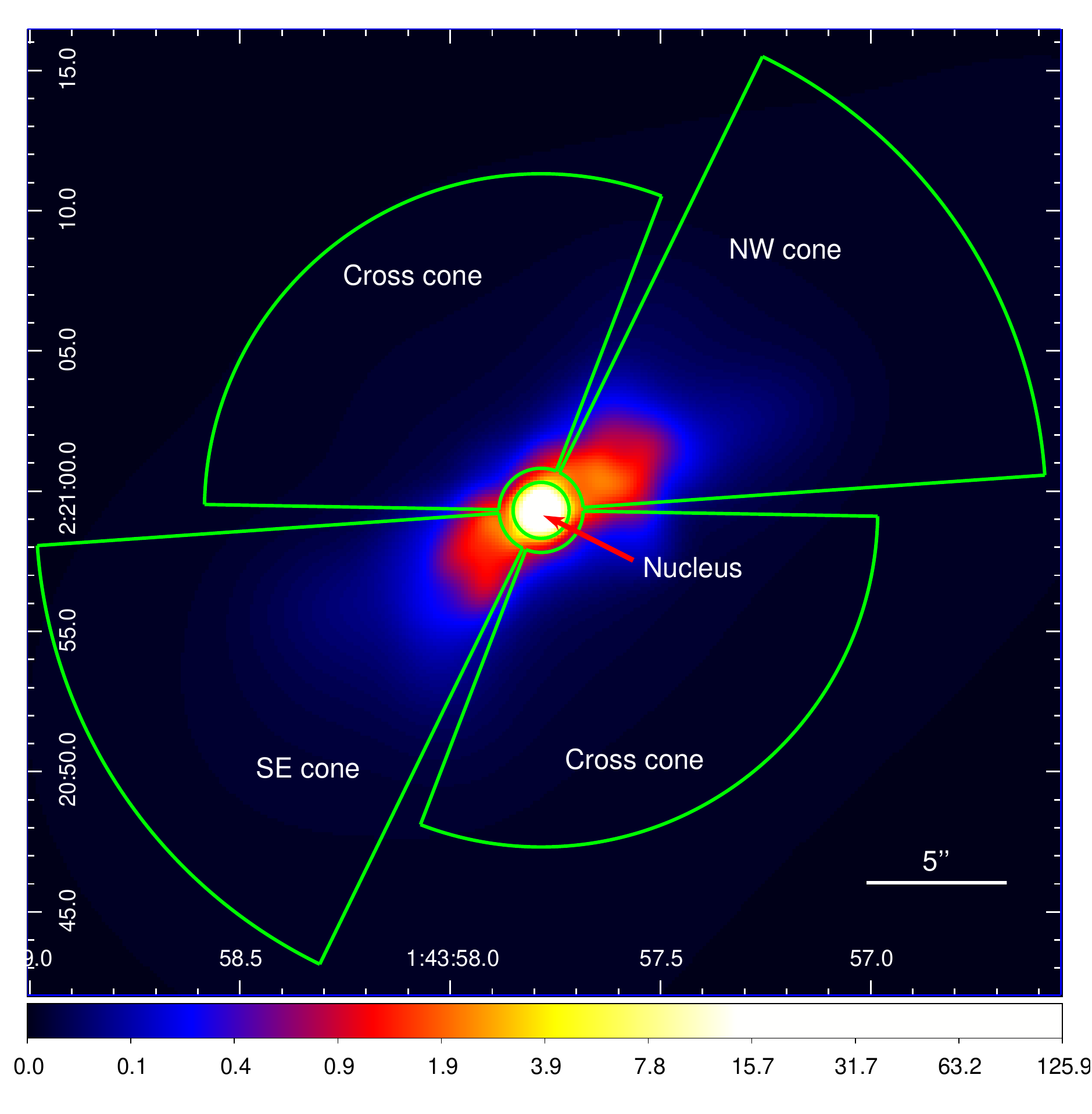}
\caption{Adaptive smoothing applied to subpixel binned ACIS-S image of the soft X-ray emission (0.3 - 2 keV) with overlaid the extraction regions considered in the text.}\label{torusregion}\medskip
\end{figure}

\begin{figure}
\centering
\includegraphics[scale=0.3]{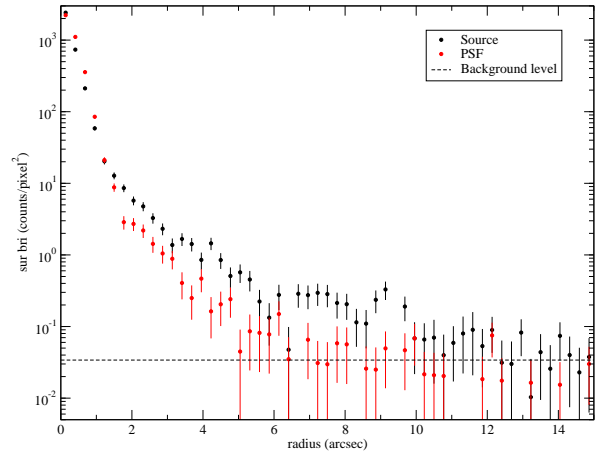}
\caption{Radial profiles of the cross cone region (black dots) in comparison with the simulated PSF (red dots) in the \(0.3-2\mbox{ keV}\) range.}\label{torusprofile}
\end{figure}

We then focused on the region perpendicular to the bicone that is supposed to be shielded by a dusty torus from seeing the central continuum in the AGN unified model; the region we studied is defined by a bicone with P.A. \(=34\degree\), a half opening angle of \(55\degree\), and inner and outer radii of \(1.5"\) and \(12"\), as shown in Figure \ref{torusregion}.
We found a significant (\({SNR}\approx 22\)) soft X-ray emission in this region, extending to \(\sim 9"\) as shown in Figure \ref{torusprofile}; we will discuss the emission from this region in detail in Sect. \ref{torusregion}.

The nuclear region has a luminosity in the 0.3 -2 keV band of \({24.16}_{-0.40}^{+0.34}\times{10}^{40}\mbox{ erg}\mbox{ s}^{-1}\), while the bi-cone region has a luminosity of \({3.26}_{-0.12}^{+0.16}\times{10}^{40}\mbox{ erg}\mbox{ s}^{-1}\) and \({2.97}_{-0.15}^{+0.16}\times{10}^{40}\mbox{ erg}\mbox{ s}^{-1}\) in the NW and in the SE direction, respectively. The cross-cone region, on the other hand, has a luminosity of \({1.60}_{-0.25}^{+0.54}\times{10}^{40}\mbox{ erg}\mbox{ s}^{-1}\).

\section{SPECTRAL ANALYSIS}

\begin{figure}
\centering
\includegraphics[scale=0.42]{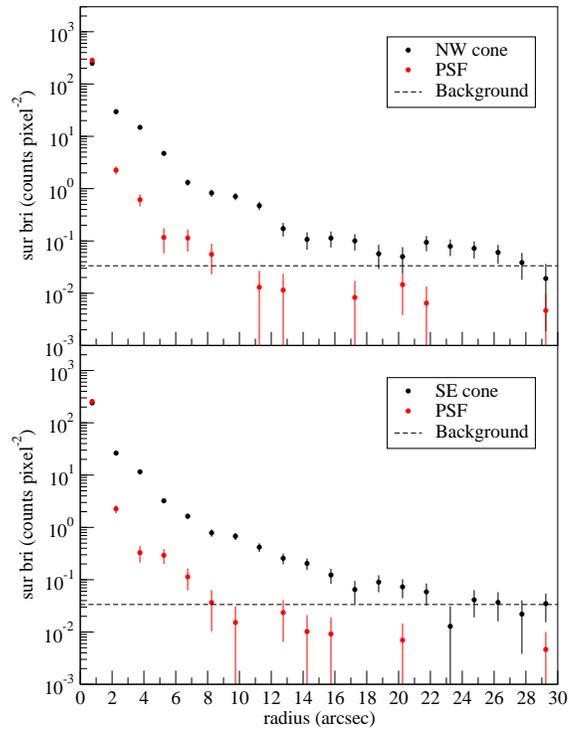}
\caption{Extended radial profiles of the NW (top panel) and SE (bottom panel) cones (black dots) in comparison with the simulated PSF (red dots) in the \(0.3-2\mbox{ keV}\) range.}\label{extradprofile}
\end{figure}

In order to study the X-ray properties of the diffuse emission we extracted spectra with the \textsc{CIAO specextract} task from three different regions centered at RA 01:43:57.78, DEC +02:20:59.32: the NW and SE cones region, both defined as cones with an inner radius of \(1.5"\) and a half opening angle of \(30\degree\), centered at P.A. \(=-56\degree\) and \(124\degree\) respectively; the nuclear region, defined as a circular region with a \(1"\) radius (see Figure \ref{torusregion}). Based on the radial profiles shown in Figure {\ref{extradprofile}} we decided to extract cones spectra out to {\(18"\)}, where the emission of the cones reaches the background level. For spectra extracted from the nuclear region we applied the point-source aperture correction to \textsc{specextract} task.

To make use of the \(\chi^2\) fit statistic we binned the spectra to obtain a minimum of \(20\) counts per bin using the \textsc{specextract} task; in the following, errors correspond to the \(1\)-\(\sigma\) confidence level for one interesting parameter (\(\Delta\chi^2 = 1\)). In all the spectral fits we included photo-electric absorption by the Galactic column density along the line of sight \(N_H = 2.52\times {10}^{20}\mbox{ cm}^{-2}\) \citep{2005A&A...440..775K}. We also tried to evaluate intrinsic absorption, however the data do not show a significant intrinsic column density.

\subsection{Cone spectra}\label{conespectra}

The soft X-ray emission from Seyfert galaxies can be effectively described in terms of several emission lines with a small power-law continuum contribution. We therefore performed the spectral analysis of emission of Mrk 573 with the \textsc{XSPEC} software (ver. 12.7.0\footnote{\href{http://heasarc.nasa.gov/xanadu/xspec/}{http://heasarc.nasa.gov/xanadu/xspec/}}, \citealt{1996ASPC..101...17A}) fitting the soft (\(0.3-2\mbox{ keV}\)) spectra of different observations in the cones region using a phenomenological model constituted by a power-law photon index fixed to \(\Gamma = 1.8\) (a typical value for Seyfert galaxies, see \citealt{2009A&A...495..421B}), plus several red-shifted emission lines adopting as a reference the lines measured in the 2006 November 18 observation (Obs. ID 07745) by \citet{2010ApJ...723.1748G} in RGS/\textit{XMM-Newton} high resolution spectra; due to low counts in these regions, we were able to obtain well constrained fits only for longest observations, namely Obs. IDs 07745 and 13124. To obtain better statistics, we also performed simultaneous fits on data from CHEERS observations (Obs. IDs 12294, 13124 and 13125) as well as for all observations (CHEERS obs. + Obs. ID 07745). To avoid contamination from the nuclear continuum we evaluated the contribution from the PSF wings in the cone regions as \(\sim 0.8\%\), and subtracted this contribution from the cones spectra. Fit results are in general agreement with previous results by \citet{2010MNRAS.405..553B} and \citet{2010ApJ...723.1748G}, and are presented in the Appendix in Table \ref{conelines} and Figure \ref{coneslinespectra}.

The strongest lines of the cones spectra are C \textsc{vi}, O \textsc{vii} triplet, O \textsc{viii}, O \textsc{vii} RRC and Ne \textsc{ix} triplet. Most lines have similar flux in both cones, except for O \textsc{viii} RRC which appears stronger in the NW cone, as already reported by \citet{2010ApJ...723.1748G}, and the Ne \textsc{ix} triplet which instead is stronger in the SE cone.

The high resolution RGS/\textit{XMM-Newton} spectroscopic analyses of \citet{2010MNRAS.405..553B} and \citet{2010ApJ...723.1748G} suggest that the soft X-ray spectrum of Mrk 573 originates from photoionized and photoexcited plasma with a contribution from collisionally excited plasma. In order to evaluate the contribution of different plasma phases to the total emission we performed spectral fitting of the two cones spectra separately, making use of self-consistent photoionization models. For this purpose we produced \textsc{xspec} grid models with the \textsc{Cloudy}\footnote{\href{http://www.nublado.org/}{http://www.nublado.org/}} c08.01 package, described by \citet{1998PASP..110..761F}. We assumed the ionization source to be an AGN continuum (with a “big bump” temperature \(T = {10}^6 \mbox{ K}\), a X-ray to UV ratio \(\alpha_{ox} = -1.30\) and an X-ray power-law component of spectral energy index \(\alpha=-0.8\)) illuminating a cloud with plane-parallel geometry and constant electron density \(n_e={10}^5\mbox{ cm}^{-3}\). Note that the fits are expected to be quite insensitive to \(n_e\) {in this density regime} \citep{2000A&AS..143..495P}. The grid of models so obtained are parametrized in terms of the ionization parameter \(U\) (varying in the range \(\log{U}=[-3.00:2.00]\) in steps of \(0.25\)) and the hydrogen column density \(N_H\) (expressed in \(\mbox{ cm}^{-2}\) varying in the range \(\log{N_H}=[19.0:23.5]\) in steps of \(0.1\)), taking into account only the reflected spectrum from the illuminated face of the cloud \citep{2010MNRAS.405..553B,2011A&A...526A..36M}.

The results of the fitting procedure are shown in the Appendix in Table \ref{conephoto} and Figure \ref{conephotospectra}. Poor fits are obtained with a single phase ionized plasma; the best fit models require two photoionized phases.
Due to low counts, the hydrogen column densities of both phases are unconstrained, so we fixed them to their best fit value of \(\log{N_{H}}=20\).

The simultaneous fit of all observations give for the NW cone \(\log{U_1}= 0.9\pm 0.3\) and  \(\log{U_2}= -0.5\pm 0.3\) for the higher and lower ionization phase, respectively; for the SE cone, on the other hand, we have \(\log{U_1}= 0.7\pm 0.2\) and \(\log{U_2}= -0.8\pm 0.2\). Comparable values are obtained for the simultaneous fit of all the CHEERS observation, as well as for obs. ID 13124 and 07745, even if for the latter the ionization parameters \(\log{U_2}\) have to be fixed to their best fit values \(-0.5\) and \(-1\) for the NW and SE cone, respectively. In all observations the two gas phases show similar fluxes \(F = (0.3\pm 0.1)\times{10}^{-13}\mbox{ erg}\mbox{ cm}^{-2}\mbox{ s}^{-1}\) in both cones.

In almost all observations the best fit models require a thermal component (included with an \textsc{apec} \textsc{xspec} model). The exception is observation 07745 on the SE cone, but this is probably due to the low statistics, this being the region with the lowest counts. All other fits require this thermal component with temperature {\(kT = 0.8 \pm 0.1\mbox{ keV}\) in the NW cone and \(kT = 1.0 \pm 0.2\mbox{ keV}\) in the SE cone}. 

We can therefore conclude that a two-phases photoionized plasma is diffused over the two cones, with the presence of a thermal component.

\subsection{SPATIAL DISTRIBUTION OF SPECTRAL FEATURES}\label{spatial}

Summarizing the results obtained in the spectral fits presented in Sect. \ref{conespectra} the two cones are characterized by mildly photoionized plasma (\(\log{U_1}=0.9\pm 0.3\) in the NW cone and \(\log{U_1}=0.7\pm 0.2\) in the SE cone) and the presence of collisionally ionized plasma {(at a temperature \(kT = 0.8 \pm 0.1\mbox{ keV}\) in the NW cone and \(kT = 1.0 \pm 0.2\mbox{ keV}\) in the SE cone}).

As recalled in Sect. \ref{imagedeconv} CHEERS observations of Mrk 573 reveal a detailed structure of the diffuse emission that is morphologically coincident with the ENLR, as mapped by [O \textsc{iii}] and suggest an interplay between radio ejecta, the optical and the soft X-ray emissions (see Figure \ref{deconv}); the inner optical arcs appear wrapped around the radio knots, as resulting from shock interactions between the radio jets and the galactic plane, and the X-ray emission is enhanced in front of these knots, suggesting the presence of collisionally ionized gas, a picture analogous to what is observed in NGC 4151 \citep{2011ApJ...736...62W}.

\begin{figure}
\centering
\includegraphics[scale=0.9]{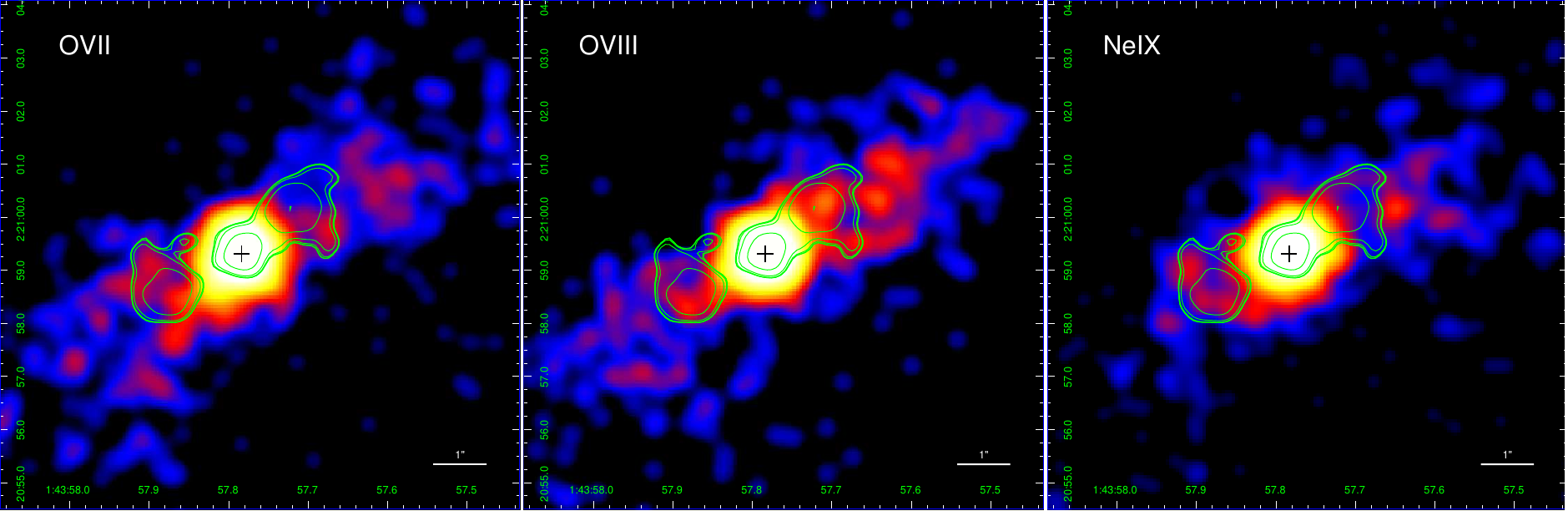}
\includegraphics[scale=0.9]{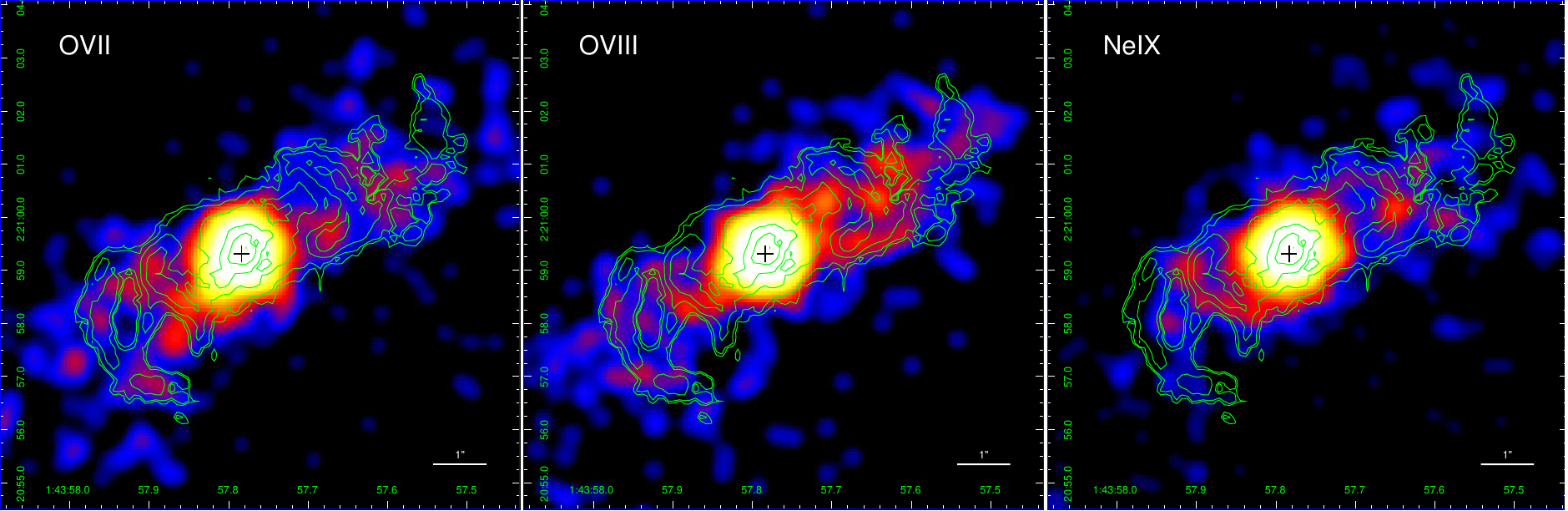}
\caption{Emission-line maps of the central 3.5 kpc of Mrk 573: from left to right are shown the O \textsc{vii} triplet (0.53 - 0.63 keV), O \textsc{viii} Ly\(\alpha\) (0.63 -0.68 keV) and Ne \textsc{ix} triplet (0.90 - 0.95 keV) maps, with overlaid the 6 cm radio contours (upper panel) and [O \textsc{iii}] contours (lower panel). The images are rebinned to 1/8 of the native pixel size and smoothed with a 3X3 FWHM gaussian filter. The nucleus position is marked with the black cross.}\label{lines}
\end{figure}

To further investigate these features, we produced maps of the X-ray emission in three energy bands corresponding to some of strongest emission lines, namely the blended O \textsc{vii} triplet (0.53 - 0.63 keV), O \textsc{viii} Ly\(\alpha\) (0.63-0.68 keV) and Ne \textsc{ix} triplet (0.90 - 0.95 keV), and present them in Figure \ref{lines}.

As already observed by \citet{2010MNRAS.405..553B} the O \textsc{vii} emission is slightly more extended in the SE cone with respect to the O \textsc{viii} Ly\(\alpha\), while in the NW cone we observe an opposite behavior.

The Ne \textsc{ix} map shows spots of enhanced emission in front of the radio lobes; these spots are also present in the ratio map of line emission shown in Figure \ref{ratios}, where we present ratios of Ne \textsc{ix} to O \textsc{vii} and to O \textsc{viii}, tracing the higher ionization gas; these maps show a nearly symmetrical structure along the two cones, characterized by regions of high ionization between the nucleus and the inner optical arcs (regions 1 and 2) and regions of increased ionization lying in front of the radio knots at the interface between the inner and outer optical arcs (regions 3 and 4).

These regions of increased ionization are also shown in Figure \ref{surbri}, where we plot variations of the ratio of [O \textsc{iii}] to 0.5 - 2 keV flux along with the distance from the nucleus, where an average conversion factor of \(8.7 \times {10}^{-12}\mbox{ erg}\mbox{ cm}^{-2}\mbox{ s}^{-1}\mbox{ counts}^{-1}\mbox{ s}\), obtained by the various spectral fits we performed, is used to convert net counts in the 0.5 - 2 keV band into flux. 
The ratio of [O \textsc{iii}] to soft X-ray flux for a single photoionized medium is expected to have an approximatively power law dependence on the radius, depending on the radial density profile \citep{2006A&A...448..499B}. The trends shown in Figure \ref{surbri} can be roughly sketched with the inner \(\sim 1"\) region corresponding to the nuclear highly photoionized phase with \(\log{U}=1\), and two outer regions corresponding to the mildly photoionized cones with \(\log{U}=0.8\) in the NW direction and \(\log{U}=0.6\) in the SE direction, respectively.

As the radio ejecta interact with the galactic plane, the regions of increased ionization shown in Figure \ref{ratios} can be interpreted as two different shock regions in which the radio ejecta interact with different regions of the ISM, with decreasing velocities as the jet slows down moving away from the nucleus; in particular, regions 3 and 4 feature significant dips in the [O \textsc{iii}] to soft X-ray flux ratio at the interface between the optical arcs (see also \citealt{2010ApJ...723.1748G}), characterized by enhanced X-ray emission due to shock heating in addition to photoionization, as already observed in NGC 4151\citep{2011ApJ...736...62W}.

In order to investigate the nature of the different plasma phases in these regions of enhanced ionization (shown in Figure \ref{ratios}) we extracted here soft X-ray spectra and fitted them adopting the photionization model described above, taking into account the nuclear contribution to these regions. In order to have enough counts and allow acceptable fits we extracted spectra from all observations and combined them with the \textsc{specextract} tool. Due to the low counts in regions 3 and 4, we 
combined their spectra. Fits results are presented in Table \ref{otheregions} and Figure \ref{otheregionspectra}.

In regions 1 and 2 the best fit model requires a highly ionized phase with \(\log{U_1} = 0.9\pm 0.1\) and \(N_{H1} = 21.5\pm 0.6\), and a low ionized phase with \(\log{U_2}= -1.2\pm 0.6\) in region 1 and \(\log{U_2}= -0.8\pm 0.3\) in region 2; due to low statistics the hydrogen column densities for the low ionization phase were fixed to their best fit values of \(\log{N_{H2}} = 20\) and \(\log{N_{H2}} = 21\) in region 1 and 2, respectively. The spectrum extracted from regions 3+4 is best fitted by a highly ionized phase with \(\log{U_1}= 1.3\pm 0.3\) and a low ionized phase with \(\log{U_2}= -0.9\pm 0.2\); in this case both hydrogen column densities were fixed to their best fit value \(\log{N_H} = 20\). For all these regions of enhanced ionization a thermal component is statistically required; in particular, region 1 needs a temperature \(kT = 1.1\pm 0.2\mbox{ keV}\), while in region 2 we have \(kT = 1.3\pm 0.2\mbox{ keV}\). The outer regions 3+4 feature a lower temperature plasma with \(kT = 0.8\pm 0.1\mbox{ keV}\), which we interpret as the origin of the enhanced X-ray emission.

To further check the location of the collisionally ionized phases we performed an analogous analysis on the [O \textsc{iii}] outer arcs. A good fit is obtained with two photoionized phases, while the inclusion of a thermal component is not statistically required, as the fitting procedure 
only yields an upper limit on its flux \(<0.02\times{10}^{-13}\mbox{ erg}\mbox{ cm}^{-2}\mbox{ s}^{-1}\) (see Table \ref{otheregions} and Figure \ref{otheregionspectra}). This is not compatible with the fluxes found in the regions of enhanced ionization (see Table \ref{otheregions}). We conclude that the thermal component found by \citet{2010MNRAS.405..553B} and \citet{2010ApJ...723.1748G} is actually composed of two physically separated plasmas, with a higher temperature phase lying between the nucleus and the inner optical arcs, and a lower temperature phase located in front of the radio knots, at the interface between the inner and outer optical arcs.

\section{DISCUSSION}\label{discussion}

CHEERS observations of Mrk 573 show soft X-ray emission characterized by a bright nucleus and a diffuse component extending along the ionizing bicone up to \(\sim 7\mbox{ kpc}\), doubling the size reported in previous works \citep{2010ApJ...723.1748G}. The spectral analysis of the diffuse emission points to the presence of a thermal component that can be interpreted as collisionally ionized gas related to radio ejecta; moreover, significant extended emission up to \(\sim 3\mbox{ kpc}\) is observed in the cross-cone direction where the obscuring torus is expected to shield the nuclear continuum according to the AGN unified model.
In this section we will discuss these two aspects of the diffuse emission, as well as the spectral properties of the nuclear emission.

\subsection{Energetic content of the collisionally ionized gas}\label{thermalgas}

Previous work on Mrk 573 \citep{2010ApJ...723.1748G,2010MNRAS.405..553B} first suggested that the soft X-ray diffuse emission of Mrk 573 can be interpreted as originating from mildly photoionized plasma with the presence of a collisionally ionized phase. With new, deep CHEERS observations we are able to conclude than the collisionally ionized phase (Sections \ref{nuclearspectra} and \ref{conespectra}) is 
composed of two physically separated plasmas, with a hotter plasma between the nucleus and the inner optical arcs, and a lower temperature plasma in front of the radio knots, at the interface between the inner and outer optical arcs.

The collisionally ionized phases can be interpreted as the result of shock interaction between the radio knots and the galactic plane, because this is the expected location of the interaction between the ionizing cones and the galaxy plane (see Figure \ref{deconv}).

\begin{table}[b]
\centering
\begin{threeparttable}
\caption{Physical parameters of the collisionally ionized gas found in regions of increased ionization represented in Figure \ref{ratios}.}\label{gasparameters}
\begin{tabular}{l|c|c|c}
\hline
\hline
Region & 1 & 2 & 3+4 \\
\hline 
\(n_e\) (\(\mbox{cm}^{-3}\)) & \(0.48\pm 0.10\) & \(0.52\pm 0.12\) & \(0.14 \pm 0.02\) \\
\(p_{{th}}\) (\({10}^{-9}\mbox{ dyne}\mbox{ cm}^{-2}\)) & \(1.78\pm 0.41\) & \(2.34\pm 0.61\) & \(0.38\pm 0.08\) \\
\(E_{{th}}\) (\({10}^{53}\mbox{ erg}\)) & \(0.79\pm 0.18\) & \(1.04\pm 0.26\) & \(2.73\pm 0.57\) \\
\(t_{{cool}}\) (\({10}^7\) yr) & \(0.76\pm 0.32\) & \(1.24\pm 0.64\) & \(1.63\pm 0.53\) \\
\(c_s\) (\(\mbox{km}\mbox{ s}^{-1}\)) & \(668\pm 103\) & \(737\pm 128\) & \(565 \pm 73\) \\
\(\varv_{{sh}}\) (\(\mbox{km}\mbox{ s}^{-1}\)) & \(905\pm 46\) & \(999 \pm 59\) & \(765\pm 53\) \\
\(t_{{cross}}\) (\({10}^5\) yr) & \(3.66\pm 0.57\) & \(3.32\pm 0.58\) & \(4.33\pm 0.56\) \\
\(E_{{th}} / t_{{cross}}\) (\({10}^{41}\mbox{erg}\mbox{ s}^{-1}\)) & \(0.69\pm 0.19\) & \(1.00\pm 0.30\) & \(2.00\pm 0.49\) \\
\hline
\hline
\end{tabular}
\end{threeparttable}
\end{table}

The electron density and thermal pressure of the collisionally ionized gas can be estimated through the normalization of the \textsc{apec} component, that is, its emission measure
\[
EM={10}^{-14}\frac{n_e\, n_H\, V\, \eta}{4\pi {\left[{D_A (1+z)}\right]}^2}\, ,
\]
where \(n_e\) and \(n_H\) are the average electron and hydrogen density, \(V\) is the emitting region volume, \(\eta\) is the filling factor and \(D_A\) is the angular diameter distance to the source. The gas parameters were obtained assuming \(n_e= 1.2\, n_H\), a filling factor \(\eta\approx 1\) 
and a spherical shape for the emitting regions. All parameters are listed in Table \ref{gasparameters}. As shown, the electron density, the thermal pressure and the temperature decrease from the inner (1 or 2) to the outer shock regions (3+4). Considering an average electron density over all regions of increased ionization \(\left<{n_{e}}\right>=0.18\mbox{ cm}^{-3}\) and a gas mass-averaged temperature \(\left<{T}\right>=0.88\mbox{ keV}\), we can evaluate an average thermal pressure \(\left<{p_{{th}}}\right> = 5.7 \times {10}^{-10}\mbox{ dyne}\mbox{ cm}^{-2}\). If the collisionally ionized gas is assumed to be diffuse over the entire cone region a thermal pressure \(1.5 \times {10}^{-9}\mbox{ dyne}\mbox{ cm}^{-2}\) is found, in agreement with the result found by \citet{2010MNRAS.405..553B}. 

To compare the thermal pressure with the pressure of radio jets we assume energy equipartition between particles and magnetic field in the jet, which provide a lower limit to the jet pressure \citep[see, e. g.,][]{2004ApJ...612..729H}. We model the two radio knots as spheres with a radius \(\sim 200\mbox{ pc}\) filled with electrons having a power law energy distribution with an index \(r=2.4\) (the average of the indices measured by \citealt{1998ApJ...502..199F}), and Lorentz factors \(2<\gamma<{10}^5\). Assuming that the 6 cm luminosities reported by \citet{2010MNRAS.405..553B} are due to synchrotron emission from this electron population we obtain values of the magnetic field of \(6.6\times{10}^{-9}\mbox{ G}\) and \(8.0\times{10}^{-9}\mbox{ G}\) for the NW and SE lobe, respectively, and jet pressures due to electrons energy density of \(3\times{10}^{-11}\mbox{ dyne}\mbox{ cm}^{-2}\) and \(4\times{10}^{-11}\mbox{ dyne}\mbox{ cm}^{-2}\) for the NW and SE lobe, respectively. These are about 10 times smaller than the thermal pressure in regions 3+4, \(\sim\) 60 times smaller than the thermal pressure in regions 1 and 2, and \(\sim\) 20 times smaller than the average thermal pressure evaluated in all regions of increased ionization. Therefore the jet pressure due to electrons is not sufficient to sustain shock heating of the hot gas \citep{2010MNRAS.405..553B}.

However, if we consider jet pressures due to the total energy density, \(7/3 U_B\) (where \(U_B\) is the magnetic field energy density), we obtain \(1.5\times{10}^{-10}\mbox{ dyne}\mbox{ cm}^{-2}\) and \(2.2\times{10}^{-10}\mbox{ dyne}\mbox{ cm}^{-2}\) for the NW and SE lobe, comparable with both the average thermal pressures \(\left<{p_{{th}}}\right>\) and the thermal pressure of the lower temperature gas found at the interface between the optical arcs (region 3+4). This suggests that shocks of radio ejecta with ISM are likely to be responsible for the gas heating we have located in front of the radio lobes. Nevertheless, due to large uncertainties in these estimates, a strong conclusion cannot be drawn.

The thermal energy associated with the collisionally ionized gas in the regions discussed above is \(E_{{th}(1)} = 7.9\times {10}^{53}\mbox{ erg}\), \(E_{{th}(2)} = 1.0\times {10}^{54}\mbox{ erg}\) \(E_{{th}(3+4)} = 2.7\times {10}^{54}\mbox{ erg}\) for regions 1, 2 and 3+4, respectively; the corresponding cooling times \(t_{{cool}} = E_{{th}} / L_C\) (where \(L_C\) is the luminosity of the thermal component) are \(\sim {10}^7\mbox{ yr}\). In these regions the local sound speed \(c_s\) is comparable with the shock velocity estimated assuming strong shock conditions with \(\varv_{{sh}}\approx 100\, {\left({kT/0.013\mbox{ keV}}\right)}^{1/2}\mbox{ km}\mbox{ s}^{-1}\) \citep{2002ApJ...576L.149R}, resulting in crossing times of the radio knots \(t_{{cross}}\sim{10}^5\mbox{ yr}\), about \(30\) times shorter than the cooling time. Thus no extra heating sources are required.

A lower limit on the fraction of the kinematic power converted to heat in the ISM can be estimated as \(L_K\gtrsim E_{{th}} / t_{{cross}}\), to yield \(L_{K1} \gtrsim 6.9\times{10}^{40}\mbox{ erg}\mbox{ s}^{-1}\), \(L_{K2} \gtrsim 1.0\times{10}^{41}\mbox{ erg}\mbox{ s}^{-1}\) and \(L_{K(3+4)} \gtrsim 2.0\times{10}^{41}\mbox{ erg}\mbox{ s}^{-1}\) for regions 1, 2 and 3+4, respectively.
This can be compared with the jet total power, that can be evaluated using the \(P_{{jet}} - P_{{radio}}\) relation in \citet{2010ApJ...720.1066C} and the \(1.4\mbox{ GHz}\) radio luminosity \(\nu L_\nu = 4.1\times{10}^{37}\mbox{ erg}\mbox{ s}^{-1}\) \citep{1998AJ....116.2682C}, to obtain \(P_{{jet}}\approx 1.3\times{10}^{42}\mbox{ erg}\mbox{ s}^{-1}\). This yield that \(\gtrsim 5\%\) of the jet power is deposited into region 1, \(\gtrsim 7\%\) is deposited in the ISM into region 2 and \(\gtrsim 15\%\) is deposited into region 3+4, so that a total fraction of \(\gtrsim 28\%\) of the jet power is deposited in the ISM, a fraction much larger than the \(0.1\%\) estimated in NGC 4151 \citep{2011ApJ...742...23W}.
{Considering the ANG bolometric luminosity evaluated from the 2-10 keV emission (see Sect. \ref{nuclearspectra}) \(L_{{bol}}=(4.4 \div 7.7)\times{10}^{44}\mbox{ erg}\mbox{ s}^{-1}\)
we can conclude that a fraction \(L_K/L_{{bol}}>0.05\%\)} of the available accretion power is thermally
coupled to the host ISM. This is substantially lower than the fraction of the accretion power of the black hole (\(\sim 5\%-100\%\) of \(L_{{bol}}\) ) that the majority of quasar feedback models require to be deposited into the host ISM \citep{2009AIPC.1201...33M}, but is somewhat compatible with a two-stage feedback
model proposed by \citet{2010MNRAS.401....7H}, which requires only \(0.5\%\) of the radiated
energy to drive the mass outflow.

\subsection{Nuclear properties}\label{nuclearspectra}

In this section we focus on the nuclear emission of Mrk 573. The region used for the spectral extraction, shown in Figure \ref{torusregion}, is a circle centered in RA 01:43:57.78, DEC +02:20:59.32 with a \(1"\) radius.

To evaluate the emission lines fluxes we adopted the same phenomenological model used in Sect. \ref{conespectra}; these results also confirm results by \citet{2010MNRAS.405..553B} and \citet{2010ApJ...723.1748G}, and are presented in the Appendix in Table \ref{nucleuslines} and Figure \ref{nucleuslinespectra}.

\begin{figure}
\centering
\includegraphics[scale=0.4]{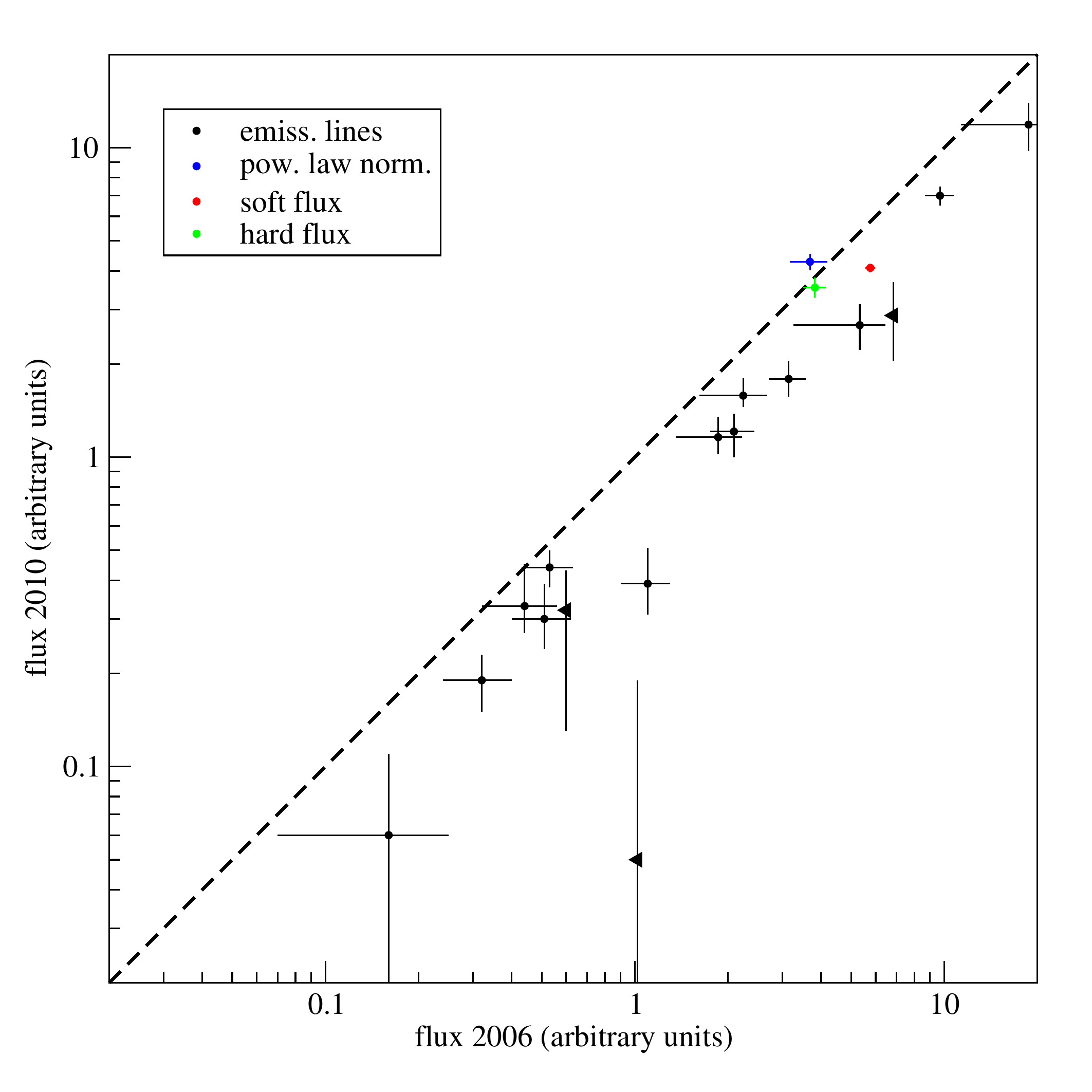}
\caption{Variations of the nuclear spectral components between 2006 obs. 07745 and 2010 CHEERS obs. Black points represent emission line fluxes (triangles indicate upper limits), the blue point represents the power-law normalization, the red point represents the soft X-ray flux and the green point represents the hard X-ray flux. Emission line fluxes and the power-law normalization are expressed in units of \({10}^{-5}\mbox{ photons} \mbox{ cm}^{-2} \mbox{ s}^{-1}\) and X-ray fluxes are expressed in units of \({10}^{-13}\mbox{ erg}\mbox{ cm}^{-2}\mbox{ s}^{-1}\).}\label{lines_fluxes}
\end{figure}

The strongest lines of the nuclear spectra are C {\scriptsize V} He\(\gamma\), 
C {\scriptsize VI} Ly\(\beta\), N {\scriptsize VII} Ly\(\alpha\), O {\scriptsize VII} triplet, O {\scriptsize VIII} Ly\(\alpha\), O {\scriptsize VII} RRC, Fe {\scriptsize XVII} and Ne {\scriptsize IX} triplet. Significant variations are observed between observations. In particular, observed continuum flux in the \(0.3-2\mbox{ keV}\) band \(f_{SX}\) is higher by a factor \(1.3\times {10}^{-13} \mbox{ erg}\mbox{ cm}^{-2}\mbox{ s}^{-1}\) \((4\sigma)\) in OBS. 07745 [\(f_{SX}=(5.4\pm 0.3)\times {10}^{-13} \mbox{ erg}\mbox{ cm}^{-2}\mbox{ s}^{-1}\)] than in the 4 years later CHEERS observations [\(f_{SX}=(4.1\pm 0.1)\times {10}^{-13} \mbox{ erg}\mbox{ cm}^{-2}\mbox{ s}^{-1}\)]. These variations are reflected in the poor fit {obtained with} merged observation (\(\chi^2=1.63\)). To show the observed nuclear flux variations we plot in Fig. \ref{lines_fluxes} fluxes of the different spectral components used in our phenomenological model. We clearly see that the nuclear flux decrease is marked by a significant decrease in several emission lines: O \textsc{vii} triplet, O \textsc{viii} Ly\(\alpha\), O \textsc{vii} RCC and Ne \textsc{x} Ly\(\alpha\), at \(3\sigma\).

We then fit the nuclear spectra with the photoionization model described in Sect. \ref{conespectra}. Since the nuclear emission significantly extends beyond the soft band, we performed nuclear spectral fits over the full \(0.3-10\mbox{ keV}\) energy range, adopting, in addition to the photoionization model: a pure neutral reflection component with the \textsc{pexrav} \textsc{xspec} model with the photon index fixed to \(1.8\), \(\cos{i}\) fixed to \(0.45\) and Si K\(\alpha\), S K\(\alpha\) and Fe K\(\alpha\) lines \citep[see][]{2010MNRAS.405..553B}. The results of the fitting procedures are presented in the Appendix in Table \ref{nucleusphoto} and Figure \ref{nucleusphotospectra}.

As for the cones, poor fits are obtained with a single phase ionized plasma; the best fit models require two photoionized phases, with higher and lower values of the ionization parameter.

\begin{figure}
\centering
\includegraphics[scale=0.9]{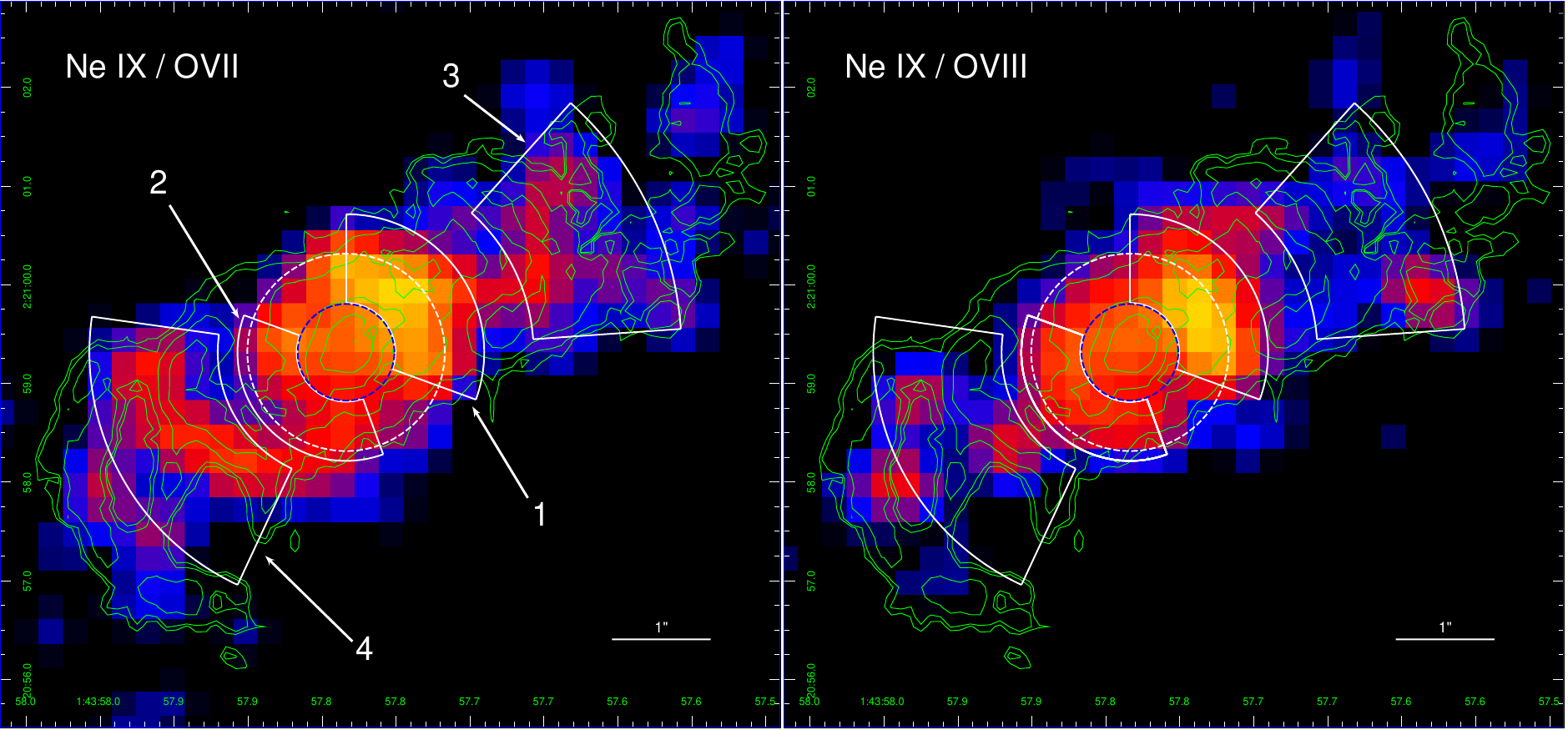}
\caption{Ratio Ne \textsc{ix} / O \textsc{vii} maps (left panel) and Ne \textsc{ix} / O \textsc{viii} (right panel) with overlaid [O \textsc{iii}] (subpixel binning 1/2 of the native pixel) smoothed with a 3X3 FWHM gaussian filter. The high ionization regions used for spectral extraction are shown with white full lines. The with dashed circle represents the 1" radius region used for the nuclear spectrum extraction, while the blue dashed circle represents the inner 0.5" radius region used for restricted nuclear fitting discussed in Sect \ref{conespectra}.}\label{ratios}
\end{figure}

\begin{figure}
\centering
\includegraphics[scale=0.3]{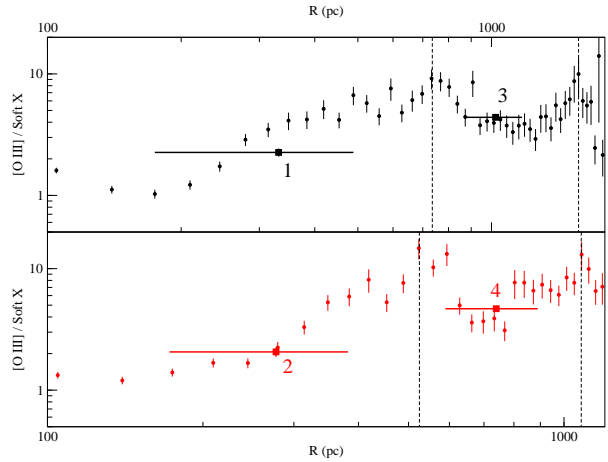}
\caption{Variation of the [O \textsc{iii}] to soft X-ray flux ratio, in the NW (upper panel) and SE (lower panel) cone as a function of the distance from the nucleus. The vertical dashed lines represent the location of the inner and outer optical arcs, and the squares represent the regions of increased ionization shown in Figure \ref{ratios}.}\label{surbri}
\end{figure}

\begin{figure}
\centering
\includegraphics[scale=0.25]{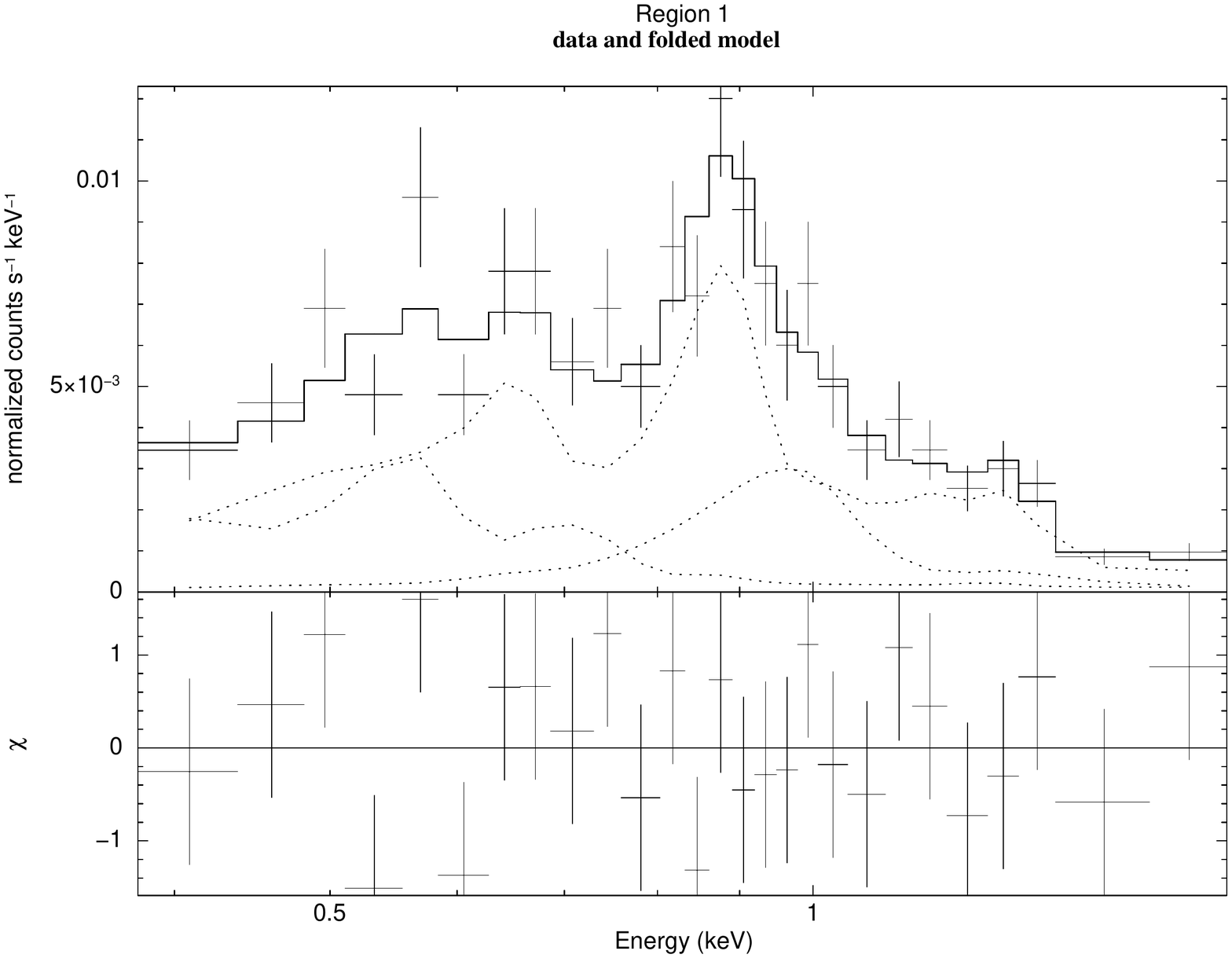}
\includegraphics[scale=0.25]{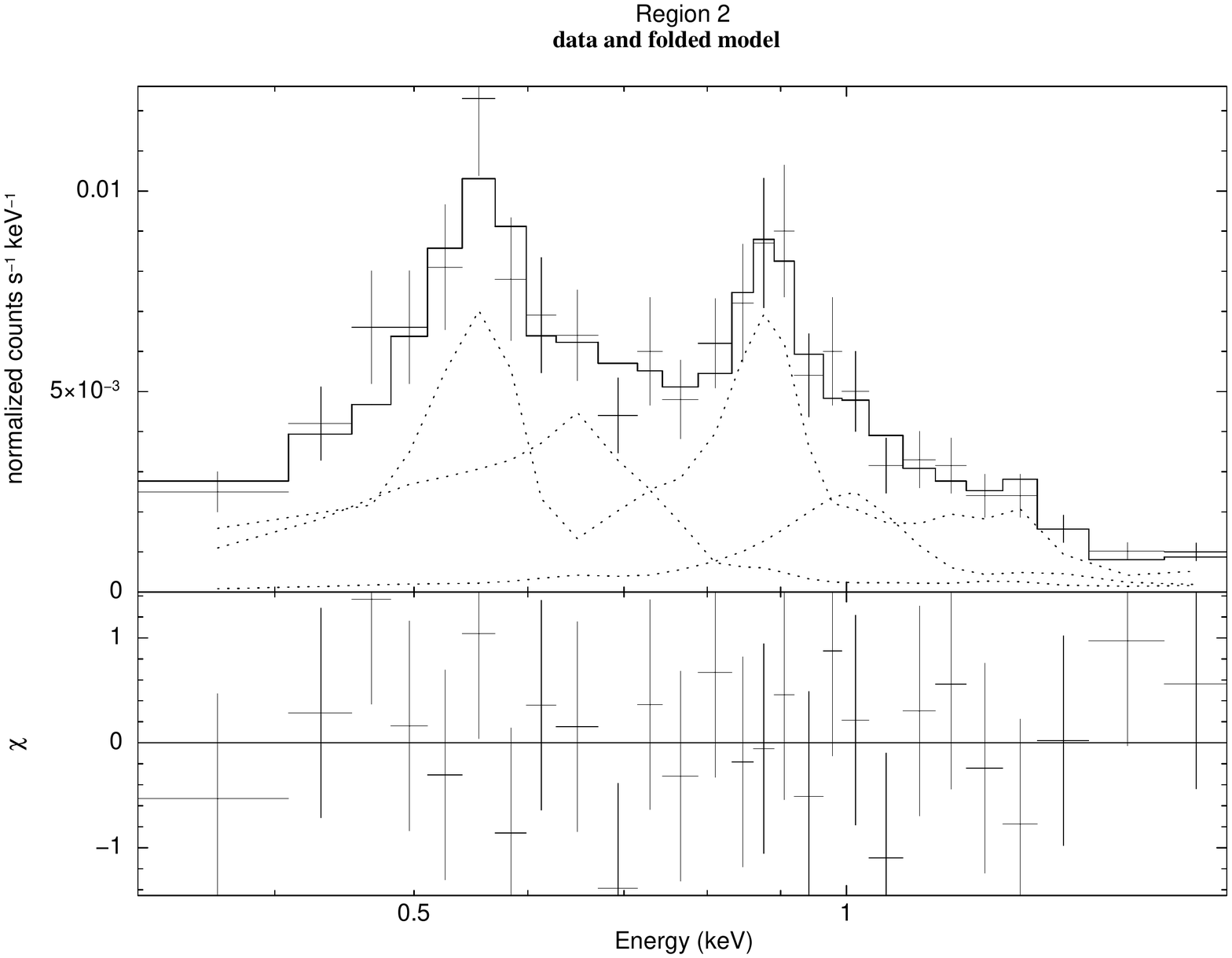}
\includegraphics[scale=0.25]{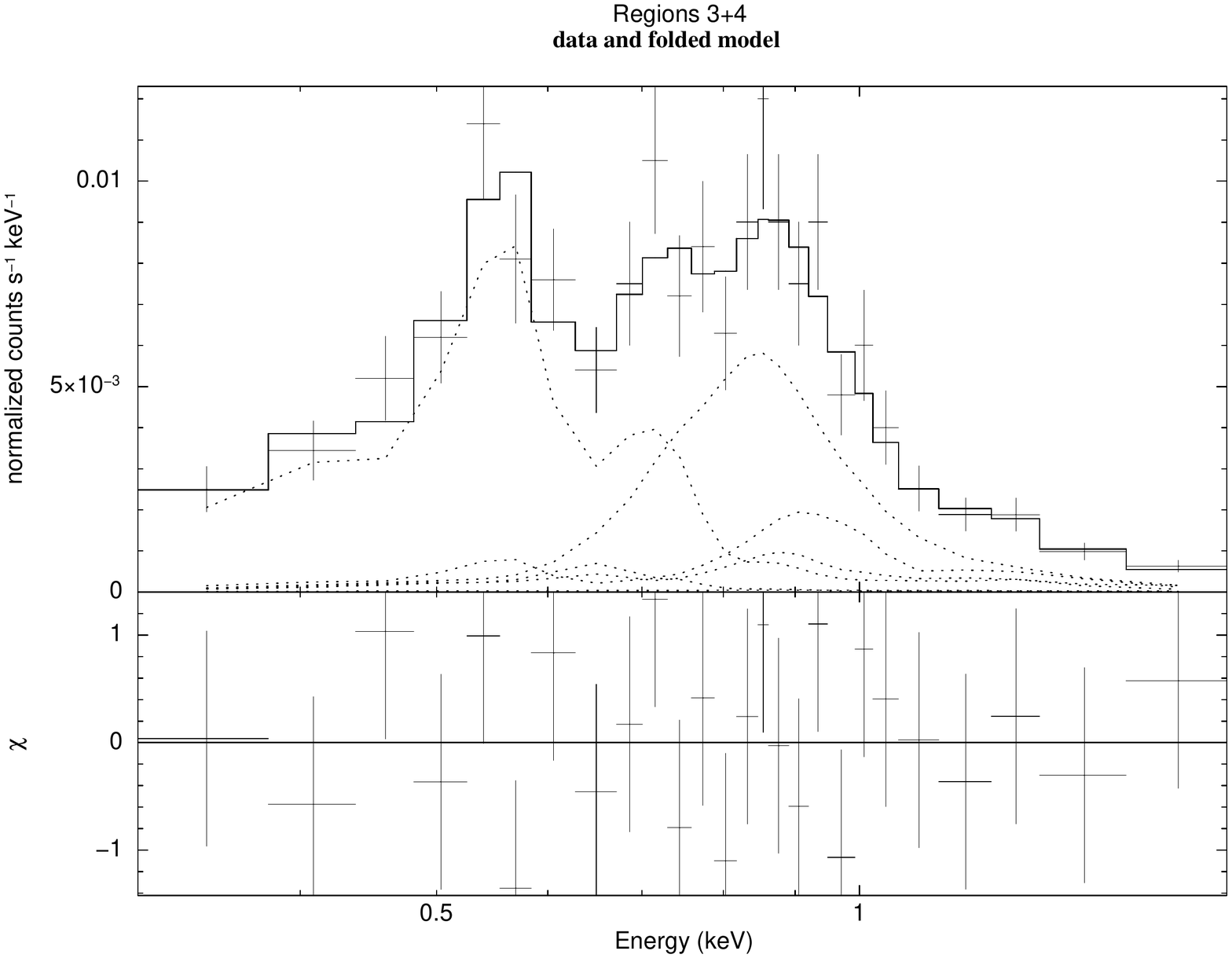}
\includegraphics[scale=0.25]{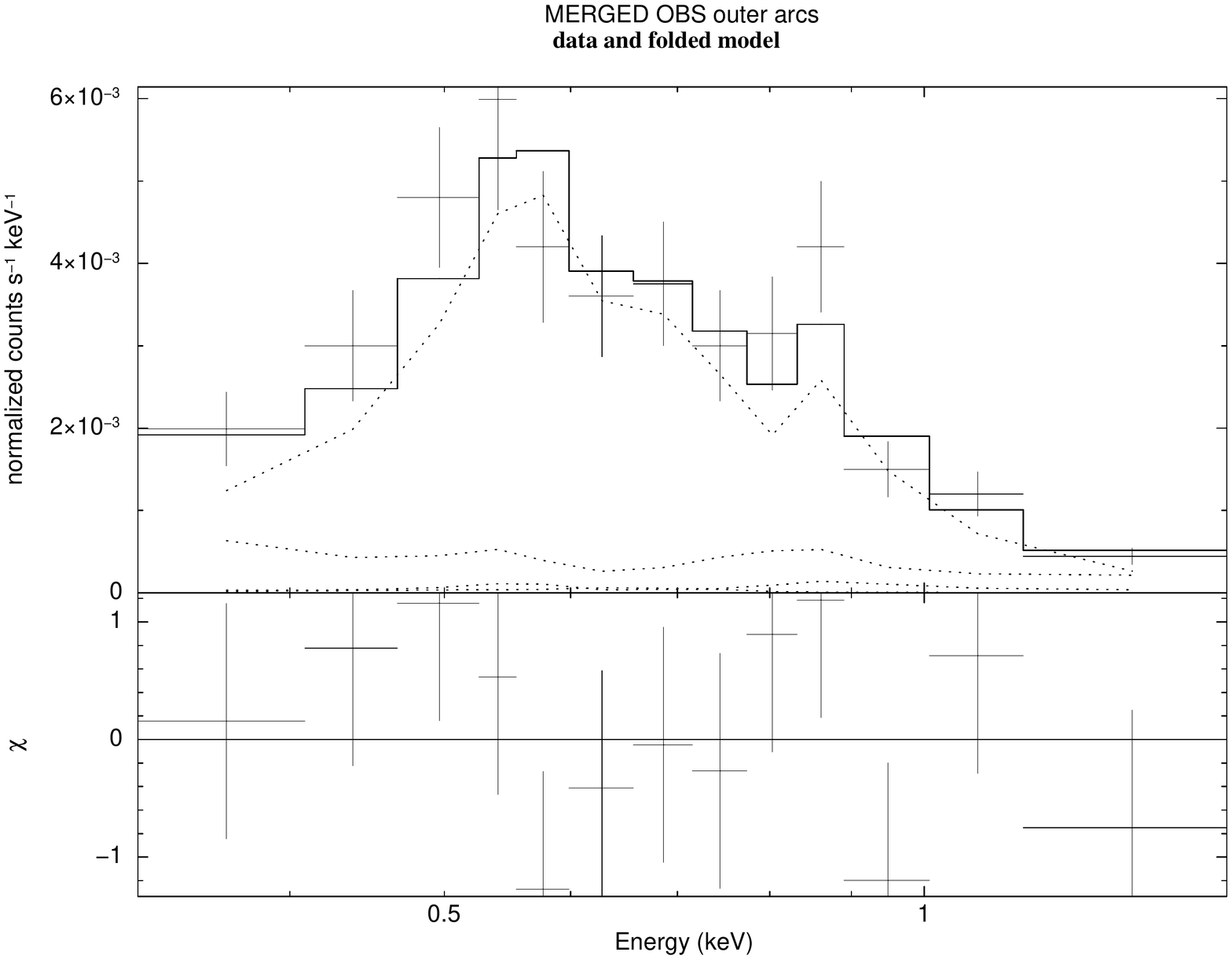}
\includegraphics[scale=0.25]{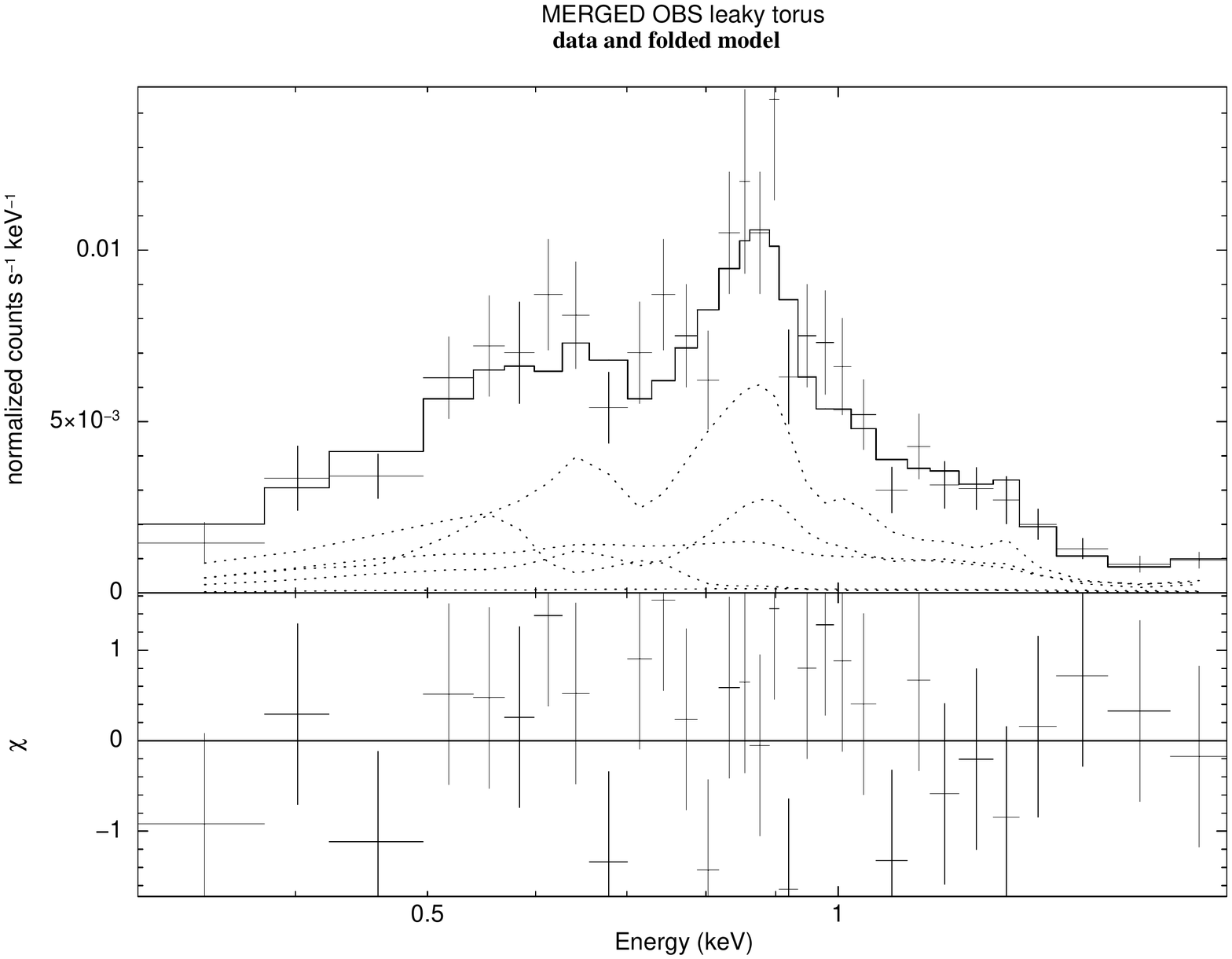}
\caption{Merged \textit{Chandra}/ACIS-S \(0.3 - 2\mbox{ keV}\) spectra of the regions discussed in Sect. \ref{spatial} and \ref{discussion}, with best fit photoionization model; for each spectrum we also show with dotted lines the additive components of the best fitting models. From top-left to bottom-right we show fits to spectra extracted in region 1, 2 and 3+4 shown in Figure \ref{ratios}, regions of outer optical arcs, and cross-cone region shown in Figure \ref{torusregion}.}\label{otheregionspectra}
\end{figure}

We also tired adding a thermal component, included with the \textsc{apec} \textsc{xspec} model, looking for the thermally ionized plasma found by \citealt{2010ApJ...723.1748G} and \citealt{2010MNRAS.405..553B}. The thermal component, however, is not statistically required by the best fit models and does not provide better \(\chi^2\) values, yielding only upper limits for the thermal component flux \(F_C\). We find for 
CHEERS obs. \(F_{C\,(0.3-10)}<0.06\times{10}^{-13}\mbox{erg}\mbox{ cm}^{-2}\mbox{ s}^{-1}\), while for all merged obs. we find \(F_{C\,(0.3-10)}<0.08\times{10}^{-13}\mbox{erg}\mbox{ cm}^{-2}\mbox{ s}^{-1}\). {These fluxes are comparable with those of} the hot collisionally ionized plasma highlighted in the ionization maps presented in Figure \ref{ratios}, {and due to the strong nuclear emission it is not possible to constrain its parameters when extracting spectra in the central 1'' region}, so we included it in the fitting model as a constant component taking into account the contribution of regions 1 and 2 to the nuclear region (see Figure \ref{ratios}).

The higher ionization phase features values of \(\log{U_1} = 1.1\pm 0.1\) and \(\log{N_{H1}} = 20.6\pm 0.2\) for the ionizing parameter and the hydrogen column density, respectively, both in obs. 07745 and in CHEERS observations; the best fit values for the less ionized phase are \(\log{U_2} = -0.75\pm 0.2\) in all observations, and \(\log{N_{H2}} = 20.2\pm 0.4\) and \(\log{N_{H2}} = 21.0\pm 0.6\) in obs. 07745 and in CHEERS observations, respectively.

The photoionization model provides soft X-ray fluxes consistent with those obtained with the phenomenological model described above: for obs. 07745 \(f_{SX}=(5.8\pm 0.2)\times {10}^{-13} \mbox{ erg}\mbox{ cm}^{-2}\mbox{ s}^{-1}\), and for the CHEERS obs \(f_{SX}=(4.1\pm 0.1)\times {10}^{-13} \mbox{ erg}\mbox{ cm}^{-2}\mbox{ s}^{-1}\). The flux in the 2 -10 keV band \(f_{HX}\) is \(f_{HX}=(3.8\pm 0.3)\times {10}^{-13} \mbox{ erg}\mbox{ cm}^{-2}\mbox{ s}^{-1}\) for obs. 07745 and \(f_{HX}=(3.5\pm 0.3)\times {10}^{-13} \mbox{ erg}\mbox{ cm}^{-2}\mbox{ s}^{-1}\) for CHEERS obs., which are consistent within uncertainties. The ionization parameter and the hydrogen column density for the two sets of observations are also consistent within the uncertainties.
{These results are} confirmed by spectral fits to a restricted nuclear region (namely, the inner 0.5'' radius of the nuclear region, shown in Figure \ref{ratios}) that does not include the regions 1-2 of increased ionization, and whose results are presented in Table \ref{innernucleusphoto} in the Appendix. These fits also show a decrease of \(f_{SX}\) and a constant \(f_{HX}\) across the observations, together with comparable parameters of the photoionization model.

{The simplest interpretation for the observed variations of the soft X-ray flux is for them to be related to intrinsic source variability, with the soft X-ray emission originating close to the line of sight. In this case the hard emission, constituted by reflected nuclear emission as expected for a Compton-thick source, must originate farther away from the line of sight. A possible site for the reflected emission to arise is the inner walls of the dusty torus \citep{1994MNRAS.267..743G,1997MNRAS.289..443I,1997A&A...325L..13M,2001MNRAS.322..669B} or warm absorber (see Sect. \ref{crossconeemission}) lying at \(r\gtrsim 4\mbox{ ly} = 1.2\mbox{ pc}\) from the nucleus, based on the time lag between observations. From \(f_{HX}\) we evaluate an observed luminosity \(L_{2-10}= (2.4\pm 0.1)\times{10}^{41}\mbox{ erg}\mbox{ s}^{-1}\), while the intrinsic luminosity can be recovered correcting this value by a factor 60 \citep{2006A&A...455..173P} or 100 \citep{2006A&A...446..459C} due to the Compton-thick nature of Mrk 573 (with a column density \(N_H>1.6\times{10}^{24}\mbox{ cm}^{-2}\), \citealt{2005A&A...444..119G}), to find \(L_{2-10}= (1.5\div 2.5)\times{10}^{43}\mbox{ erg}\mbox{ s}^{-1}\). Then we can evaluate the AGN bolometric luminosity using a bolometric correction of 30 \citep{2004ASSL..308..187R}, to find \(L_{{bol}} = (4.4 \div 7.7)\times{10}^{44}\mbox{ erg}\mbox{ s}^{-1}\), consistent with \(L_{{bol}} = 5.2\times{10}^{44}\mbox{ erg}\mbox{ s}^{-1}\) evaluated by \citet{2010ApJ...723.1748G} modeling the optical and infrared spectrum of Mrk 573 with a clumpy torus model, and slightly smaller than the \(L_{{bol}} = 7.9\times{10}^{44}\div 1.3\times{10}^{46}\mbox{ erg}\mbox{ s}^{-1}\) derived by \citet{2009ApJ...698..106K} using the correlation between the [O\textsc{iv}] and \(L_{2-10}\). The bolometric luminosity can be used to set a lower limit on the torus inner radius \(r_{0}=0.4\, {L_{45}}^{1/2}\,{T_{1500}}^{-2.6}\mbox{ pc}\), where \(L_{45}\) is the bolometric AGN luminosity in units of \({10}^{45}\mbox{ erg}\mbox{ s}^{-1}\) and \(T_{1500}\) is the dust sublimation temperature in units of \(1500\mbox{ K}\) \citep{2008ApJ...685..160N}. Using our estimate of the bolometric luminosity and setting \(r=r_0\) we obtain a sublimation temperature \(\sim 1000\mbox{ K}\) characteristic of silicate dust grains.}

\subsection{Emission from the cross cone region}\label{crossconeemission}

As discussed in Sect. \ref{imagedeconv} we found a significant emission in the cross cone direction. 
This emission is not expected in the presence of a continuous obscuring dusty torus, and may therefore be due to nuclear radiation leaking through a clumpy structure \citep{2008ApJ...685..147N}. Cross-cone emission was also found in NGC 4151 by \citet{2011ApJ...742...23W}, where the authors interpreted it as the result of leaking nuclear continuum through a filtering structure with an uncovered fraction \(\sim 1\%\).

\citet{2010AJ....140..577F} developed a kinematic model of Mrk 573 NLR, and concluded that the apparent bicone half opening angle of \(\sim 30\degree\) is likely due to the intersection with the galactic plane of a much wider ionizing bicone, with an intrinsic half opening angle \(\sim 53\degree\), close to the \(60\degree\) for the warm absorber structure presented by \citet{2000ApJ...545...63E}.

In order to evaluate the uncovered fraction of this filtering structure we fitted the spectrum of the soft emission in the cross-cone region shown in Figure \ref{torusregion} with a photoionization model, taking into account contamination from nuclear emission {evaluated from the contribution of the PSF wings in the cross-cone regions to be \(\sim 2.3\%\).}

We used the same grid models produced with the \textsc{Cloudy} package used above, adopting the same model adopted for fitting the cone emission with two photoionized phases; the result of this fit is presented in Table \ref{otheregions} and Figure \ref{otheregionspectra}.

The hydrogen column densities of both phases have been fixed to their best fit value of \(\log{N_{H}}=20\); the higher ionization component has a ionization factor \(\log{U_1}={0.8}\pm{0.1}\) similar to the value obtained for the cone region, while for the lower ionization component we put an upper limit on the ionization factor \(\log{U_2}<-2.4\). The latter value is significantly lower than the one we obtain in the cone region, indicating that a smaller fraction of the nuclear continuum leaks in the cross-cone region. 
{In order to estimate the uncovered fraction of the warm absorber we can use the ionization parameters obtained for the cone regions and for the cross-cone region. Since \(U \sim Q/r^2\), where \(Q\) is the number of nuclear ionizing photons per second illuminating the clouds, and assuming the same distance \(r\) for the cones and the cross-cone region from the nucleus, the ratio of the ionization parameters gives us the ratio of the nuclear ionizing photons seen by the different regions. So, if we assume the cone regions to see the unshielded nuclear radiation, this ratio gives} for the shielding structure an uncovered fraction \(<3\%\), consistent with the fraction evaluated in NGC 4151.

\section{CONCLUSIONS}

We have presented results from new ACIS/\textit{Chandra} observations of Seyfert 2 galaxy Mrk 573 obtained as part of CHEERS survey: these add to another existing observation performed in 2006 (OBSID: 07745, PI: Bianchi) to reach an exposure time of \(\sim 110\mbox{ ks}\), allowing us to study specific features of the soft X-ray emission (\(0.3-2\mbox{ keV}\)) from this source. 

\begin{figure}
\centering
\includegraphics[scale=0.45]{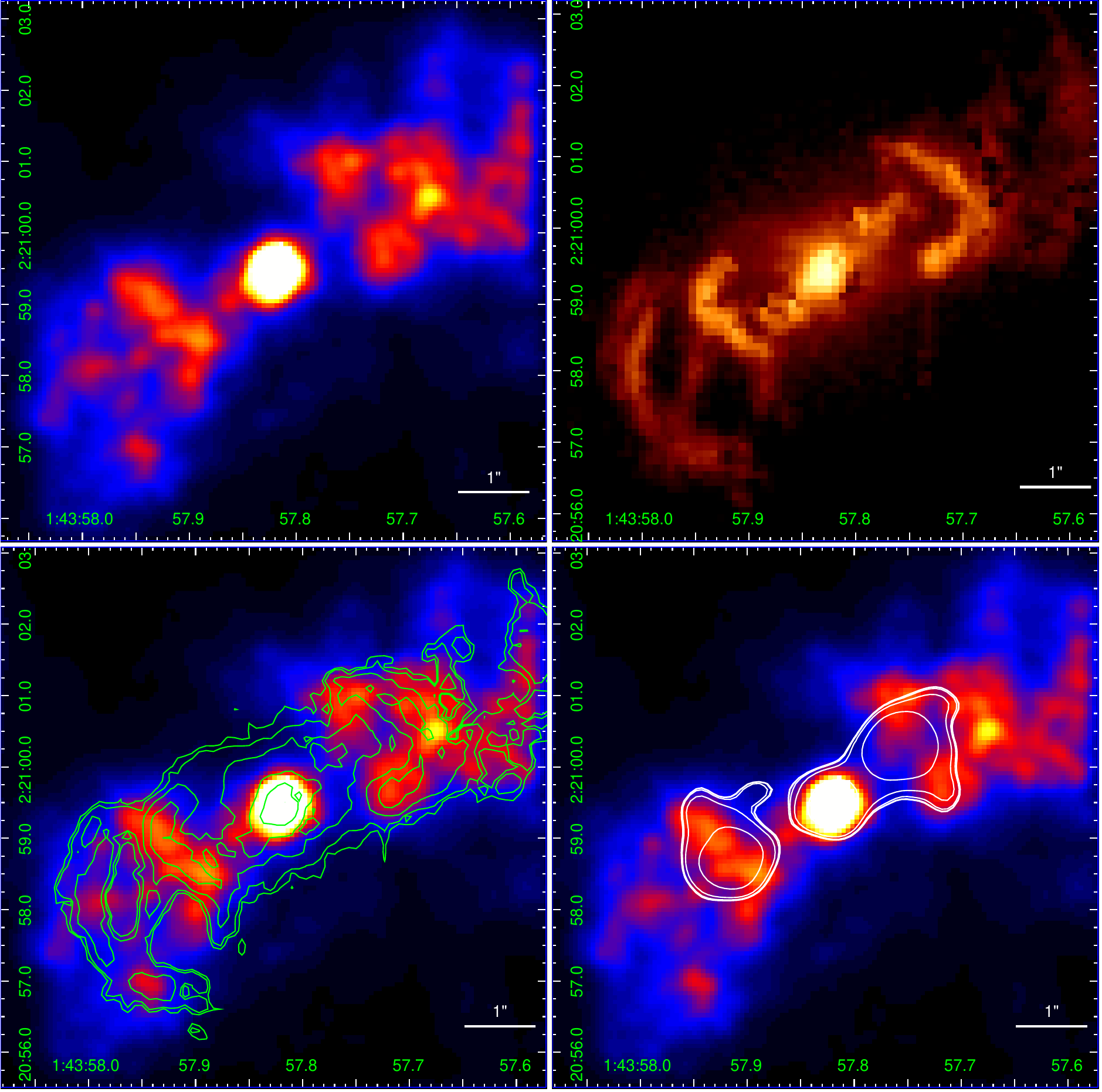}
\caption{(Upper-left panel) PSF-deconvolved ACIS image (\(0.3-2\mbox{ keV}\)) using the EMC2 method, with subpixel binning (1/8 of the native pixel size) and a 2X2 FWHM gaussian filter smoothing. The same image is presented in the lower-left and lower-right panels with overlaid the \textit{HST} [O \textsc{iii}] and radio \textit{VLA} 6 cm contours, respectively. (Upper-right panel) \textit{HST} [O \textsc{iii}] image.}\label{deconv}\medskip
\end{figure}

{There is a remarkable correspondence between the soft X-ray and the bi-conical NLR and ENLR as mapped by [O \textsc{iii}] emission, and the radio jets, as originally reported by \citet{2010MNRAS.405..553B} and \citet{2010ApJ...723.1748G}, suggesting a physical interplay among the three components.} This is even more evident in the comparison of the merged image of all available \textit{Chandra} observations with the \textit{HST} and \textit{VLA} images presented in Figure \ref{deconv}.

We have performed spectral analyses on different region of the source, and modeled the extracted spectra with different fitting models. We find that the nuclear emission is best fitted with a highly photoionized plasma (\(\log{U}=1.1\pm 0.1\)) and shows a significant (\(4\sigma\)) decrease of the observed soft X-ray flux that can be interpreted as the result of intrinsic variability of the nuclear continuum. The extended cone emission is effectively modeled with a mildly photoionized plasma (\(\log{U}=0.9\pm 0.3\) for the NW cone and \(\log{U}=0.7\pm 0.2\) for the SE cone) diffused over the bicone. A collisionally ionized gas at a temperature \(\sim 0.8\mbox{ keV}\)  appears to be located ahead of the observed radio knots, at the interface between the optical arcs. Another denser, hotter thermal component at \(\sim 1.1\mbox{ keV}\) is located between the nucleus and the radio knots. The thermal components can be interpreted in terms of shock interaction of the radio ejecta with the galaxy plane, with the jet depositing \(\gtrsim 30\%\) of its power into the ISM. This represents a \(>0.05\%\) fraction of the available accretion power that is thermally coupled to the ISM, substantially lower than what required by the majority of quasar feedback models, but compatible with a two-stage feedback.

We detect diffuse, soft X-ray emission extending up to \(\sim 13\mbox{ kpc}\) along the bicone axis. In the region perpendicular to this axis, where the obscuring torus is supposed to prevent nuclear X-ray reaching the ISM, we find significant emission extending up to \(\sim 9"\) from the nucleus. We model this emission with a plasma featuring a lower ionization parameter with respect to the emission extracted in the bicone region, an so we interpret the former as originating from the nuclear continuum filtered by a warm absorber with an uncovered factor \(< 3\%\).
Deeper observations are however required in order to better constrain the model parameter in this region of faint emission.

\acknowledgments

{We acknowledge useful comments and suggestions by our anonymous referee.}
{This work is supported by NASA grant GO1-12009X and HST
GO-12365.01-A (PI: Wang). We acknowledge support from the CXC, which is operated by the Smithsonian Astrophysical Observatory (SAO) for and on behalf of NASA under Contract NAS8-03060.}
A. P. thanks Andrea Marinucci, Francesco Massaro and Guido Risaliti for useful discussions. {This research has made use of data obtained from the Chandra Data Archive, and software provided by the CXC in the application packages CIAO and Sherpa. Some of the data presented in this paper were obtained from the Multimission Archive at the Space Telescope Science Institute (MAST). STScI is operated by the Association of Universities for Research in Astronomy, Inc., under NASA contract NAS5-26555.}

\begin{table*}
\centering
\begin{threeparttable}
\caption{Observations properties}\label{obsprop}
\begin{tabular}{lccccc}
\hline
\hline
Obs. ID                & 07745                    & 12294                    & 13124                    & 13125                    & merged                   \\
Obs. Date            & 2006-11-18               & 2010-09-16               & 2010-09-17               & 2010-09-19               & -                        \\ 
Exposure (ks)        & 35.07                    & 9.92                     & 52.37                    & 16.76                    & 114.12                   \\
Net Counts 0.3 - 2 keV (error) \tnote{a} & 3173(57)                     & 739(27)                       & 3649(61)                      & 1216(35)                     & 8777(95)                     \\
Net Counts 2 - 10 keV (error) \tnote{a} & 345(23)                     & 79(12)                      & 443(27)                      & 160(15)                      & 1027(41)                     \\
\hline
\hline
\end{tabular}
\begin{tablenotes}[para]
                 \item {Notes:}\\
                 \item[a] Counts evaluated in a circular region of \(15"\) radius.
\end{tablenotes}
\end{threeparttable}
\end{table*}

\begin{table}\small
\centering
\begin{threeparttable}
\caption{Best fit photoionization models for other regions (merged observations).}\label{otheregions}
\begin{tabular}{l|c|c|c|c|c}
\hline
\hline
Region                      
& 1 & 2 & 3+4                & [O \textsc{iii}] outer arcs & cross-cone region     \\
Net Counts 0.3 - 2 keV (error)	& 1002(32) & 1248(36) &   691(26)                 & 308(18)                & 547(24)              \\ 
\hline 
Model Parameter				&		& &								&										&										\\
\hline
\(\log {U_1}\)			& \({0.87}_{-0.12}^{+0.16}\) &	\({0.85}\pm{0.12}\)	& \({1.33}\pm{0.31}\)	& \({0.18}_{-0.12}^{+0.09}\)	& \({0.81}_{-0.05}^{+0.08}\)	\\
\(\log {N_{H\,1}}\)    & \({21.41}_{-0.57}^{+0.76}\) &  \({21.55}\pm{0.58}\) 
& \(20^*\)                 & \({20.07}_{-0.37}^{+0.38}\)   & \(20^*\)\\
\(F_{1\,(0.3-2)}\)\tnote{a} & \({0.22}\pm{0.06}\) & \({0.19}_{-0.03}^{+0.04}\)
& \({0.04}\pm{0.03}\) & \({0.17}\pm{0.02}\) &  \({0.16}\pm{0.03}\)\\
\(\log {U_2}\)           & \({-1.17}_{-0.33}^{+0.60}\) & \({-0.75}\pm{0.28}\)
& \({-0.85}_{-0.14}^{+0.25}\)                & \(-2^*\)   & \(<-2.42\)\\
\(\log {N_{H\,2}}\)    & \(20^*\) & \(21^*\) & \(20^*\)                 & \(20^*\)           & \(20^*\)\\ 
\(F_{2\,(0.3-2)}\)\tnote{a}         & \({0.13}\pm{0.07}\) & \({0.16}_{-0.03}^{+0.04}\)
& \({0.26}_{-0.06}^{+0.03}\) & \({0.06}\pm{0.03}\)   &  \({0.04}_{-0.02}^{+0.03}\)     \\
\(kT\)\tnote{b}           & \({1.06}_{-0.11}^{+0.26}\) & \({1.29}\pm{0.15}\)
& \({0.76}\pm{0.11}\) & -                    & -                   \\
\(F_{C\,(0.3-2)}\)\tnote{a} & \({0.05}\pm{0.02}\) &  \({0.04}_{-0.01}^{+0.02}\)
& \({0.08}_{-0.02}^{+0.03}\) & -                       &-                     \\
\hline
\(\chi^2\)(dof)          & 0.98(20) & 0.59(20) & 0.72(20)                 & 0.98(9)        & 1.01(20)                 \\
\(F_{0.3-2}\)\tnote{a}    & \({0.39}\pm{0.04}\) & \({0.39}\pm{0.02}\)
& \({0.42}\pm{0.03}\) & \({0.23}\pm{0.02}\) & \({0.29}_{-0.02}^{0.03}\)\\
\hline
\hline
\end{tabular}
       \begin{tablenotes}[para]
                 \item {Notes:}\\
                 \item[a] Unabsorbed flux in units of \({10}^{-13}\mbox{ erg}\mbox{ cm}^{-2}\mbox{ s}^{-1}\).\\
                 \item[b] Plasma temperature in keV.\\
       \end{tablenotes}
\end{threeparttable}
\end{table}

\appendix

\section{Spectral fits results}

\begin{figure}
\centering
\includegraphics[scale=0.22]{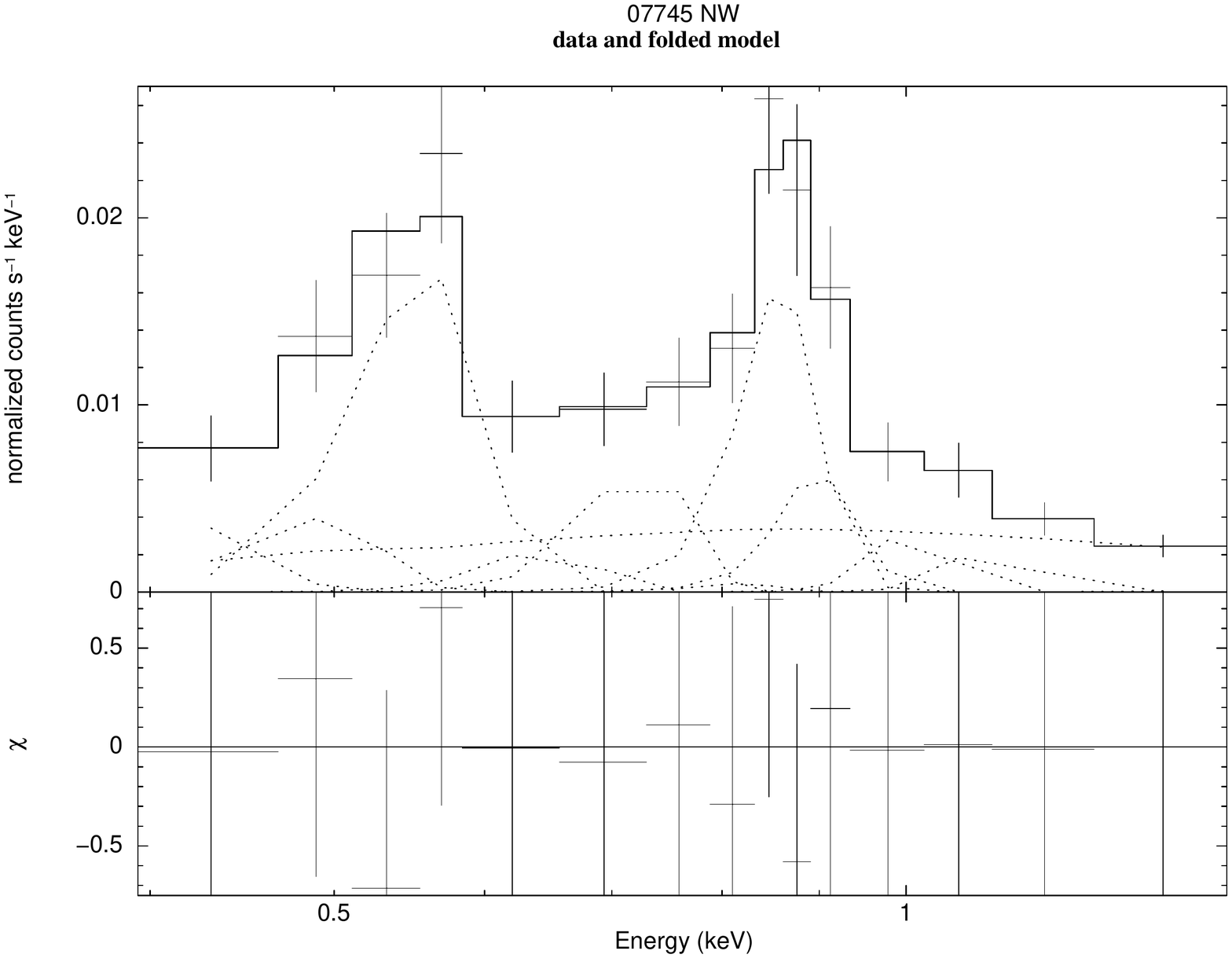}
\includegraphics[scale=0.22]{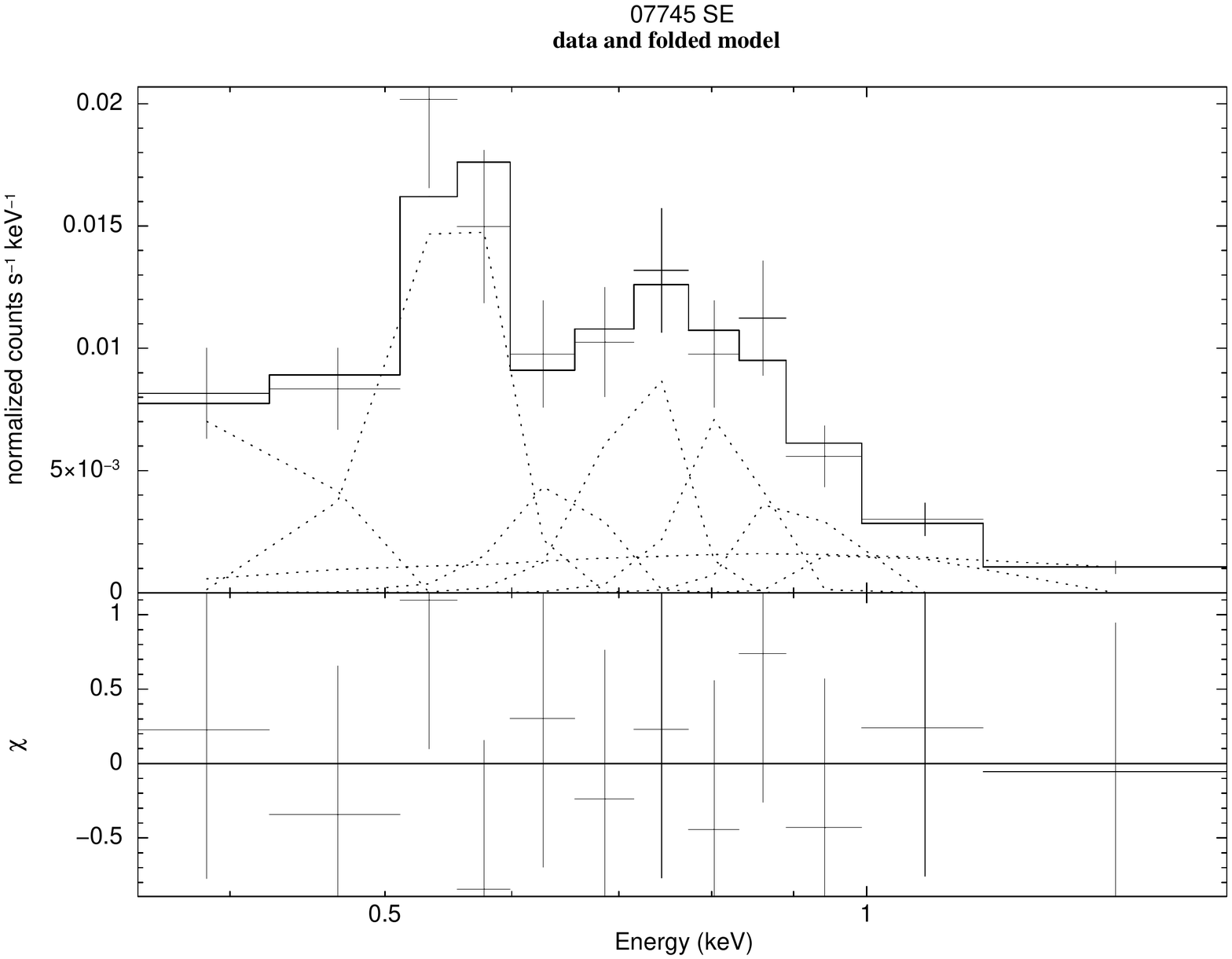}
\includegraphics[scale=0.22]{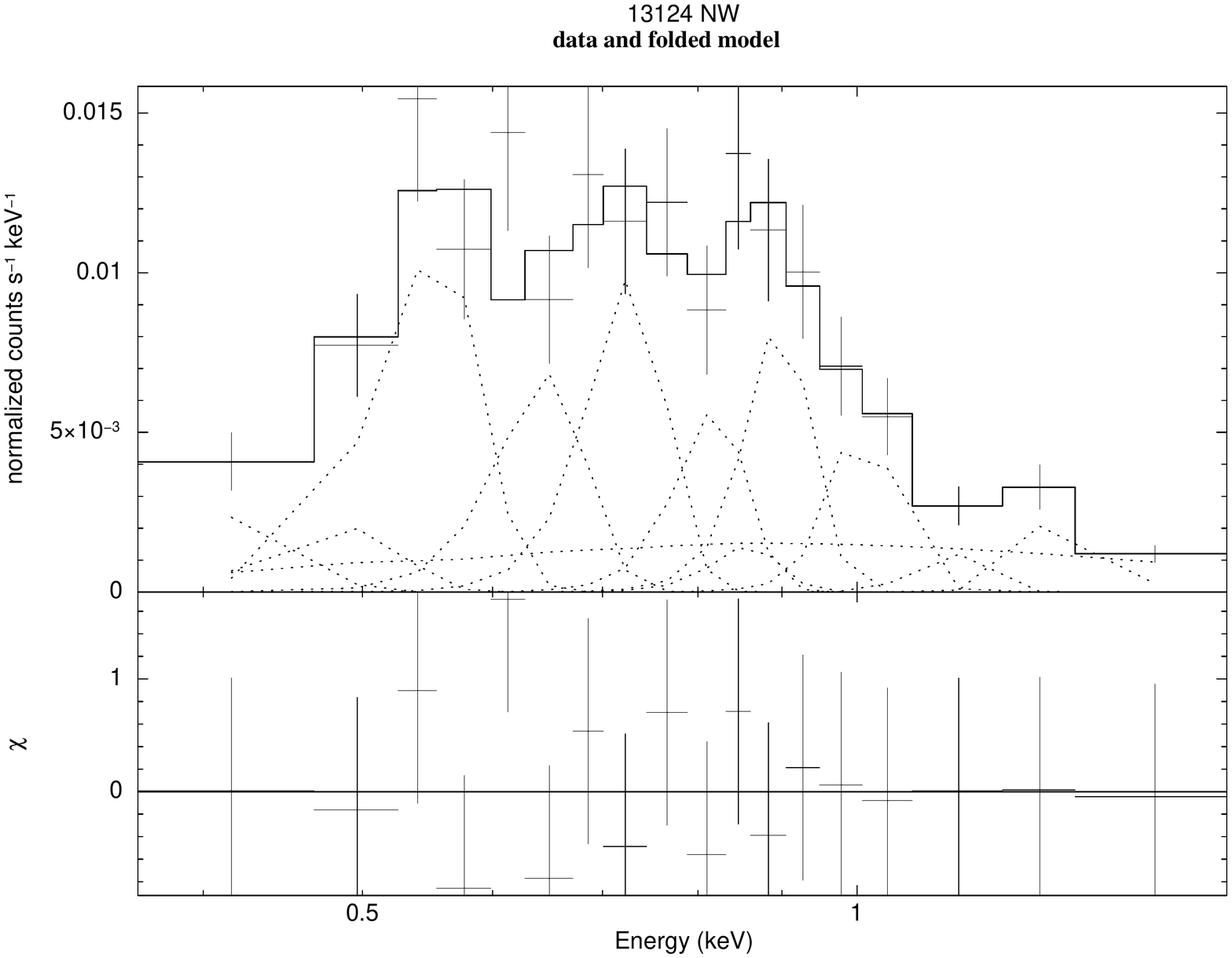}
\includegraphics[scale=0.22]{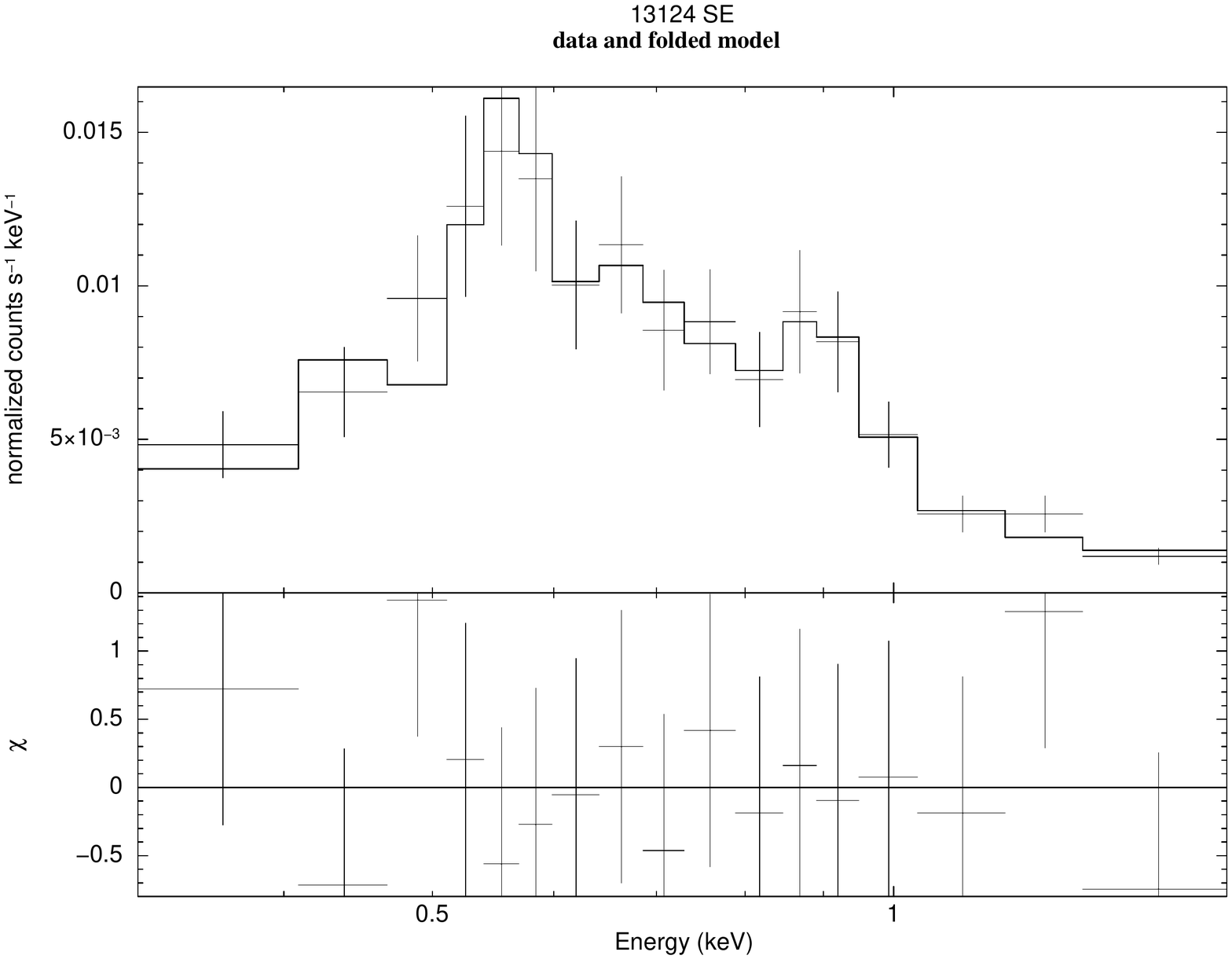}
\includegraphics[scale=0.22]{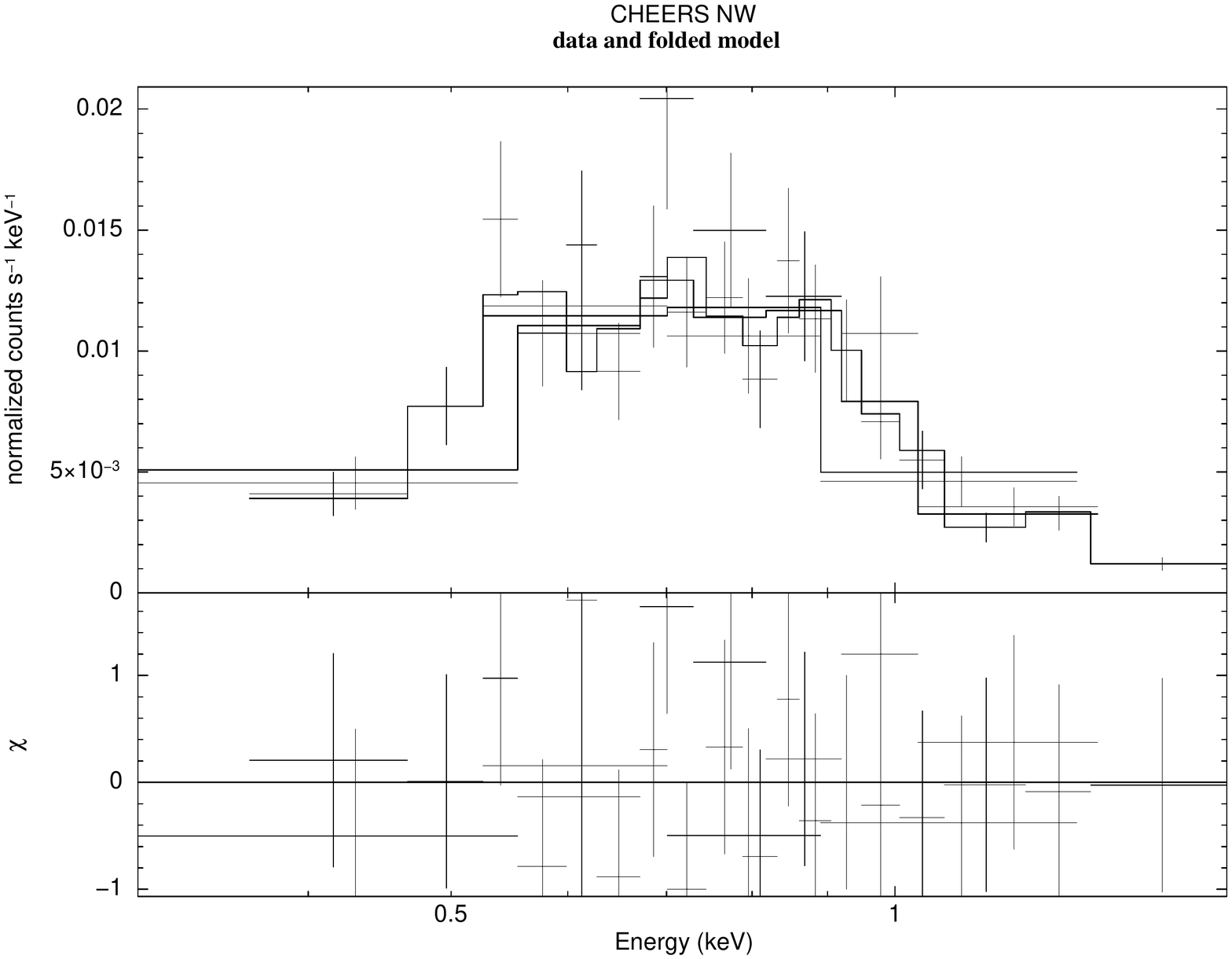}
\includegraphics[scale=0.22]{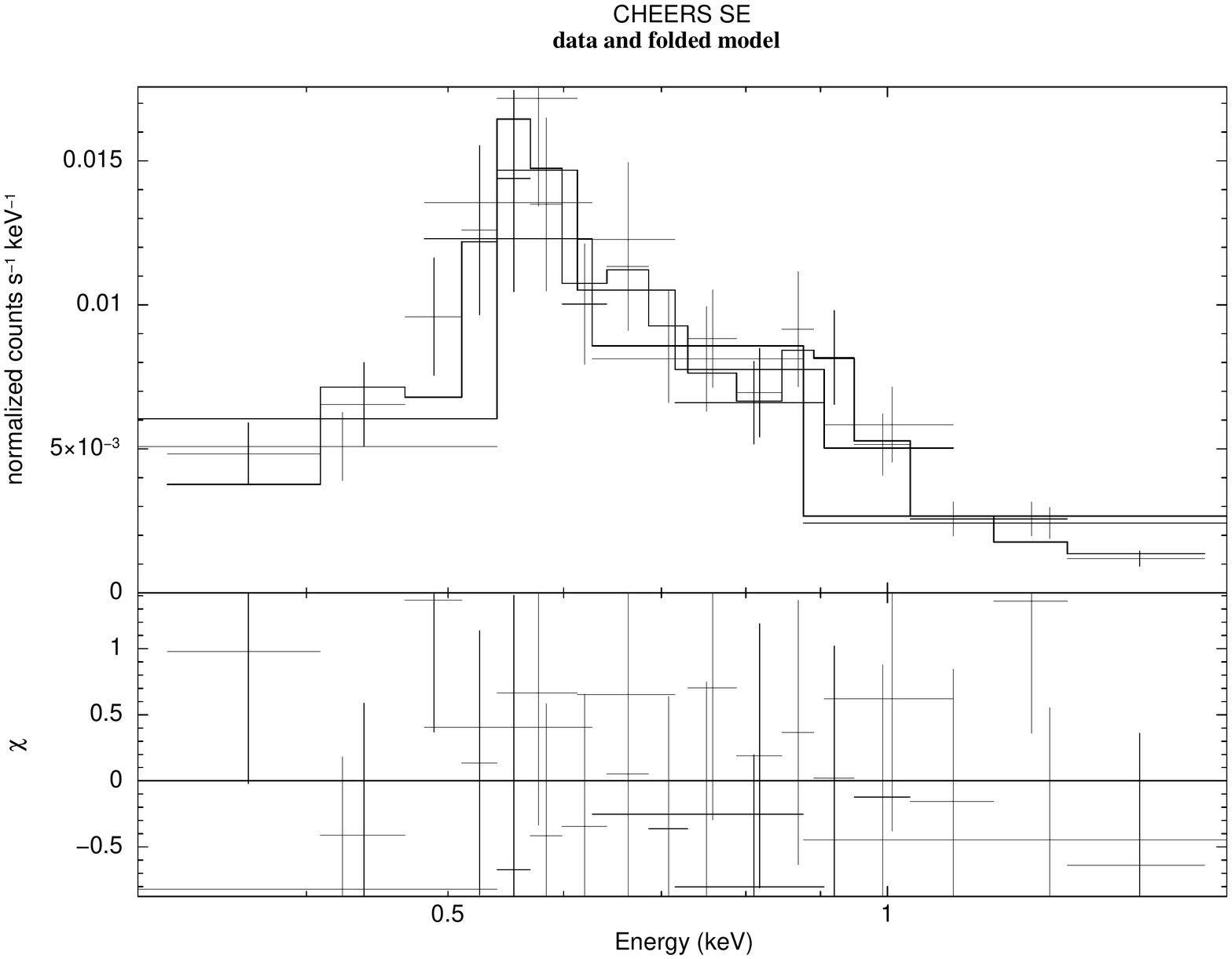}
\includegraphics[scale=0.22]{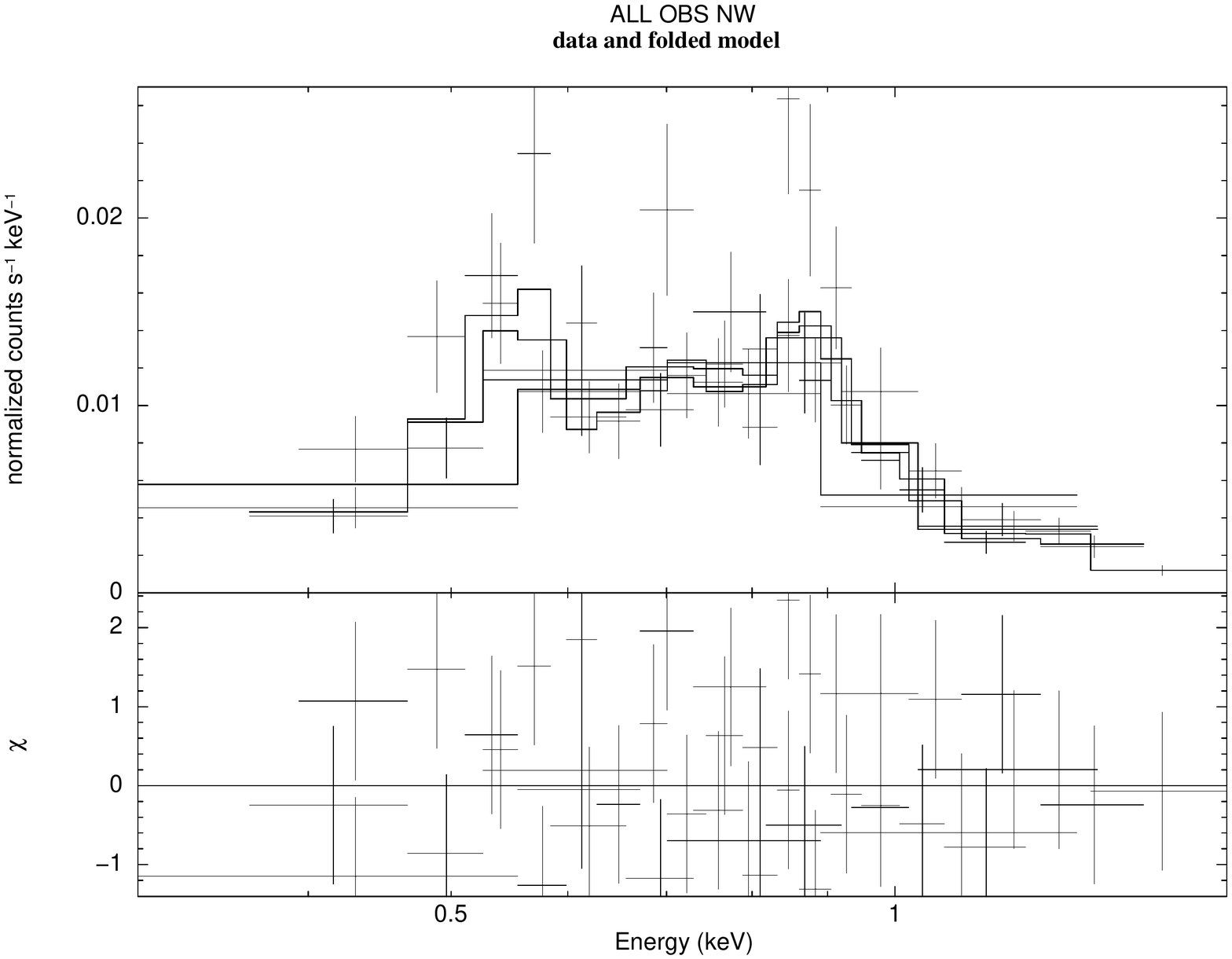}
\includegraphics[scale=0.22]{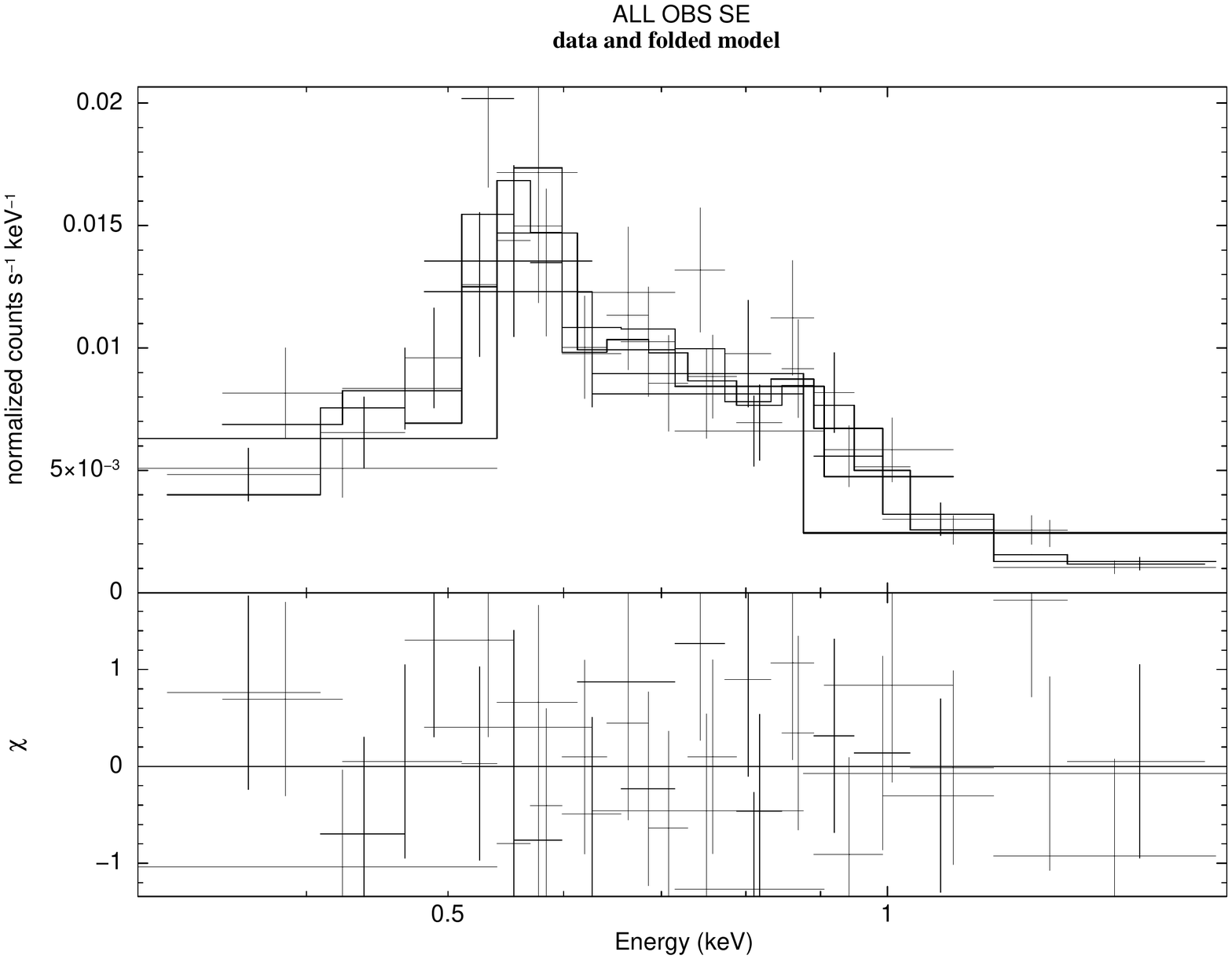}
\caption{\textit{Chandra}/ACIS-S \(0.3 - 2\mbox{ keV}\) spectra of the NW (left column) and SE (right column) cones emission (extracted from the region shown in Figure \ref{torusregion}) in different observations, with best fit emission lines models. From top to bottom we show fits to OBS. 07745, 13124, merged data from CHEERS observations and merged data from all observations. For spectra of single observations we also show with dotted lines the additive components of the best fitting models.}\label{coneslinespectra}
\end{figure}

\begin{figure}
\centering
\includegraphics[scale=0.22]{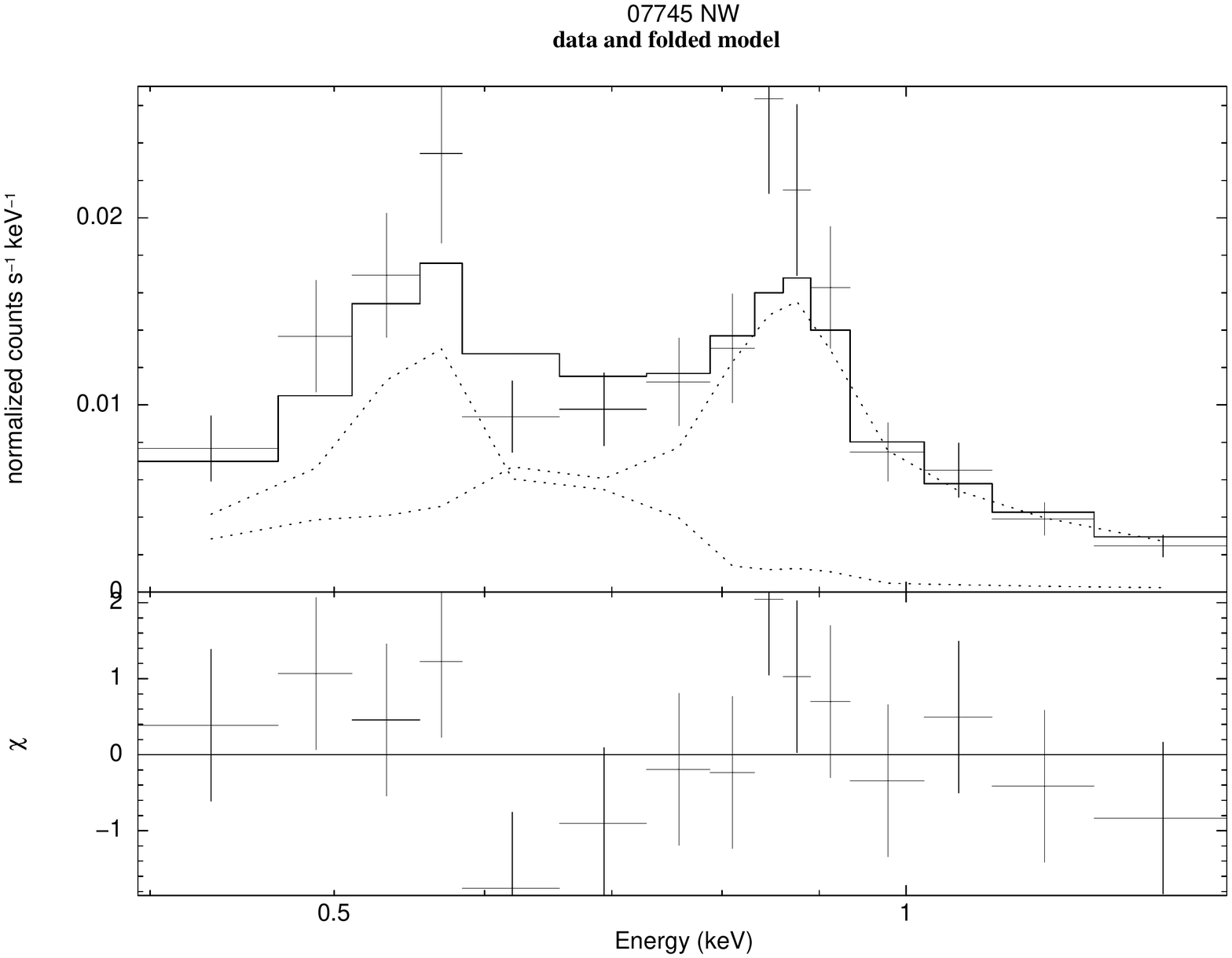}
\includegraphics[scale=0.22]{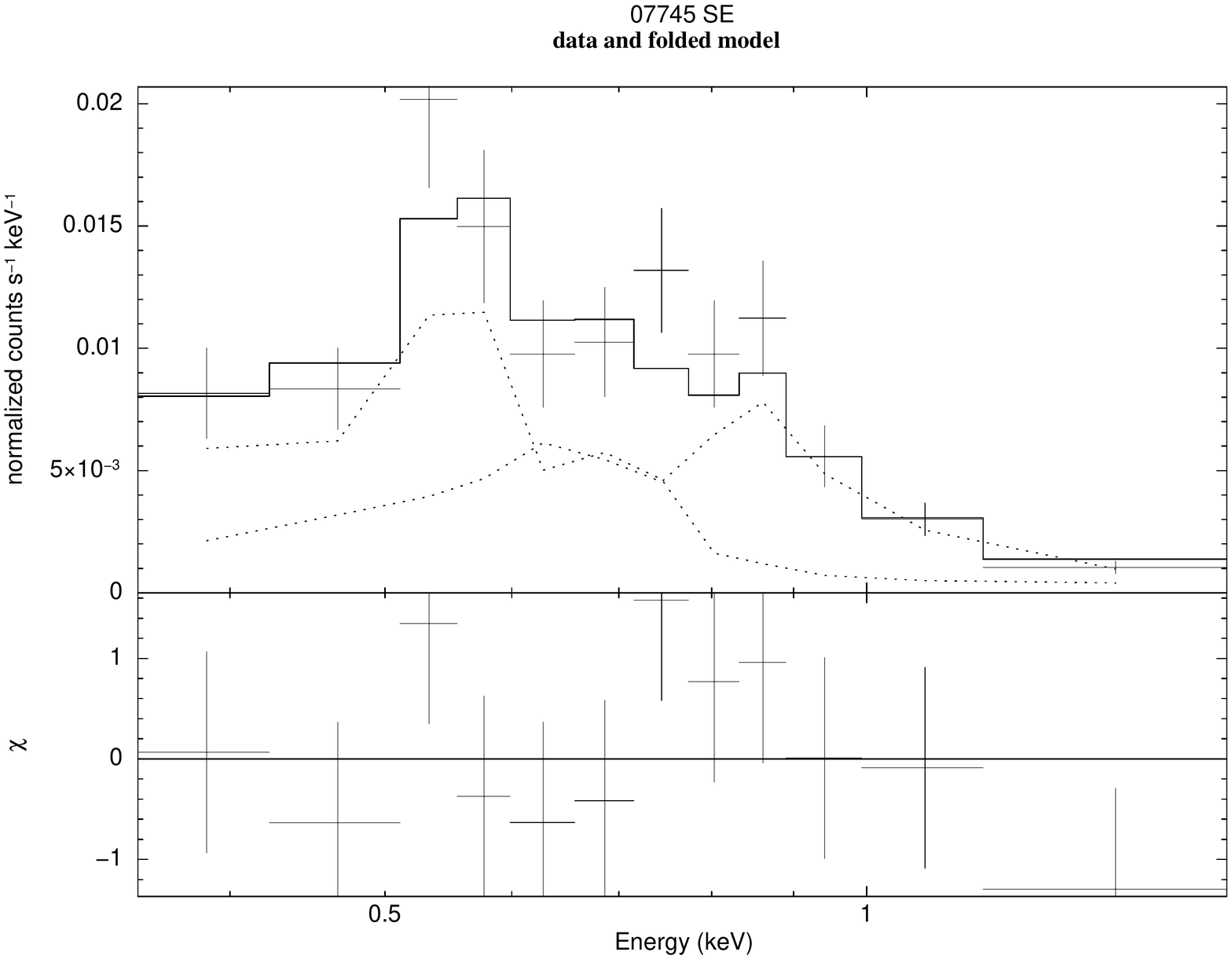}
\includegraphics[scale=0.22]{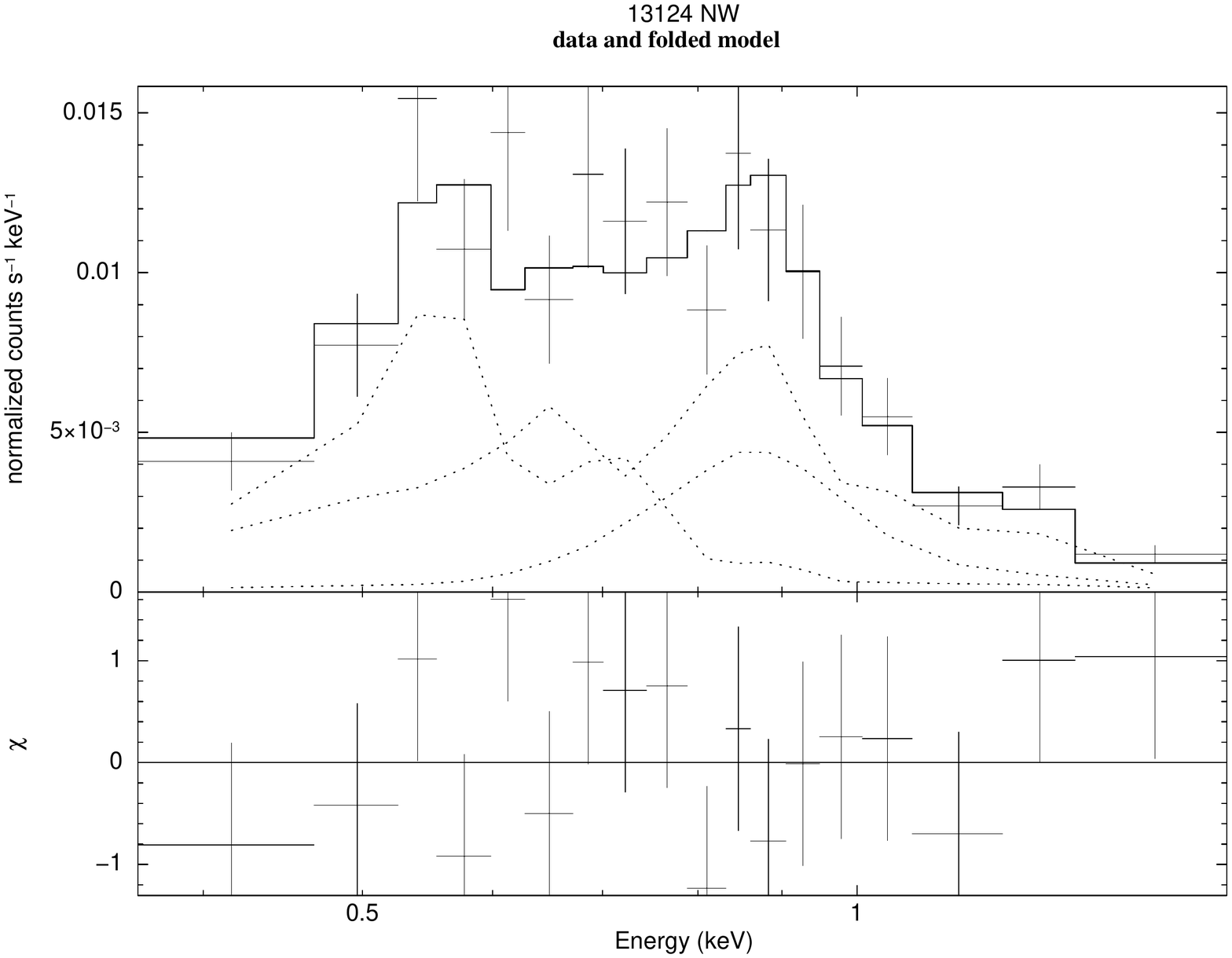}
\includegraphics[scale=0.22]{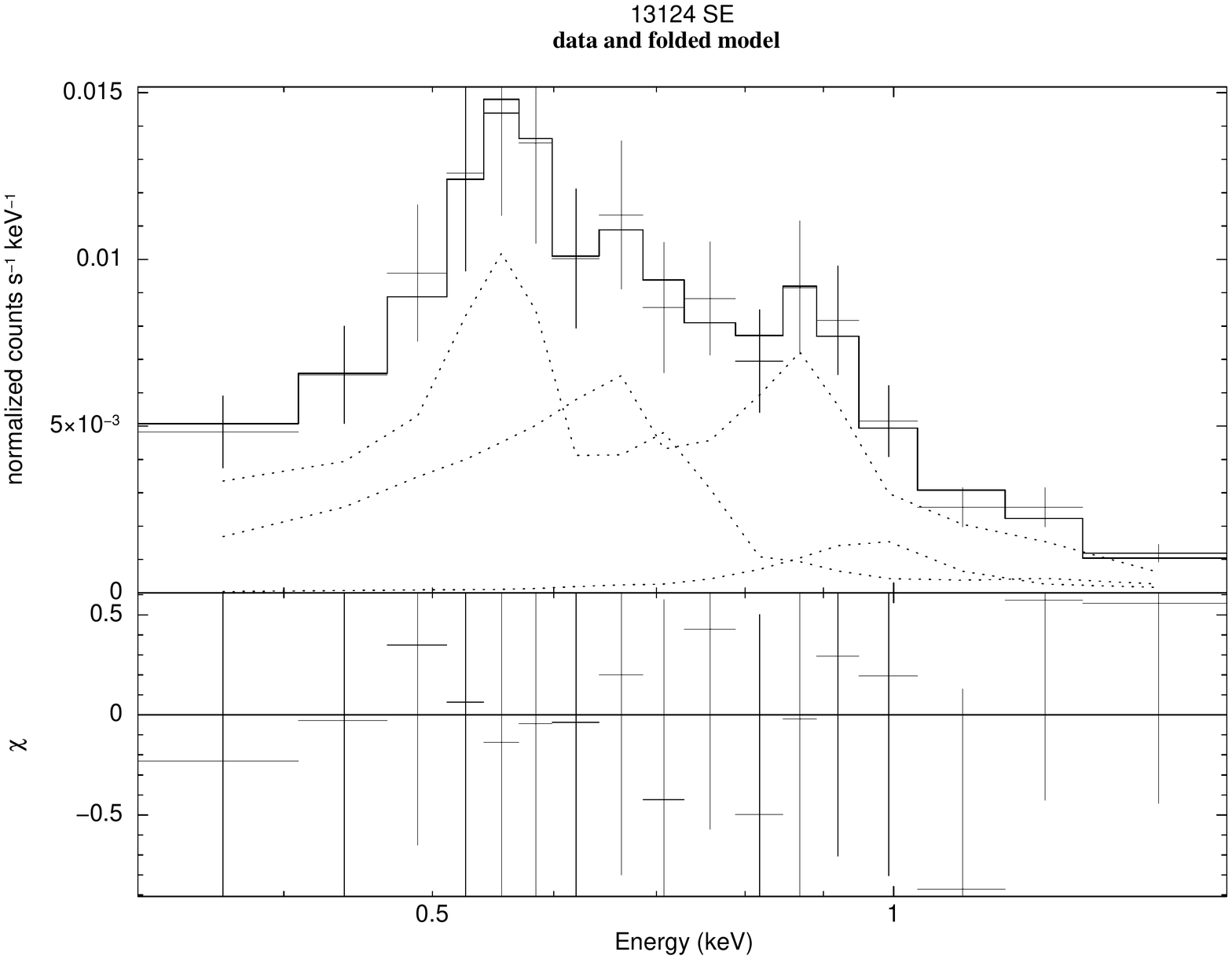}
\includegraphics[scale=0.22]{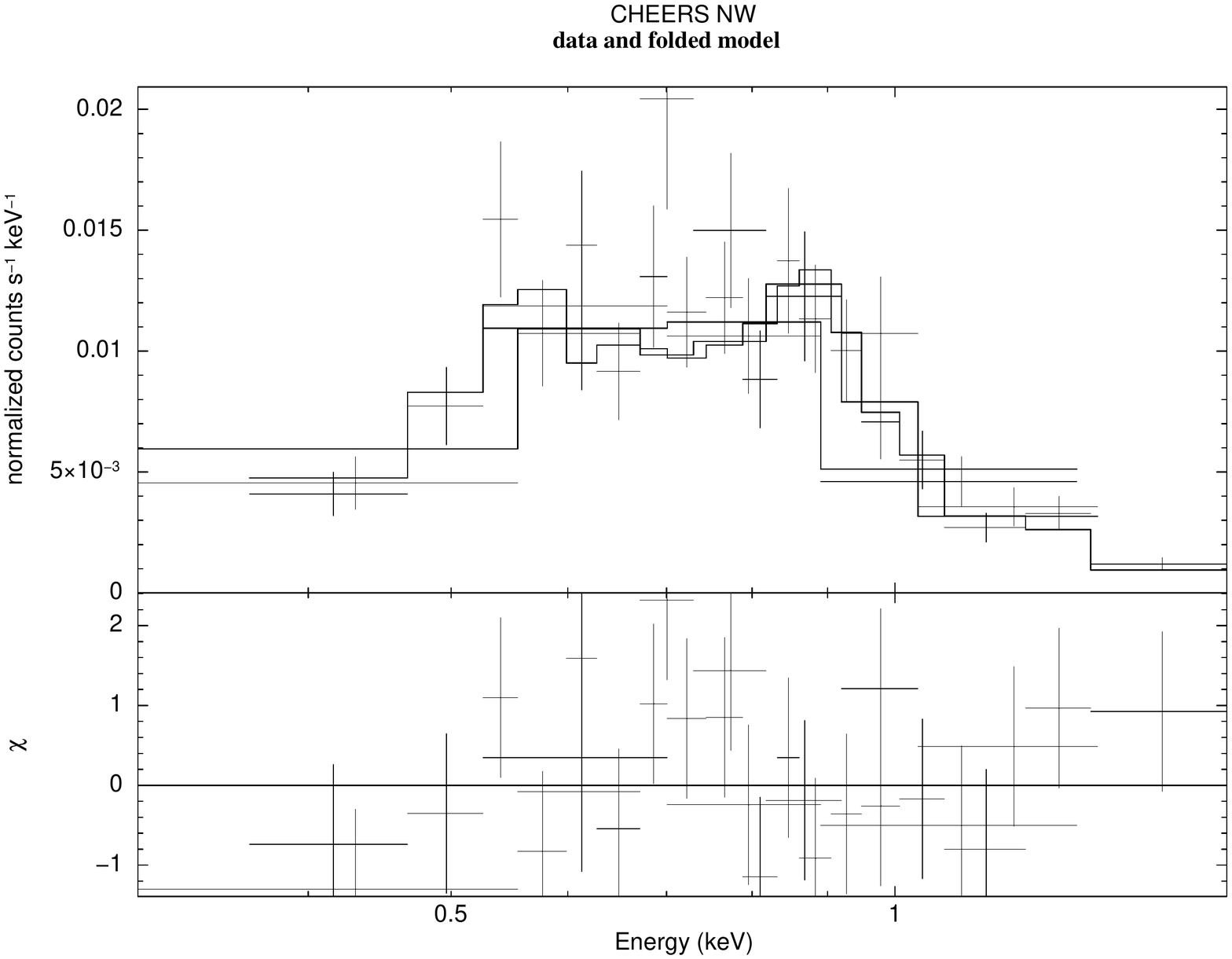}
\includegraphics[scale=0.22]{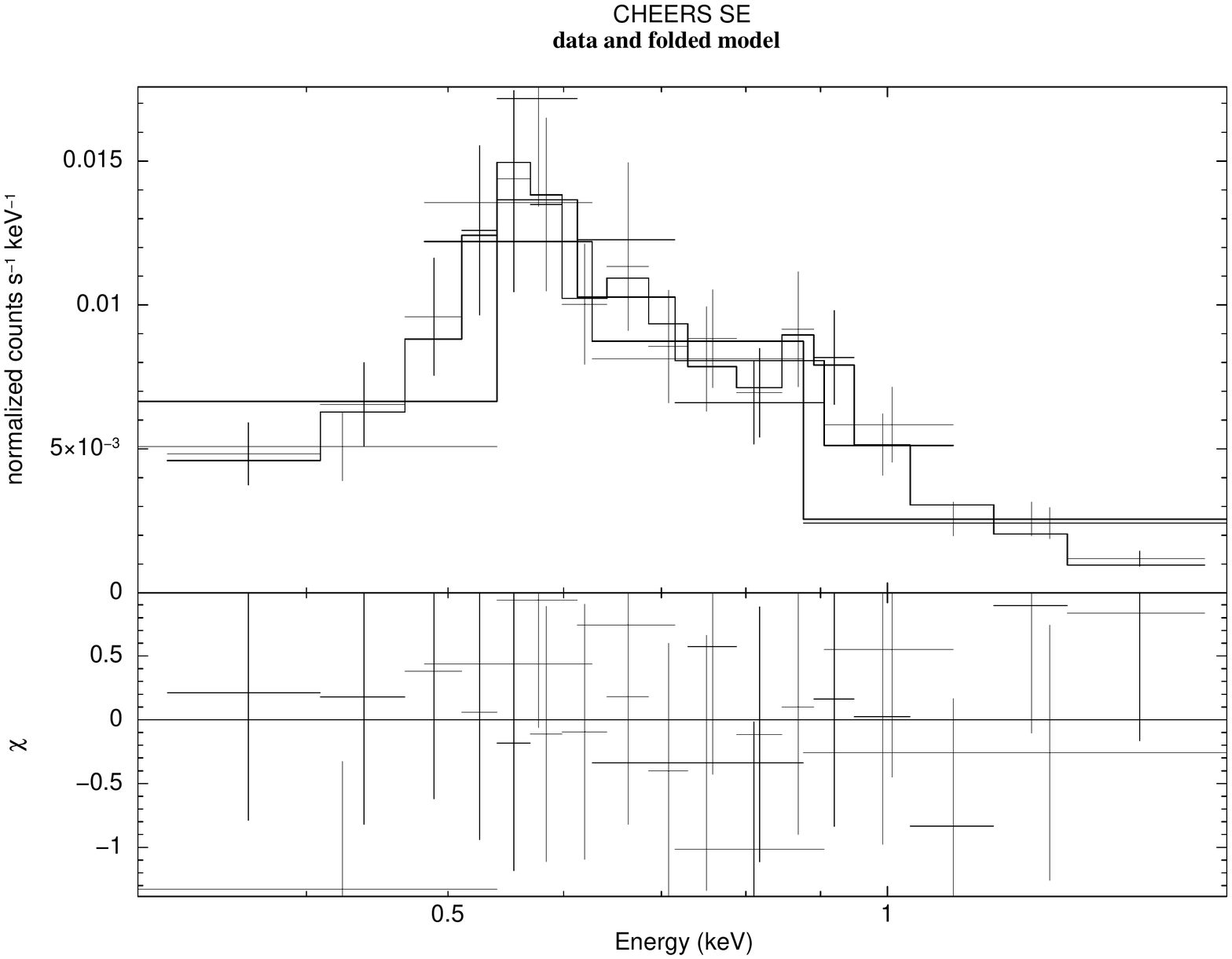}
\includegraphics[scale=0.22]{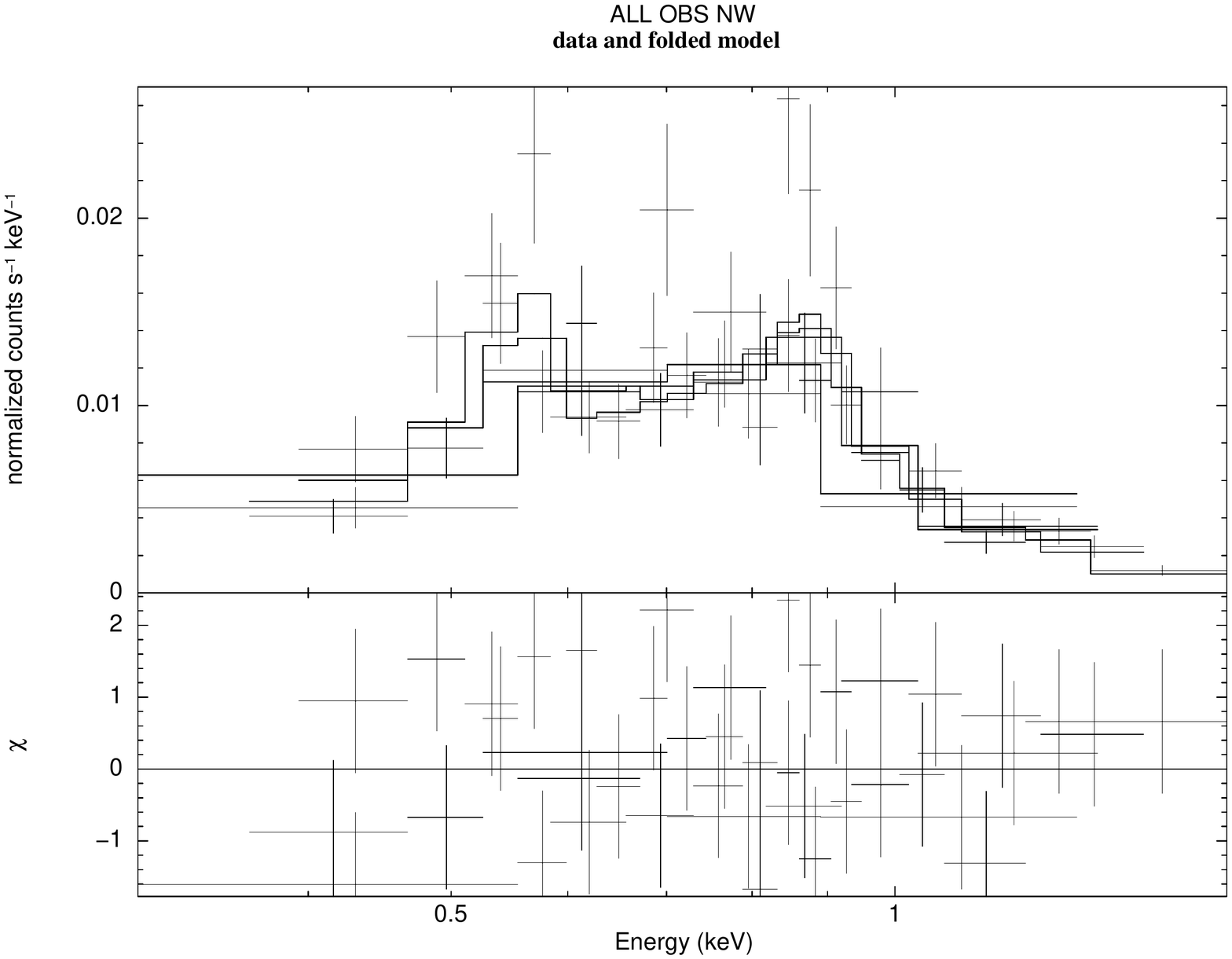}
\includegraphics[scale=0.22]{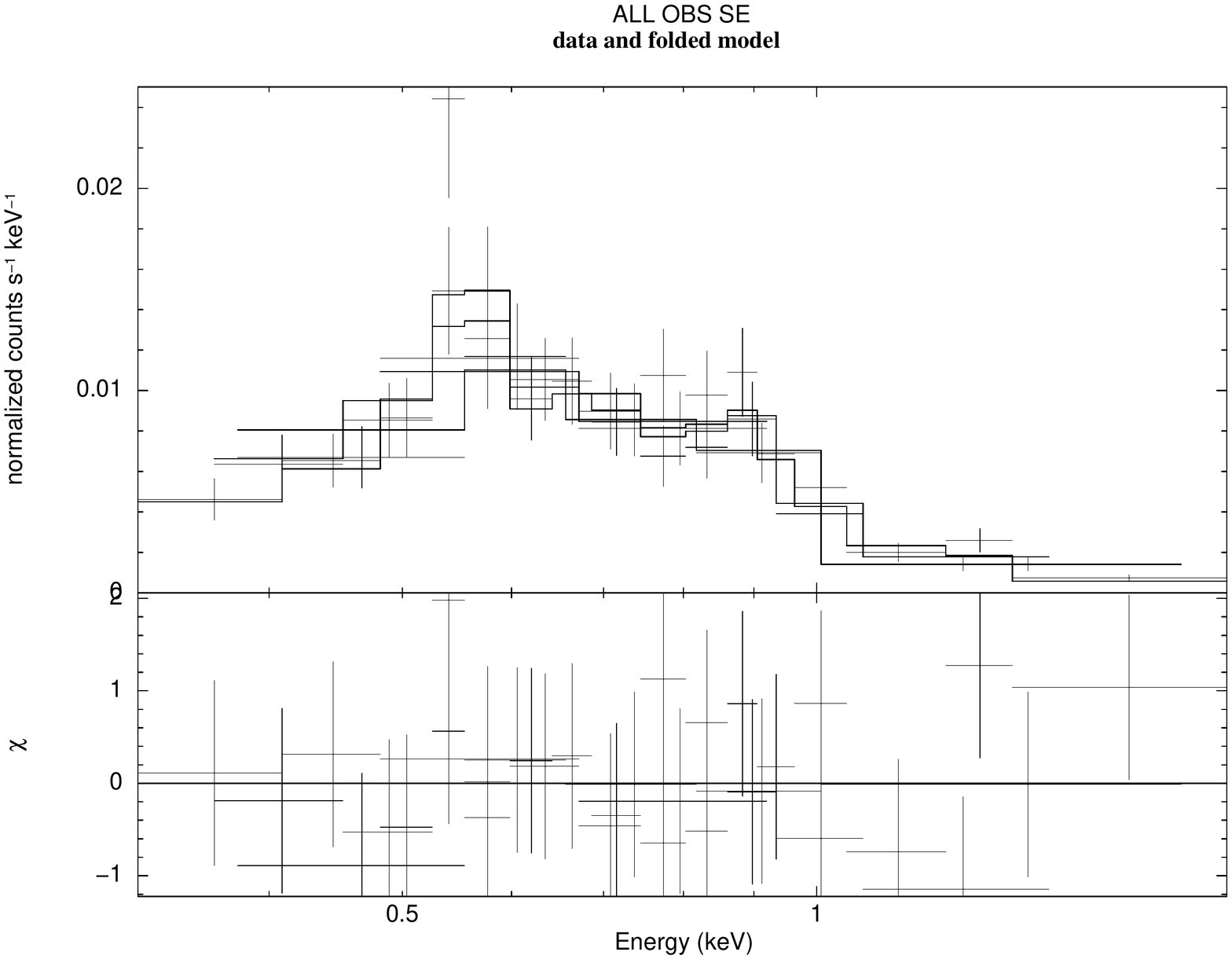}
\caption{\textit{Chandra}/ACIS-S \(0.3 - 2\mbox{ keV}\) spectra of the NW (left column) and SE (right column) cones emission (extracted from the region shown in Figure \ref{torusregion}) in different observations, with best fit photoionization models. From top to bottom we show fits to OBS. 07745, 13124, merged data from CHEERS observations and merged data from all observations. For spectra of single observations we also show with dotted lines the additive components of the best fitting models.}\label{conephotospectra}
\end{figure}

\begin{figure}
\centering
\includegraphics[scale=0.25]{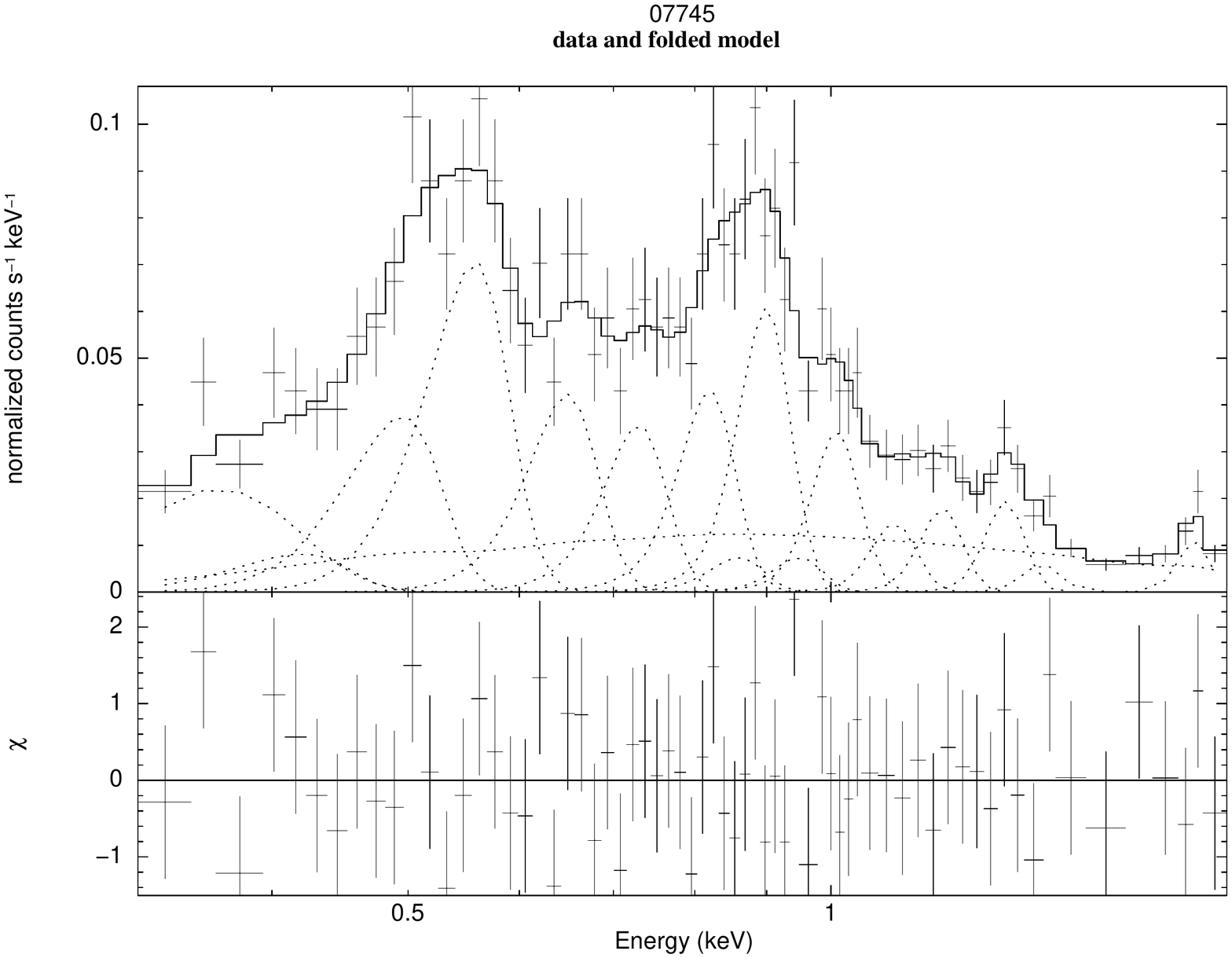}
\includegraphics[scale=0.25]{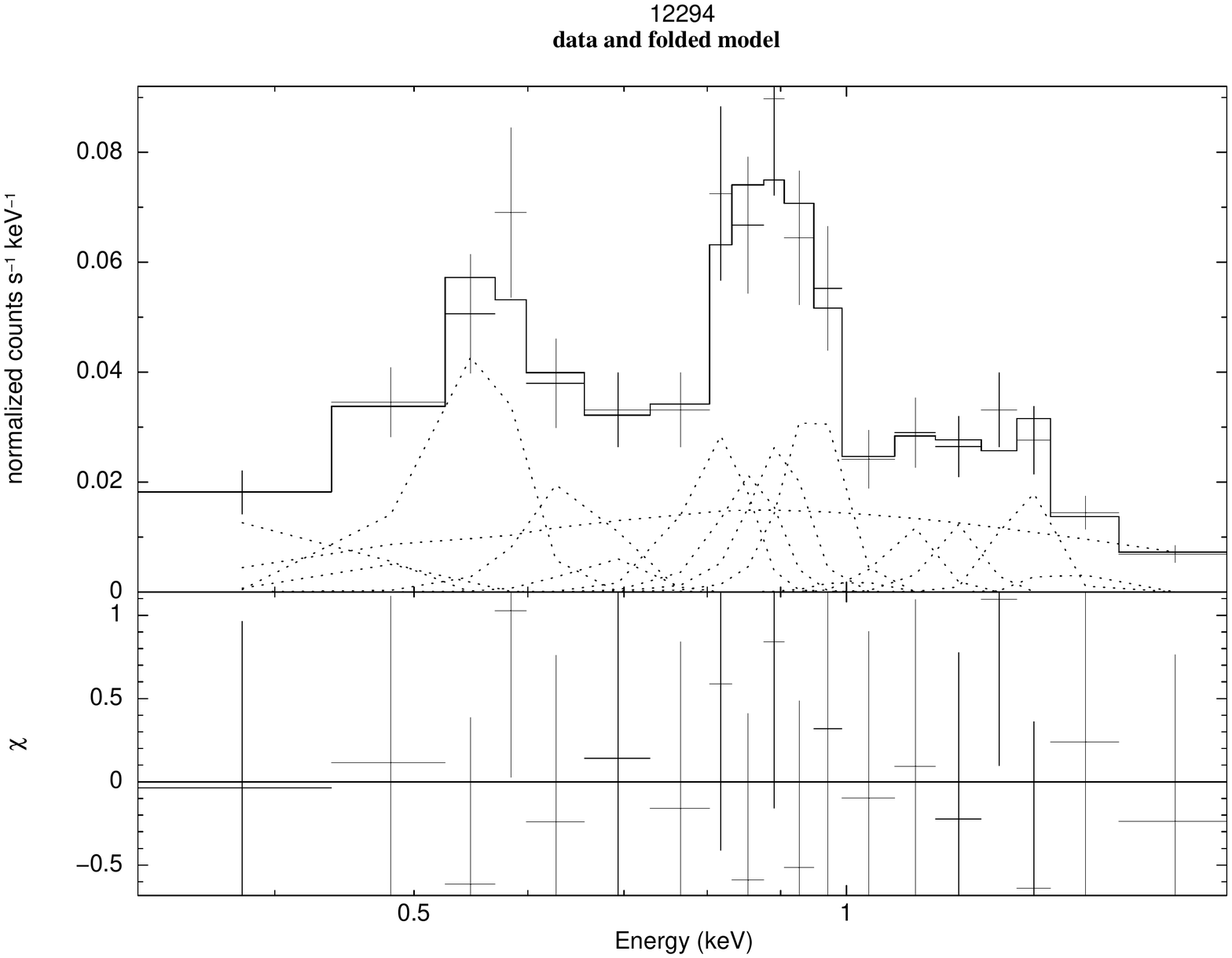}
\includegraphics[scale=0.25]{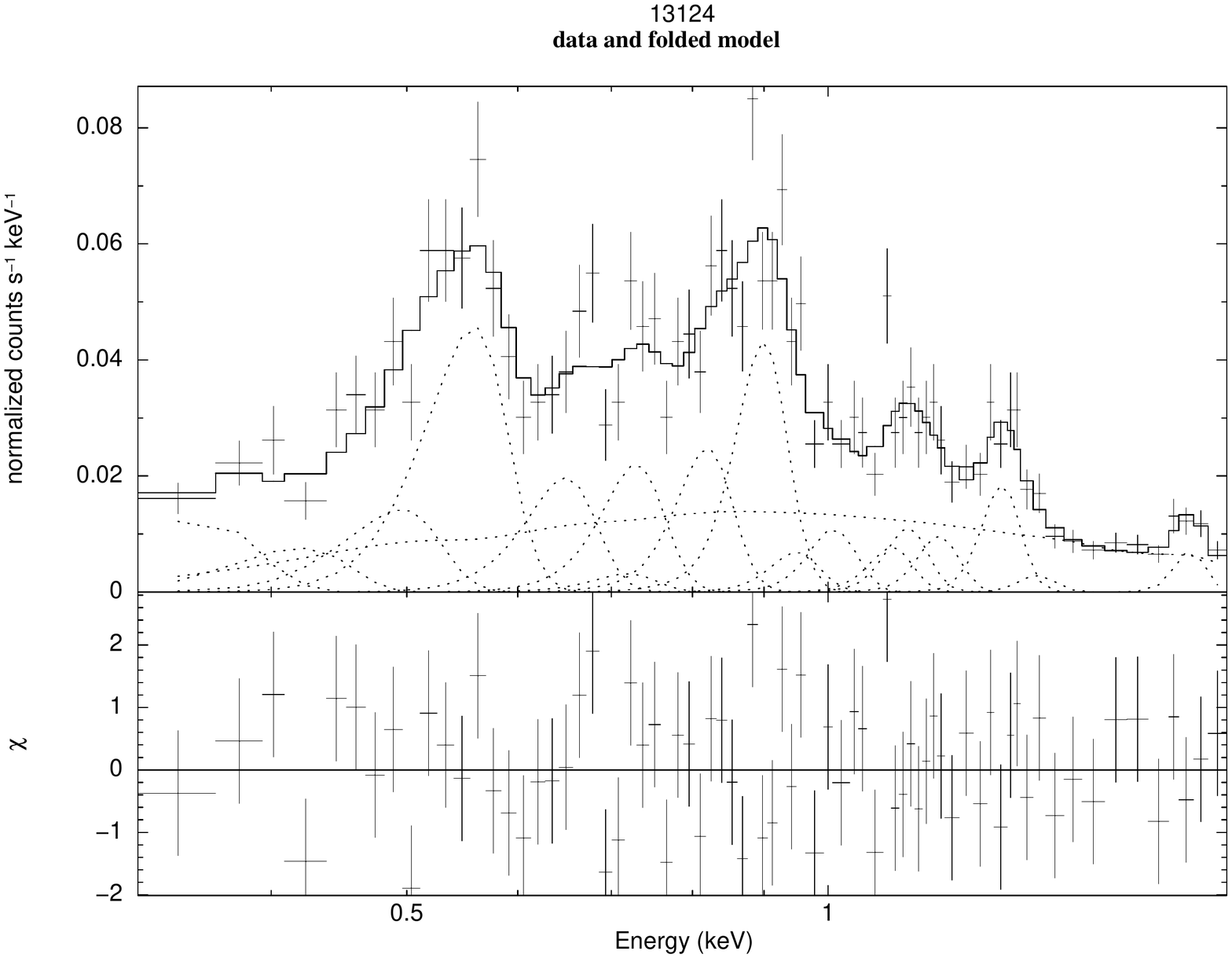}
\includegraphics[scale=0.25]{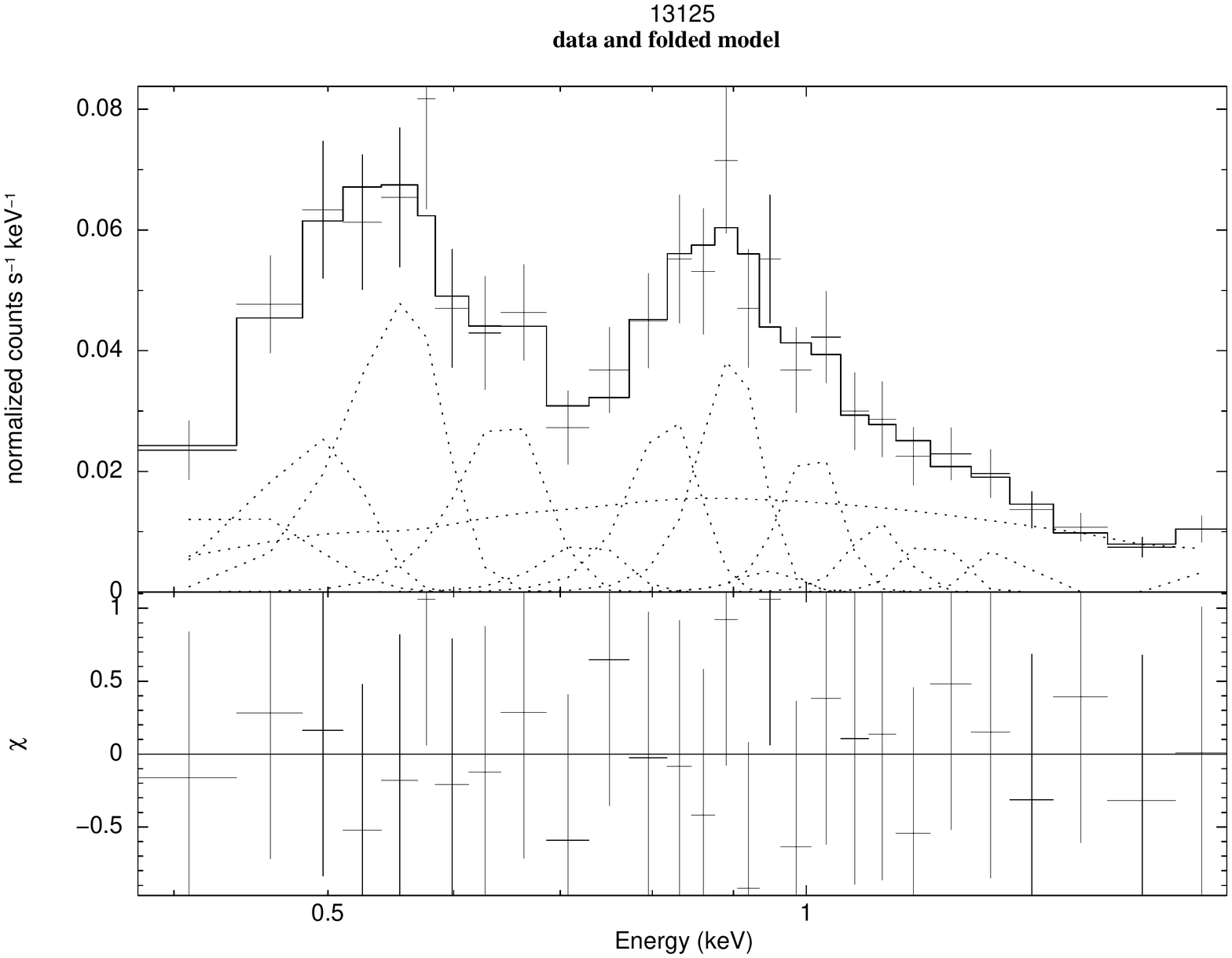}
\includegraphics[scale=0.25]{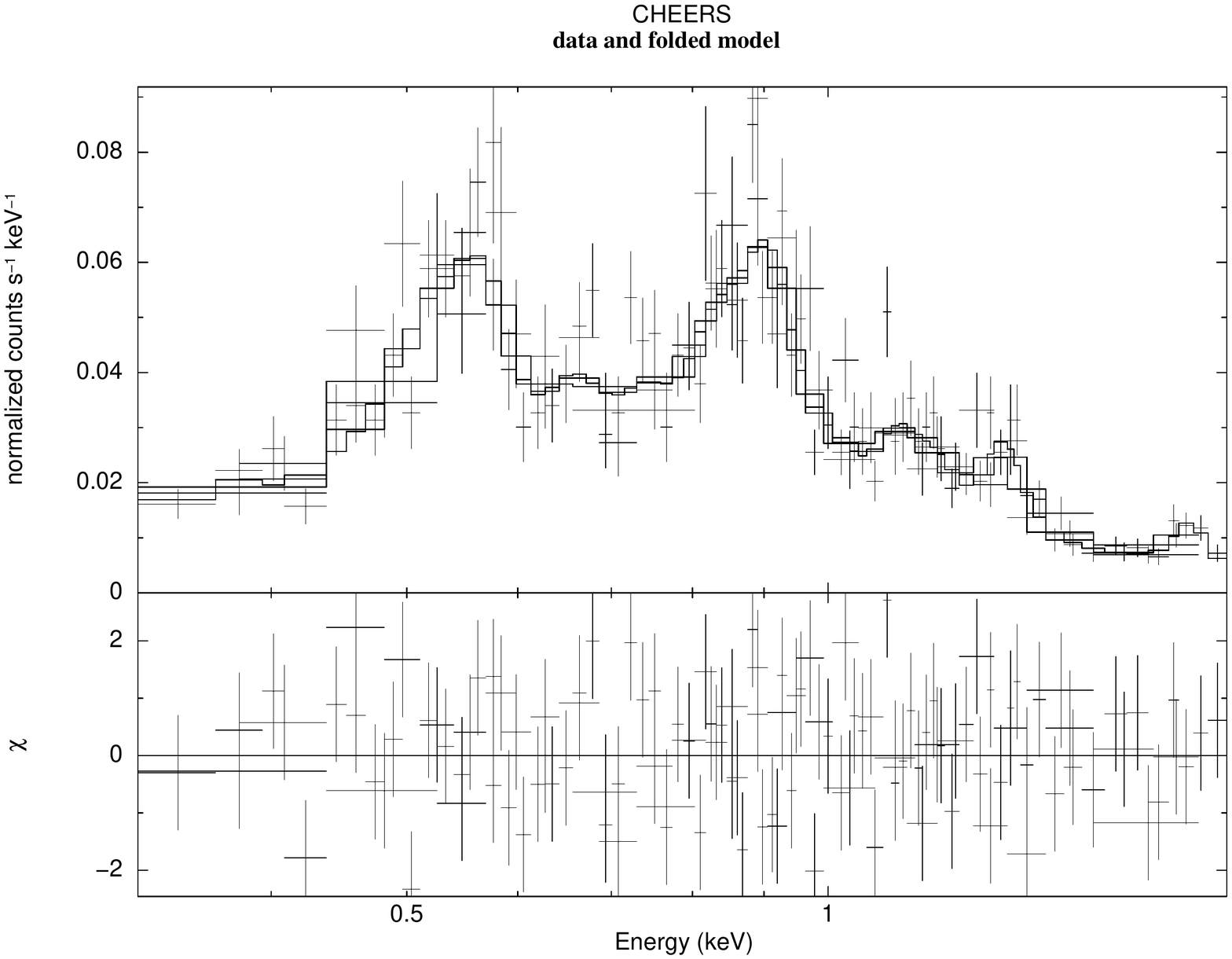}
\includegraphics[scale=0.25]{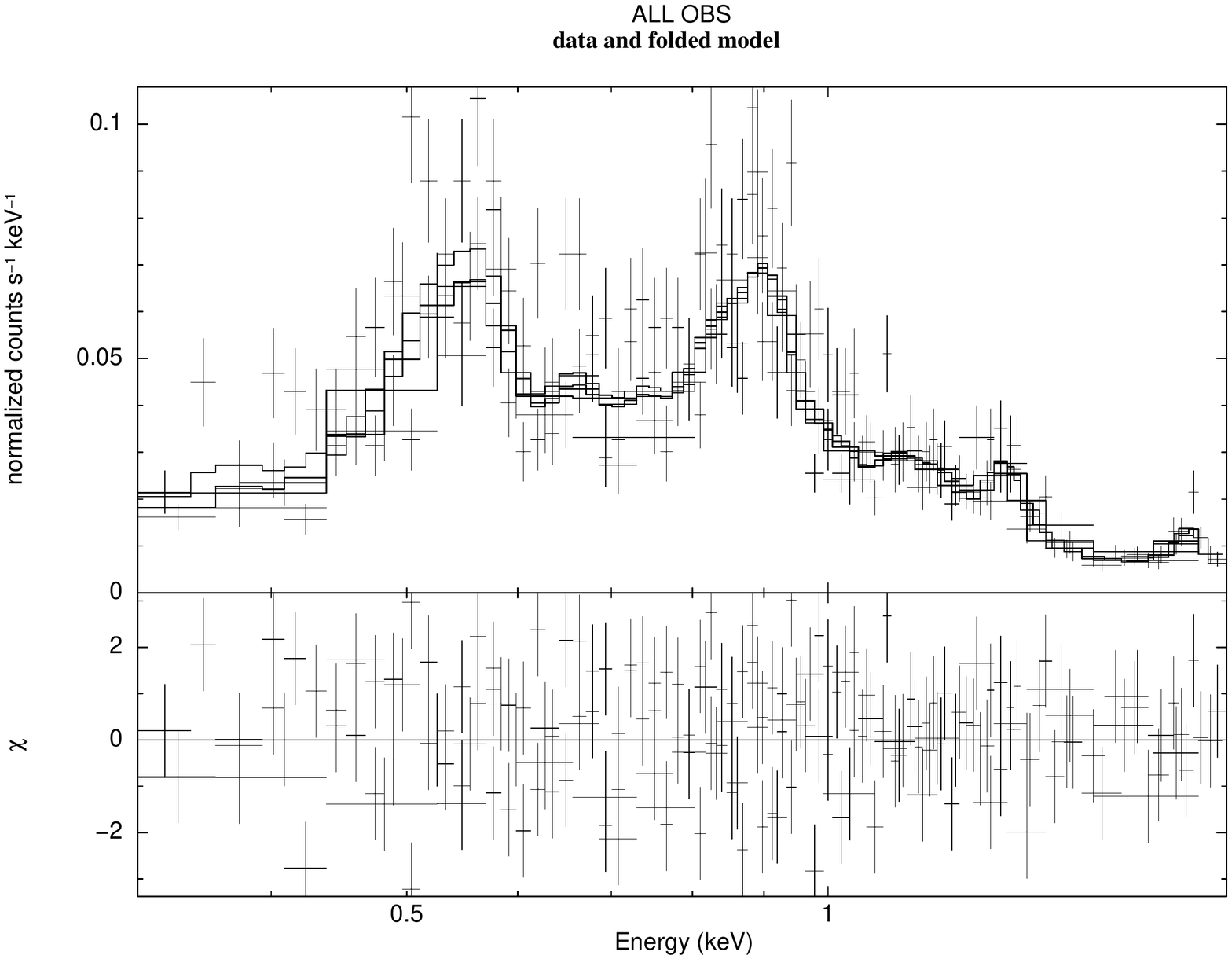}
\caption{\textit{Chandra}/ACIS-S \(0.3 - 2\mbox{ keV}\) spectra of the nuclear emission (extracted from the region shown in Figure \ref{torusregion}) in different observations, with best fit emission lines models. From top left to bottom right we show fits to OBS. 07745, 12294, 13124, 13125, merged data from CHEERS observations and merged data from all observations. For spectra of single observations we also show with dotted lines the additive components of the best fitting models.}\label{nucleuslinespectra}
\end{figure}

\begin{figure}
\centering
\includegraphics[scale=0.25]{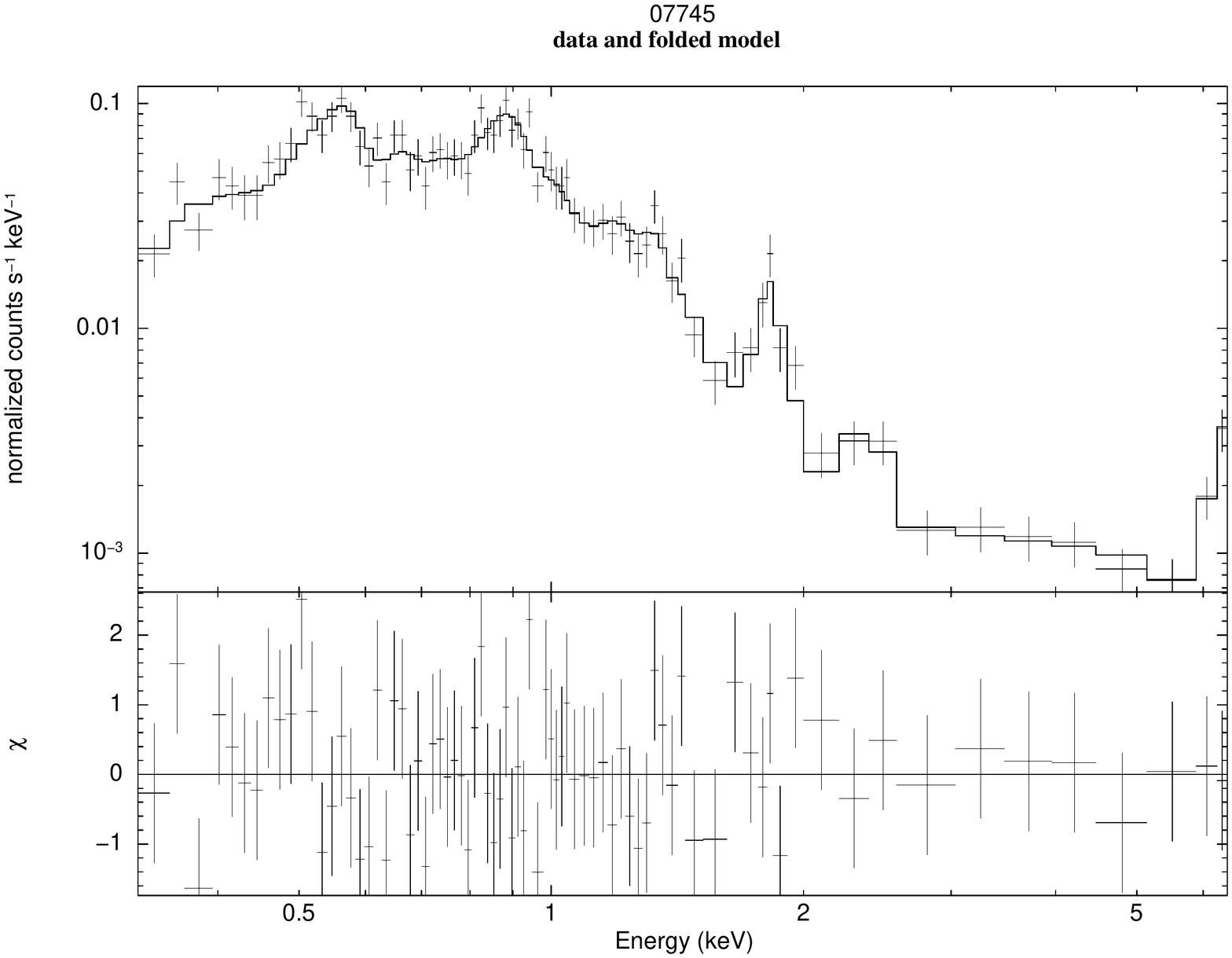}
\includegraphics[scale=0.25]{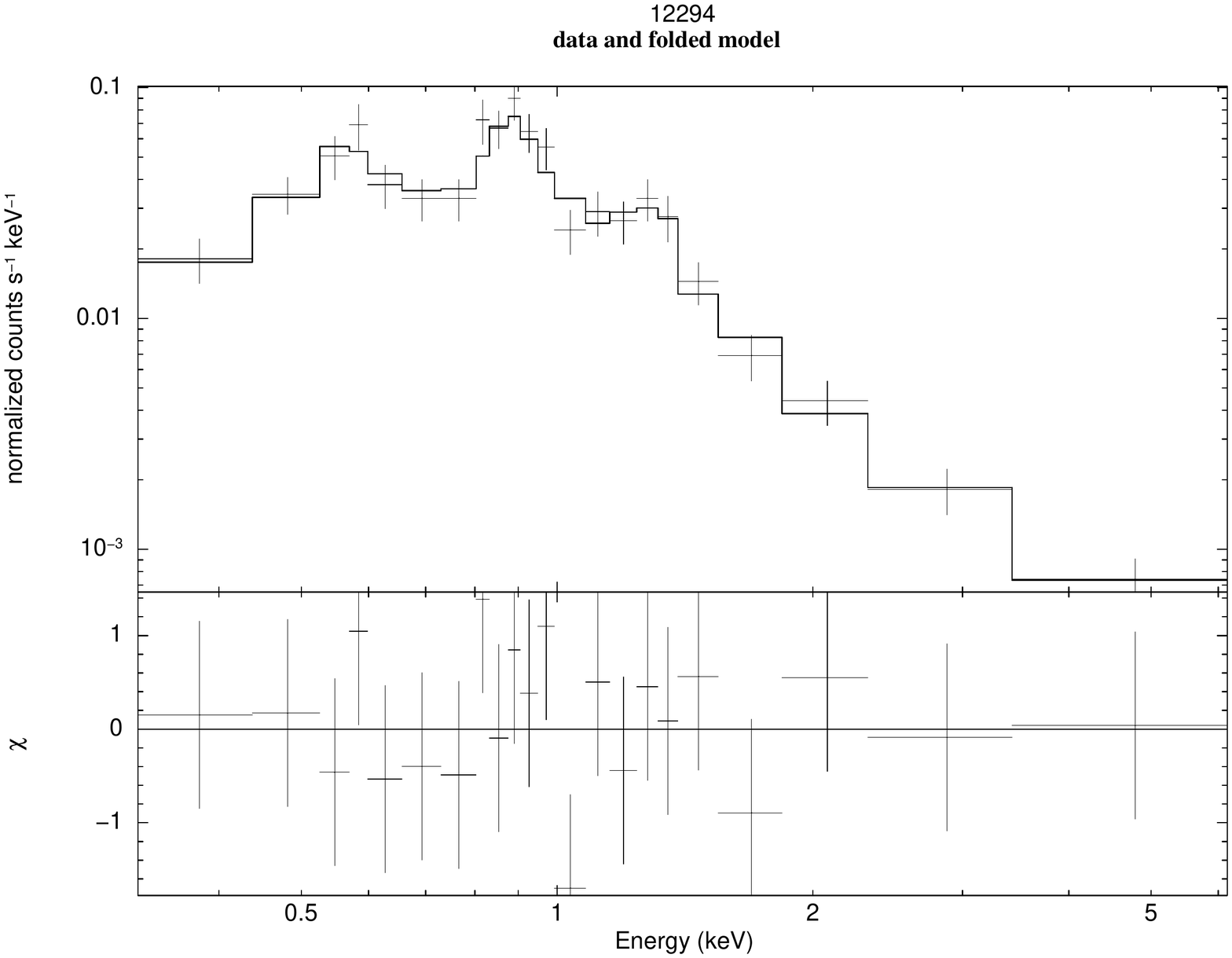}
\includegraphics[scale=0.25]{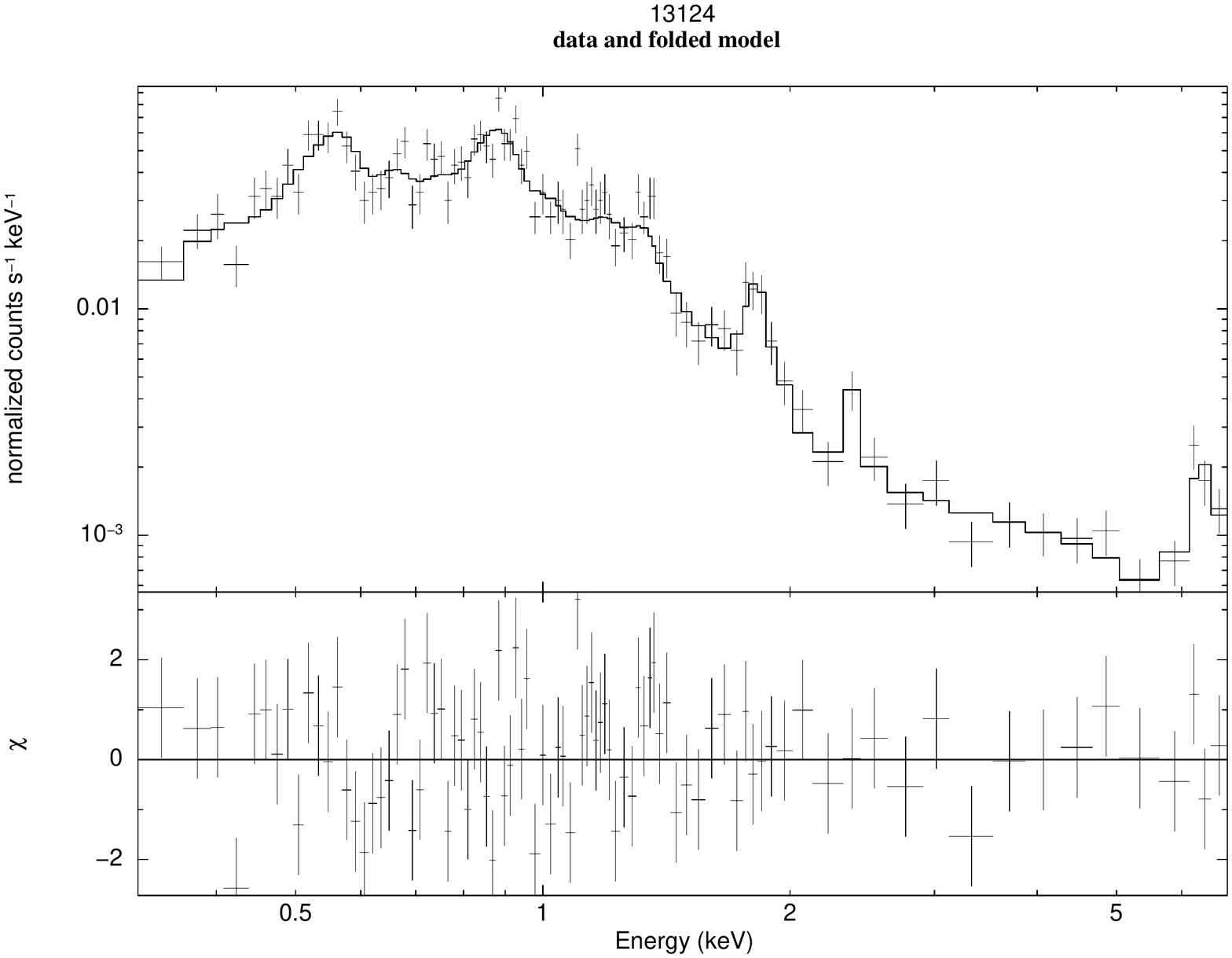}
\includegraphics[scale=0.25]{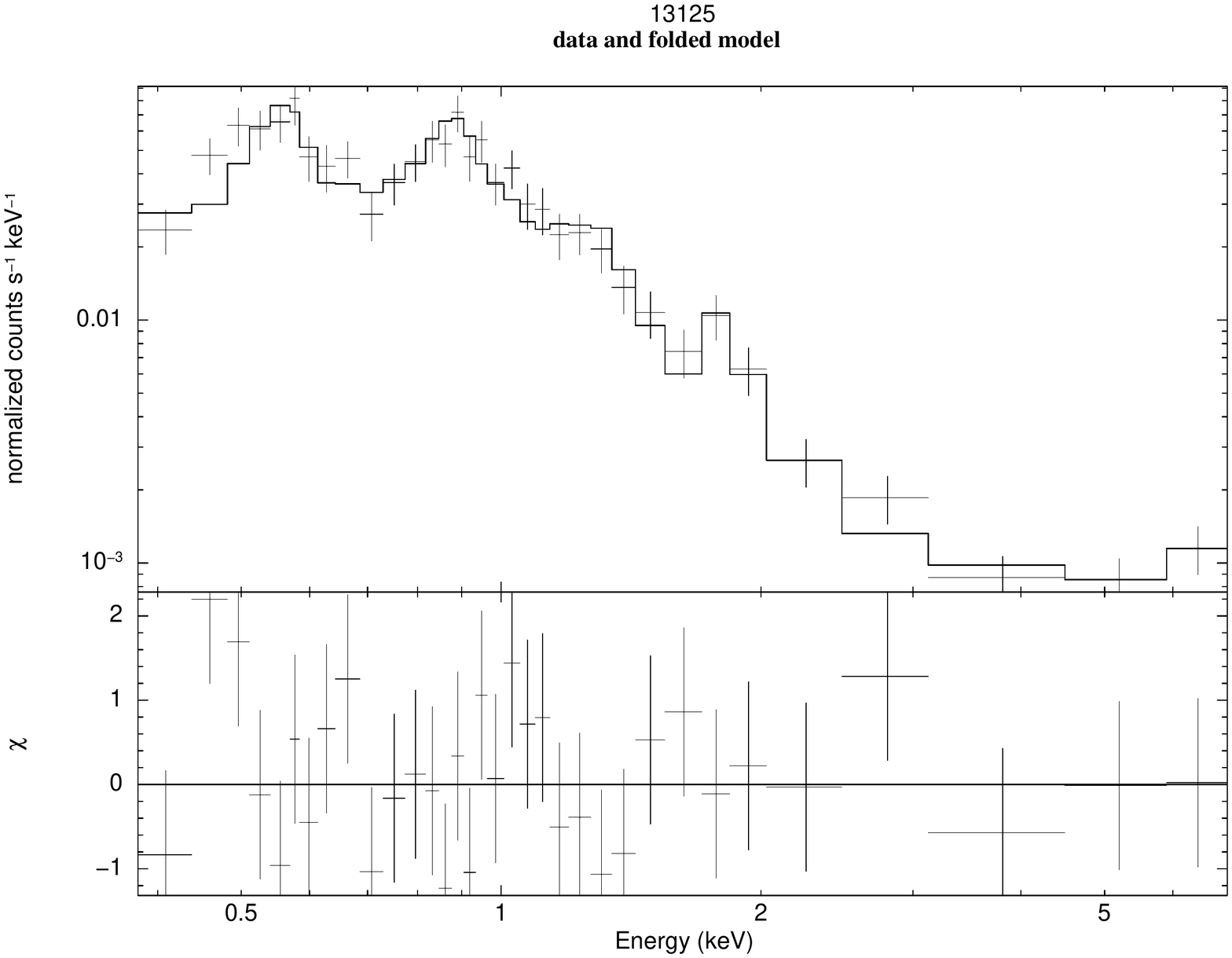}
\includegraphics[scale=0.25]{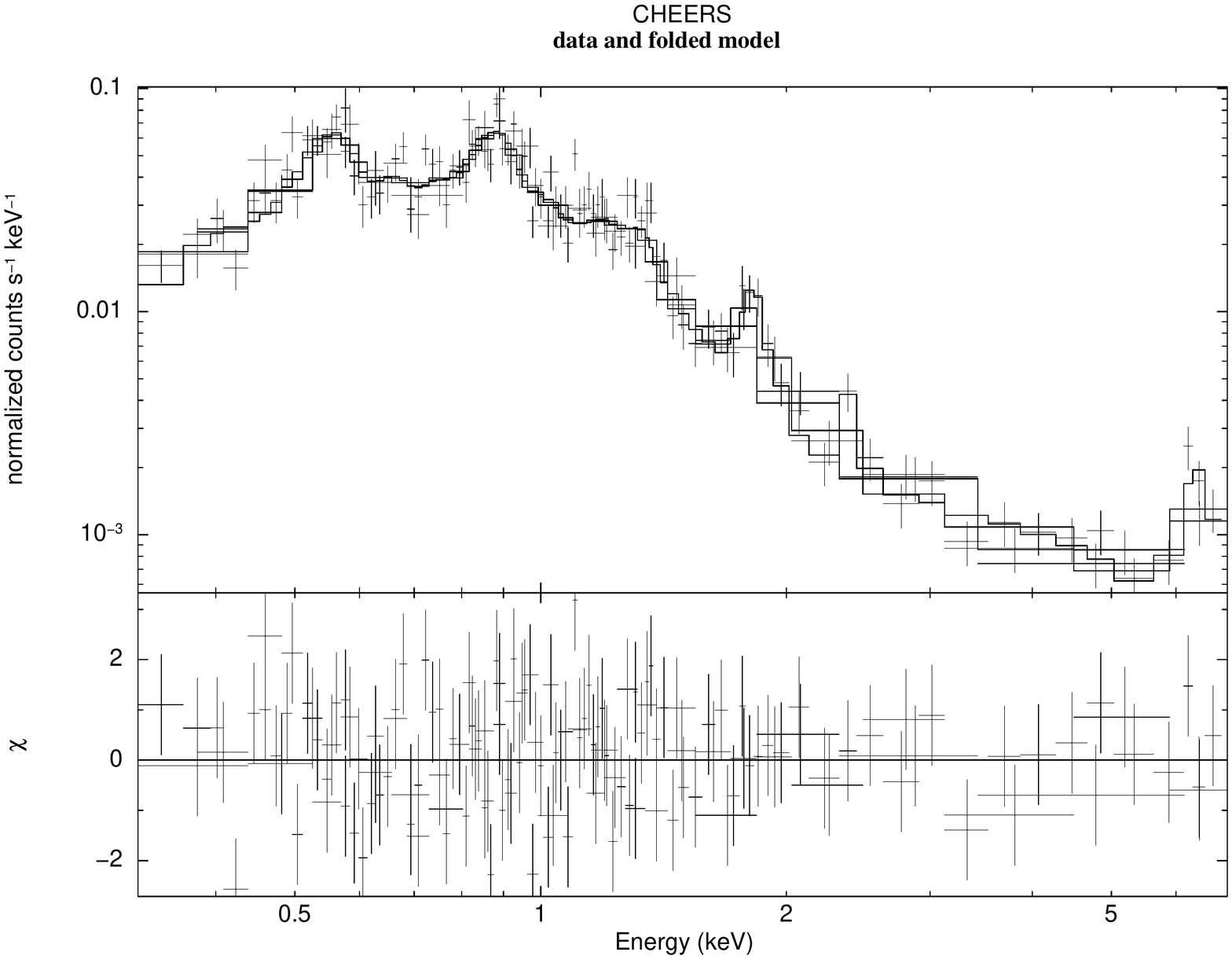}
\includegraphics[scale=0.25]{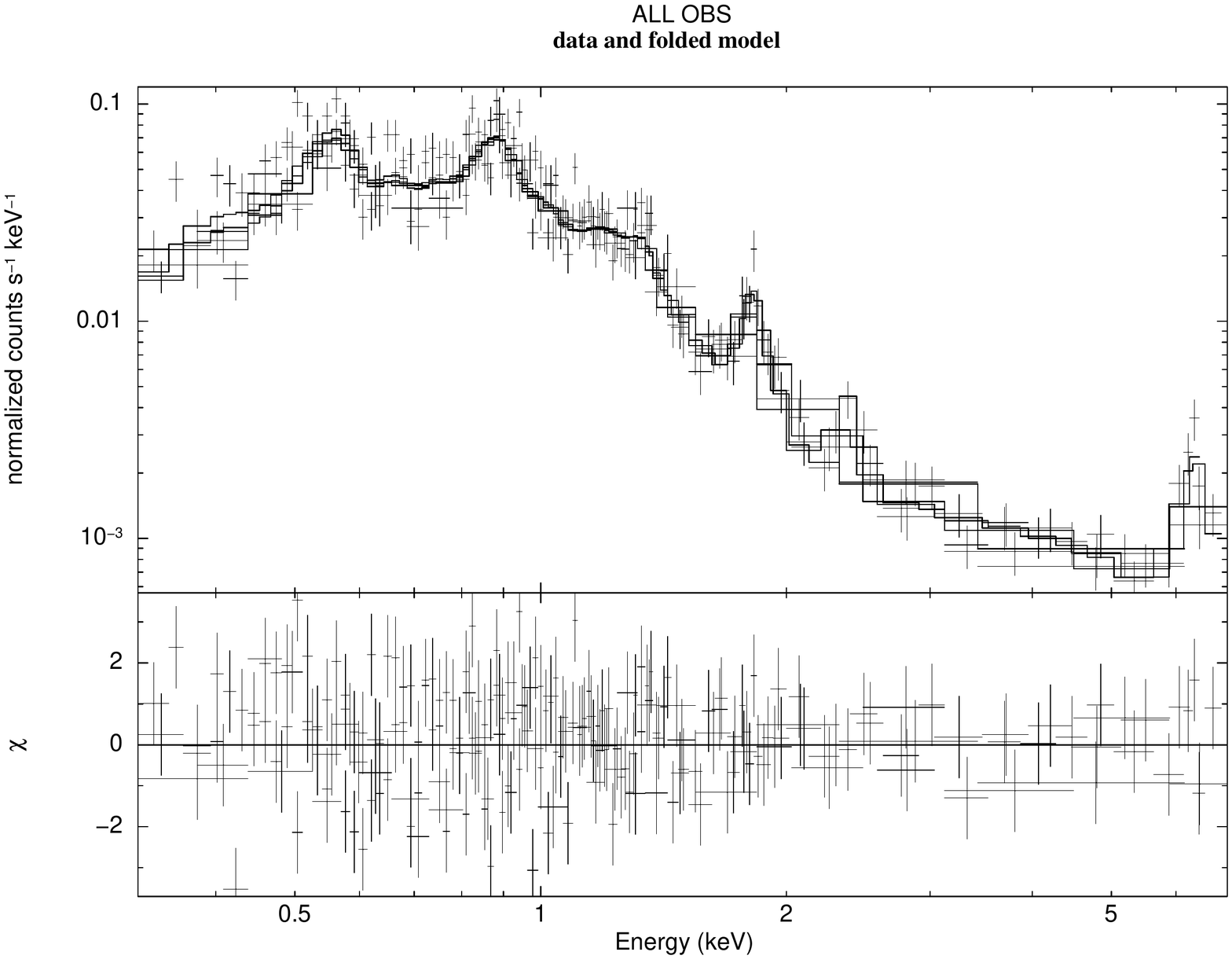}
\caption{\textit{Chandra}/ACIS-S \(0.3 - 10\mbox{ keV}\) spectra of the nuclear emission (extracted from the region shown in Figure \ref{torusregion}) in different observations, with best fit photoionization models. From top left to bottom right we show fits to OBS. 07745, 12294, 13124, 13125, merged data from CHEERS observations and merged data from all observations.}\label{nucleusphotospectra}
\end{figure}

{In this Appendix we present results of spectral fits for the models presented in Sects. \ref{conespectra} and \ref{nuclearspectra}, both for the cone (Tables \ref{conelines}, \ref{conephoto} and Figures \ref{coneslinespectra}, \ref{conephotospectra}) and the nuclear (Tables \ref{nucleuslines}, \ref{nucleusphoto}, \ref{innernucleusphoto} and Figures \ref{nucleuslinespectra}, \ref{nucleusphotospectra}) regions. When possible, fits were performed each observation as well as on merged observations.}

\begin{table}\tiny
\centering
\begin{threeparttable}
\caption{Measured line fluxes of the cone regions}\label{conelines}
\begin{tabular}{lc|cc|cc|cc|cc}
\hline
\hline
Obs. ID                     &                 & \multicolumn{2}{|c|}{07745}                         & \multicolumn{2}{|c|}{13124}                         & \multicolumn{2}{|c|}{CHEERS OBS}                    & \multicolumn{2}{|c}{ALL OBS}                        \\
\cline{3-10}
Cone                        &                 & NW                       & SE                       & NW                       & SE                       & NW                       & SE                       & NW                       & SE                       \\
\multicolumn{2}{l|}{Net Counts 0.3 - 2 keV (error)}                   & 358(19)                  & 277(17)                  &  424(21)                 & 371(19)                  & 656(26)                  & 550(24)                  & 1013(32)                 & 826(29)                  \\
\hline 
Line                        & Energy\tnote{a} & \multicolumn{8}{|c}{Flux\tnote{b}} \\
\hline
C \textsc{V} He\(\gamma\)    & 0.371           & -                        &  -                       & -                        & -                        & -                        & -                        & -                        & -                        \\
N \textsc{VI} triplet        & 0.426           & -                        &  -                       & -                        & -                        & -                        & -                        & -                        & -                        \\
C \textsc{VI} Ly\(\beta\)    & 0.436           & \({0.86}_{-0.54}^{+0.62}\) & \({2.03}\pm{0.46}\) & \({0.75}\pm{0.36}\) & \({1.73}\pm{0.35}\) & \({0.71}\pm{0.36}\) & \({1.60}\pm{0.32}\) & \({0.74}\pm{0.31}\) & \({1.75}\pm{0.26}\) \\
N \textsc{VII} Ly\(\alpha\)  & 0.500           & \({0.40}_{-0.30}^{+0.40}\) & -                        & \(<0.48\) & -                        & \(<0.46\) & -                        & \({0.30}\pm{0.20}\) & -                        \\
O \textsc{VII} triplet       & 0.569           & \({1.65}_{-0.30}^{+0.40}\) & \({1.65}\pm{0.27}\) & \({1.16}\pm{0.25}\) & \({1.42}\pm{0.21}\) & \({1.14}\pm{0.24}\) & \({1.45}\pm{0.19}\) & \({1.30}\pm{0.20}\) & \({1.52}\pm{0.15}\) \\
O \textsc{VIII} Ly\(\alpha\) & 0.654           & \({<0.32}\) & \({0.26}\pm{0.13}\) & \({0.38}\pm{0.11}\) & \({0.40}\pm{0.11}\) & \({0.38}\pm{0.10}\) & \({0.44}\pm{0.10}\) & \({0.32}\pm{0.08}\) & \({0.38}\pm{0.08}\) \\
O \textsc{VII} He\(\gamma\)  & 0.698           & -                        & -                        & -                        & -                        & -                        & -                        & -                        & -                        \\
Fe \textsc{XVII} \(3s2\)     & 0.727           & -                        & -                        & -                        & -                        & -                        & -                        & -                        & -                        \\
O \textsc{VII} RRC           & 0.739           & \({0.31}_{-0.12}^{+0.14}\) & \({0.40}\pm{0.11}\) & \({0.43}\pm{0.09}\) & \({0.27}\pm{0.08}\) & \({0.48}_{-0.08}^{+0.09}\) & \({0.26}\pm{0.08}\) & \({0.42}\pm{0.07}\) & \({0.31}\pm{0.06}\) \\
Fe \textsc{XVII} \(3d2p\)    & 0.826           & \(<{0.21}\) & \({0.23}\pm{0.07}\) & \({0.18}\pm{0.11}\) & \({0.11}\pm{0.05}\) & \({0.20}_{-0.11}^{+0.06}\) & \({0.09}\pm{0.05}\) & \({0.17}\pm{0.09}\) & \({0.13}\pm{0.04}\) \\
O \textsc{VIII} RRC          & 0.871           & \({0.43}_{-0.25}^{+0.11}\) &  -                       & \(<{0.19}\) & -                        & \(<{0.15}\) & -                        & \({<0.23}\) & -                        \\
Ne \textsc{IX} triplet       & 0.915           & \({0.16}_{-0.09}^{+0.18}\) & \({0.13}\pm{0.06}\) & \({0.22}_{-0.10}^{+0.07}\) & \({0.18}\pm{0.05}\) & \({0.24}_{-0.09}^{+0.05}\) & \({0.17}\pm{0.04}\) & \({0.23}\pm{0.08}\) & \({0.16}\pm{0.04}\) \\
Fe \textsc{XX} \(3d2p\)      & 0.965           & -                        & -                        & -                        & -                        & -                        & -                        & -                        & -                        \\
Ne \textsc{X} Ly\(\alpha\)   & 1.022           & \({0.08}_{-0.05}^{+0.06}\) & \({0.09}\pm{0.05}\) & \({0.13}\pm{0.03}\) & \({0.06}\pm{0.03}\) & \({0.14}\pm{0.03}\) & \({0.07}\pm{0.03}\) & \({0.14}\pm{0.03}\) & \({0.07}\pm{0.03}\) \\
Ne \textsc{IX} He\(\gamma\)  & 1.127           & \({0.06}_{-0.05}^{+0.10}\) & -                        & -                        & -                        & -                        & -                        & -                        & -                        \\
Ne \textsc{IX} He\(\delta\)  & 1.152           & -                        & -                                       & \({0.04}\pm{0.02}\) & \({<0.03}\) & \({0.04}\pm{0.02}\) & \(<{0.03}\) & \({0.05}\pm{0.02}\) & \(<{0.03}\) \\
Ne \textsc{X} Ly\(\beta\)    & 1.211           & -                        & -                        & -                        & -                        & -                        &  -                       & -                        & -                        \\
Mg \textsc{XI} triplet       & 1.352           & \(<{0.08}\) & -  & \({0.06}\pm{0.03}\) &             - & \({0.05}\pm{0.03}\) & - & \({0.05}\pm{0.02}\) & - \\
Fe \textsc{XXII} \(4p2p\)    & 1.425           & -                        & -                        & -                        & -                        & -                        & -                        & - & -                        \\
Si \textsc{XIII} triplet     & 1.839           & -                        & -                                     & - & - & - & - & - & - \\
Power-Law norm.             &        & \({0.66}_{-0.55}^{+0.16}\) & \({0.31}\pm{0.08}\) & \({0.31}\pm{0.09}\)  & \({0.44}_{-0.08}^{+0.07}\) & \({0.30}\pm{0.11}\)  & \({0.43}\pm{0.07}\) & \({0.32}\pm{0.11}\) & \({0.37}\pm{0.05}\)  \\
\hline
\(\chi^2\)(dof)             &                 & 0.72(3)                  & 0.82(4)                  & 1.19(6)                  & 0.77(8)                  & 0.88(16)                 & 0.62(16)                 & 1.29(31)                 & 0.75(28)                 \\
\({F_{0.3-2}}\)\tnote{c}    &                 & \({0.58}_{-0.06}^{+0.04}\) & \({0.51}\pm{0.04}\) & \({0.45}\pm{0.03}\) & \({0.49}\pm{0.03}\) & \({0.45}\pm{0.03}\) & \({0.48}\pm{0.03}\) & \({0.47}\pm{0.02}\) & \({0.49}\pm{0.02}\) \\ 
\hline
\hline
\end{tabular}
\begin{tablenotes}[para]
                 \item {Notes:}\\
                 \item[a] Line rest-frame energy in keV.\\
                 \item[b] Line fluxes in units of \({10}^{-5}\mbox{ photons} \mbox{ cm}^{-2} \mbox{ s}^{-1}\).\\                
                 \item[c] Unabsorbed flux in the \(0.3-2\mbox{ keV}\) band in units of \({10}^{-13}\mbox{ erg}\mbox{ cm}^{-2}\mbox{ s}^{-1}\).
       \end{tablenotes}
\end{threeparttable}
\end{table}

\begin{table}\tiny
\centering
\begin{threeparttable}
\caption{Best fit photoionization models for the cone regions}\label{conephoto}
\begin{tabular}{l|cc|cc|cc|cc}
\hline
\hline
Obs. ID                       & \multicolumn{2}{|c|}{07745}                               & \multicolumn{2}{|c|}{13124}                               & \multicolumn{2}{|c|}{CHEERS OBS}                          & \multicolumn{2}{|c}{ALL OBS}                              \\
\hline 
Model Parameter               & NW                          & SE                          & NW                          & SE                          & NW                          & SE                          & NW                          & SE                          \\
\hline
\(\log {U_1}\)               & \({1.00}\pm{0.08}\)    & \({0.75}_{-0.19}^{+0.20}\)    & \({0.78}\pm{0.15}\)    & \({0.62}_{-0.23}^{+0.29}\)    & \({0.75}_{-0.19}^{+0.34}\)    & \({0.55}_{-0.18}^{+0.23}\)    & \({0.93}\pm{0.28}\)    & \({0.69}_{-0.19}^{+0.31}\)    \\
\(\log {N_{H\,1}}\)         & \({20}^*\) & \({20}^*\) & \({20}^*\) & \({20}^*\) & \({20}^*\) & \({20}^*\) & \({20}^*\) & \({20}^*\) \\
\(F_{1\,(0.3-2)}\)\tnote{a}   & \({0.36}_{-0.05}^{+0.04}\)    & \({0.26}_{-0.09}^{+0.08}\)    & \({0.25}_{-0.09}^{+0.13}\)    & \({0.27}_{-0.11}^{+0.09}\)    & \({0.24}_{-0.07}^{+0.11}\)    & \({0.27}_{-0.08}^{+0.09}\)    & \({0.21}_{-0.07}^{+0.10}\)    & \({0.22}\pm{0.08}\)    \\
\(\log {U_2}\)              & \(-0.5^*\)                    & \(-1^*\)                     & \({-0.55}_{-0.39}^{+0.34}\) & \({-0.95}_{-0.26}^{+0.32}\)  & \({-0.50}_{-0.33}^{+0.55}\) & \({-0.85}_{-0.27}^{+0.40}\)  & \({-0.51}_{-0.31}^{+0.27}\) & \({-0.80}_{-0.18}^{+0.21}\) \\
\(\log {N_{H\,2}}\)         & \(20^*\)                    & \(20^*\)                     & \(20^*\)                    & \(20^*\)                    & \(20^*\)                    & \(20^*\)                    & \({20}^*\)                    & \(20^*\)                    \\
\(F_{2\,(0.3-2)}\)\tnote{a}   & \({0.32}_{-0.05}^{+0.06}\)    & \({0.41}\pm{0.11}\)    & \({0.26}_{-0.08}^{+0.12}\)    & \({0.34}_{-0.11}^{+0.12}\)    & \({0.25}_{-0.07}^{+0.11}\)    & \({0.30}_{-0.10}^{+0.09}\)    & \({0.31}_{-0.07}^{+0.09}\)    & \({0.37}\pm{0.08}\)    \\
\(kT\)\tnote{b}               & -    & -    & \({0.82}\pm{0.13}\)    & \({1.04}_{-0.17}^{+0.23}\)    & \({0.89}\pm{0.11}\) & \({1.09}_{-0.11}^{+0.41}\)    & \({0.80}\pm{0.09}\)    & \({0.97}\pm{0.17}\)    \\
\(F_{C\,(0.3-2)}\)\tnote{a}   & -    & -    & \({0.07}_{-0.04}^{+0.05}\)    & \({0.04}\pm{0.03}\)    & \({0.07}_{-0.05}^{+0.03}\)    & \({0.04}\pm{0.02}\)    & \({0.08}_{-0.03}^{+0.05}\)    & \({0.04}\pm{0.03}\)    \\
\hline
\(\chi^2\)(dof)               & 1.16(12)   &  0.96(9)   & 1.04(12) &  0.47(11)   & 1.10(22)   & 0.42(19)    &  1.20(37)   & 0.57(31)   \\
\(F_{0.3-2}\)\tnote{a}       & \({0.68}_{-0.05}^{+0.04}\)    & \({0.68}\pm{0.05}\)    & \({0.57}_{-0.04}^{+0.05}\)    &   \({0.64}_{-0.05}^{+0.06}\)     & \({0.57}\pm{0.03}\)  & \({0.61}_{-0.05}^{+0.04}\)    & \({0.60}\pm{0.04}\)    & \({0.62}\pm{0.04}\)    \\
\hline
\hline
\end{tabular}
       \begin{tablenotes}[para]
                 \item {Notes:}\\
                 \item[a] Unabsorbed flux in the \(0.3-2\mbox{ keV}\) band in units of \({10}^{-13}\mbox{ erg}\mbox{ cm}^{-2}\mbox{ s}^{-1}\).    \\
                 \item[b] Plasma temperature in keV.          
       \end{tablenotes}
\end{threeparttable}
\end{table}

\begin{table}\footnotesize
\centering
\begin{threeparttable}
\caption{Measured line fluxes of the nuclear region}\label{nucleuslines}
\begin{tabular}{lc|c|c|c|c|c|c}
\hline
\hline
Obs. ID.                       &        & 07745                      & 12294                      & 13124                      & 13125                    & CHEERS OBS                    & ALL OBS                      \\
\multicolumn{2}{l|}{Net Counts 0.3 - 2 keV (error)}     & 2109(46)                       & 460(21)                        & 2308(48)                       & 765(28)                      & 3533(59)                       & 5642(75)                        \\
\hline 
Line                        & Energy\tnote{a} & \multicolumn{6}{c}{Flux\tnote{b}}                                                                                                                                                                                                             \\
\hline
C \textsc{V} He\(\gamma\)    & 0.371  & \({18.73}_{-7.40}^{+10.23}\) & -                          & \({12.22}_{-2.31}^{+2.25}\) & -                        & \({11.88}_{-2.12}^{+2.11}\) & \({12.71}_{-0.53}^{0.55}\) \\
N \textsc{VI} triplet        & 0.426  & -    & -                          & -                          & -                        & -                          & -                           \\
C \textsc{VI} Ly\(\beta\)    & 0.436  & \(<6.84\)                          & \({7.57}_{-2.31}^{+2.51}\)  & \({2.80}_{-0.95}^{+0.86}\)   & \({6.18}_{-3.86}^{+16.49}\) & \({2.87}_{-0.83}^{+0.81}\)   & \({3.65}_{-0.23}^{+0.18}\)    \\
N \textsc{VII} Ly\(\alpha\)  & 0.500  & \({5.33}_{-2.08}^{+1.11}\)   & \(<2.26\)   & \({2.29}\pm{0.53}\)   & \({4.10}_{-3.64}^{+2.49}\) & \({2.67}\pm{0.45}\)   & \({3.28}_{-0.12}^{+0.13}\)    \\
O \textsc{VII} triplet       & 0.569  & \({9.69}_{-1.03}^{+1.07}\)  & \({7.02}_{-1.38}^{+1.69}\)   & \({6.90}\pm{0.61}\)   & \({7.40}_{-4.03}^{+1.47}\) & \({7.00}\pm{0.50}\)   & \({7.76}\pm{0.13}\)    \\
O \textsc{VIII} Ly\(\alpha\) & 0.654  & \({3.14}\pm{0.43}\)   & \({1.74}_{-0.50}^{+1.13}\)   & \({1.56}_{-0.28}^{+0.32}\)   & \({2.47}\pm{0.60}\) & \({1.79}_{-0.22}^{+0.25}\)   & \({2.14}_{-0.07}^{+0.05}\)    \\
O \textsc{VII} He\(\gamma\)  & 0.698  & -                          & \({0.61}_{-0.54}^{+0.47}\)   & -                          & -                        & -                          & -                           \\
Fe \textsc{XVII} \(3s2\)     & 0.727  & -                          & -                          & \(<0.43\)   & -                        & -                          & -                           \\
O \textsc{VII} RRC           & 0.739  & \({2.09}\pm{0.34}\)   & -                          & \({1.34}_{-0.21}^{+0.43}\)   & \({0.55}\pm{0.41}\) & \({1.21}_{-0.22}^{+0.17}\)   & \({1.45}_{-0.06}^{+0.03}\)    \\
Fe \textsc{XVII} \(3d2p\)    & 0.826  & \({1.86}_{-0.50}^{+0.36}\)   & \({1.36}_{-0.51}^{+0.39}\)   & \({1.10}_{-0.20}^{+0.17}\)   & \({1.34}\pm{-0.36}\) & \({1.16}_{-0.14}^{+0.19}\)   & \({1.33}_{-0.03}^{+0.04}\)    \\
O \textsc{VIII} RRC          & 0.871  & \(<1.02\)   & \({0.80}_{-0.43}^{+0.48}\)   & -                          & -                        & \(<0.19\)   & \({0.11}_{-0.06}^{+0.01}\)    \\
Ne \textsc{IX} triplet       & 0.915  & \({2.24}_{-0.62}^{+0.44}\)   & \({1.08}_{-1.05}^{+0.47}\)   & \({1.61}_{-0.15}^{+0.18}\)   & \({1.49}_{-0.44}^{+0.36}\) & \({1.58}_{-0.13}^{+0.22}\)   & \({1.75}_{-0.03}^{+0.04}\)    \\
Fe \textsc{XX} \(3d2p\)      & 0.965  & \(<0.60\)   & \({1.28}_{-0.34}^{+0.56}\)   & \({0.23}_{-0.11}^{+0.14}\)   & \(<{0.53}\) & \({0.32}_{-0.19}^{+0.11}\)   & \({0.32}_{-0.03}^{+0.02}\)    \\
Ne \textsc{X} Ly\(\alpha\)   & 1.022  & \({1.10}\pm{0.20}\)   & \(<0.32\)   & \({0.35}_{-0.09}^{+0.10}\)   & \({0.82}_{-0.29}^{+0.25}\) & \({0.39}_{-0.08}^{+0.12}\)   & \({0.55}_{-0.03}^{+0.02}\)    \\
Ne \textsc{IX} He\(\gamma\)  & 1.127  & \({0.44}\pm{0.12}\)   & \({0.41}_{-0.21}^{+0.22}\)   & \({0.22}_{-0.08}^{+0.10}\)   & \({0.35}_{-0.17}^{+0.18}\) & \({0.33}_{-0.07}^{+0.12}\)   & \({0.40}\pm{0.02}\)    \\
Ne \textsc{IX} He\(\delta\)  & 1.152  & -                          & -                          & \({0.31}_{-0.08}^{+0.15}\)   & -                        & \({0.15}_{-0.07}^{+0.20}\)   & \({0.07}\pm{0.02}\)    \\
Ne \textsc{X} Ly\(\beta\)    & 1.211  & \({0.51}\pm{0.11}\)   & \({0.41}_{-0.10}^{+0.18}\)   & \({0.27}_{-0.08}^{+0.07}\)   & \({0.28}\pm{0.15}\) & \({0.30}_{-0.06}^{+0.09}\)   & \({0.38}_{-0.02}^{+0.01}\)    \\
Mg \textsc{XI} triplet       & 1.352  & \({0.53}\pm{0.10}\)   & \({0.59}\pm{0.18}\)   & \({0.49}\pm{0.07}\)   & \({0.23}\pm{0.13}\) & \({0.44}\pm{0.06}\)   & \({0.47}_{-0.02}^{+0.01}\)    \\
Fe \textsc{XXII} \(4p2p\)    & 1.425  & \({0.16}\pm{0.09}\)   & \({0.18}_{-0.17}^{+0.15}\)   & \({0.08}_{-0.06}^{+0.05}\)   & -                        & \({0.06}_{-0.04}^{+0.05}\)   & \({0.09}_{-0.01}^{+0.02}\)    \\
Si \textsc{XIII} triplet     & 1.839  & \({0.32}\pm{0.08}\)   & -                          & \({0.21}\pm{0.05}\)   & \({0.15}\pm{0.12}\) & \({0.19}\pm{0.04}\)   & \({0.23}\pm{0.01}\)    \\
Power-Law norm.             &        & \({3.68}\pm{0.51}\)   & \({4.48}_{-0.72}^{+0.75}\)   & \({4.20}_{-0.30}^{+0.32}\)   & \({4.71}\pm{0.74}\) & \({4.28}_{-0.26}^{+0.25}\)   & \({4.06}_{-0.06}^{+0.08}\)    \\
\hline
\(\chi^2\)(dof)             &        & 0.93(49)                   & 1.28(4)                    & 1.28(55)                   & 0.54(13)                 & 1.21(102)                  & 1.63(169)                   \\
\({F_{(0.3-2)}}\)\tnote{c}           &        & \({5.40}_{-0.23}^{+0.29}\)   & \({3.69}\pm{0.22}\)   & \({4.01}\pm{0.15}\)   & \({3.79}_{-0.68}^{+0.82}\) & \({4.05}_{-0.14}^{+0.13}\)   & \({4.37}\pm{0.11}\)    \\
\hline
\hline
\end{tabular}
       \begin{tablenotes}[para]
                 \item {Notes:}\\
                 \item[a] Line rest-frame energy in keV.\\
                 \item[b] Line fluxes in units of \({10}^{-5}\mbox{ photons} \mbox{ cm}^{-2} \mbox{ s}^{-1}\).\\                
                 \item[c] Unabsorbed flux in the \(0.3-2\mbox{ keV}\) band in units of \({10}^{-13}\mbox{ erg}\mbox{ cm}^{-2}\mbox{ s}^{-1}\).      
       \end{tablenotes}
\end{threeparttable}
\end{table}

\begin{table}\footnotesize
\centering
\begin{threeparttable}
\caption{Best fit photoionization models for the nuclear region}\label{nucleusphoto}
\begin{tabular}{l|c|c|c|c|c|c}
\hline
\hline
Obs. ID.                            & 07745                & 12294 &    13124 & 13125                      & CHEERS OBS                     & ALL OBS                      \\
Net Counts 0.3 - 10 keV (error)            & 2352(48)              &  508(23)        & 2663(52) &  840(29)                     & 4063(64)                          & 6415(80)                           \\
\hline 
Model Parameter                  &                             &          &                   & &                            &              \\
\hline
\(\log {U_1}\)        & \({1.09}_{-0.04}^{+0.05}\)       & \({1.11}_{-0.07}^{+0.09}\)        &    \({1.03}_{-0.03}^{0.04}\)       & \({1.09}_{-0.06}^{+0.09}\)       & \({1.05}_{-0.03}^{+0.04}\)       & \({1.06}\pm{0.03}\) \\
\(\log {N_{H\,1}}\) & \({20.62}_{-0.22}^{+0.20}\) & \({21.20}_{-0.49}^{+1.02}\)  & \({20.99}_{-0.22}^{+0.34}\) & \({20.91}_{-0.32}^{+0.44}\) & \({20.99}_{-0.39}^{+0.38}\) & \({20.88}_{-0.15}^{+0.17}\) \\
\(F_{1\,(0.3-2)}\)\tnote{a} & \({2.77}_{-0.18}^{+0.19}\) & \({2.75}_{-0.45}^{+0.40}\) & \({1.94}_{-0.14}^{+0.12}\) & \({2.19}\pm{0.28}\) & \({2.03}_{-0.15}^{+0.12}\) & \({2.21}_{-0.13}^{+0.14}\) \\
\(F_{1\,(2-10)}\)\tnote{a}  & \({0.41}_{-0.05}^{+0.05}\) & \({0.64}_{-0.21}^{+0.33}\) & \({0.30}\pm{0.05}\) & \({0.37}_{-0.09}^{+0.10}\) & \({0.32}_{-0.05}^{+0.04}\) & \({0.34}_{-0.04}^{+0.05}\) \\
\(\log {U_2}\)                  & \({-0.75}_{-0.17}^{+0.13}\) & \({-0.89}\pm{0.63}\) & \({-0.73}_{-0.19}^{+0.28}\) & \({-0.88}\pm{0.23}\) & \({-0.72}_{-0.16}^{+0.26}\) & \({-0.77}_{-0.12}^{+0.13}\) \\
\(\log {N_{H\,2}}\)             & \({20.22}_{-0.35}^{+0.37}\) & \({21.86}\pm{1.60}\) & \({20.82}_{-0.38}^{+0.42}\) & \({21.66}_{-1.29}^{+0.73}\) & \({21.04}\pm{0.60}\) & \({20.70}_{-0.32}^{+0.34}\) \\
\(F_{2\,(0.3-2)}\)\tnote{a} & \({2.86}_{-0.28}^{+0.15}\) & \({1.17}_{-0.53}^{+0.56}\) & \({1.47}_{-0.17}^{+0.22}\) & \({2.10}_{-0.53}^{+0.52}\) & \({1.51}_{-0.20}^{+0.21}\) & \({1.86}_{-0.18}^{+0.20}\) \\
\(F_{2\,(2-10)}\)\tnote{a} & \({0.24}_{-0.05}^{+0.07}\) & \({0.34}_{-0.22}^{+0.74}\) & \({0.16}_{-0.04}^{+0.07}\) & \({0.48}_{-0.26}^{+0.83}\) & \({0.18}_{-0.03}^{0.07}\) & \({0.20}_{-0.04}^{+0.06}\) \\
Refl\tnote{c}                                          & \({1.20}_{-0.17}^{+0.16}\)  & \({0.70}_{-0.51}^{+0.37}\) & \({0.69}\pm{0.19}\) & \({0.73}\pm{0.65}	\)  & \({0.69}_{-0.10}^{+0.16}\) & \({0.82}\pm{0.14}\)  \\
\(E_{{Si K\alpha}}\)\tnote{d}    & \(1.839^*\)                 & \(1.839^*\)      & \(1.839^*\)            & \(1.839^*\)                 & \(1.839^*\)            & \(1.839^*\)     \\
\(F_{{Si K\alpha}}\)\tnote{e}    & \({0.14}\pm{0.07}\)  & -  & \({0.10}\pm{0.05}\)  & \(<0.22\)  & \({0.08}\pm{0.05}\)    & \({0.10}\pm{0.04}\)    \\
\(E_{{S K\alpha}}\)\tnote{d}     & \(2.430^*\)                 & \(2.430^*\)   & \(2.430^*\)              & \(2.430^*\)                 & \(2.430^*\)     & \(2.430^*\)              \\
\(F_{{S K\alpha}}\)\tnote{e}     & \({0.18}\pm{0.08}\) & \(<0.41\)   & \({0.13}\pm{0.06}\)  & \(<0.22\)  & \({0.11}\pm{0.05}\)    & \({0.13}\pm{0.04}\)    \\
\(E_{{Fe K\alpha}}\)\tnote{d}    & \({6.35}\pm{0.01}\)  & -  & \({6.53}_{-0.06}^{+0.09}\) & \({6.03}_{-0.03}^{+1.00}\)   & \({6.53}_{-0.07}^{+0.10}\)    & \({6.48}_{-0.05}^{+0.06}\)    \\
\(F_{{Fe K\alpha}}\)\tnote{e}    & \({0.89}_{-0.18}^{+0.17}\)  & -  & \({1.20}_{-0.23}^{+0.26}\)  & \({0.93}_{-0.43}^{+0.52}\)  & \({1.13}_{-0.20}^{+0.24}\)    & \({1.14}_{-0.17}^{+0.19}\)    \\
\hline
\(\chi^2\)(dof)                  & 0.94(68)     & 0.79(14)               & 1.42(76)    & 1.10(23)            & 1.21(132)                   & 1.55(211)                   \\
\(F_{(0.3-2)}\)\tnote{a} & \({5.77}_{-0.21}^{+0.22}\)  & \({4.00}_{-0.32}^{+0.29}\) & \({4.06}\pm{0.14}\) & \({4.40}_{-0.25}^{+0.35}\)  & \({4.09}\pm{0.13}\)    & \({4.54}_{-0.06}^{+0.11}\)   \\
\(F_{(2-10)}\)\tnote{a} & \({3.84}\pm{0.32}\) & \({2.36}_{-0.57}^{+0.53}\) & \({3.64}_{-0.29}^{+0.31}\) & \({3.32}_{-0.60}^{+0.46}\) & \({3.53}_{-0.25}^{+0.28}\) & \({3.68}\pm{0.21}\) \\
\hline
\hline
\end{tabular}
       \begin{tablenotes}[para]
                 \item {Notes:}\\
                 \item[a] Unabsorbed flux in units of \({10}^{-13}\mbox{ erg}\mbox{ cm}^{-2}\mbox{ s}^{-1}\).\\
                 \item[b] Plasma temperature in keV.\\
                 \item[c] Normalization of the reflection component in units of \({10}^{-3}\mbox{ photons}\mbox{ keV}^{-1}\mbox{ cm}^{-2}\mbox{ s}^{-1}\)\\
                 \item[d] Line rest-frame energy in keV.\\
                 \item[e] Line fluxes in units of \({10}^{-5}\mbox{ photons} \mbox{ cm}^{-2} \mbox{ s}^{-1}\).\\                
       \end{tablenotes}
\end{threeparttable}
\end{table}

\begin{table}
\centering
\begin{threeparttable}
\caption{Best fit photoionization models for the inner nuclear region}\label{innernucleusphoto}
\begin{tabular}{l|c|c|c}
\hline
\hline
Obs. ID.                            & 07745 & 13124                     & CHEERS OBS                      \\
Net Counts 0.3 - 10 keV (error)            & 1307(36)              &  1441(38)        & 2030(45)        \\
\hline 
Model Parameter                                  &                 &            &           \\
\hline
\(\log {U_1}\)        & \({1.10}\pm{0.06}\)       & \({1.02}\pm{0.05}\)        &    \({1.05}_{-0.05}^{+0.04}\)   \\
\(\log {N_{H\,1}}\) & \({20.51}_{-0.42}^{+0.32}\) & \({20.95}_{-0.24}^{+0.21}\)  & \({20.99}_{-0.29}^{+0.16}\)  \\
\(F_{1\,(0.3-2)}\)\tnote{a} & \({1.55}\pm{0.17}\) & \({1.10}\pm{0.09}\) & \({1.12}_{-0.10}^{+0.11}\) \\
\(F_{1\,(2-10)}\)\tnote{a}  & \({0.22}_{-0.05}^{+0.06}\) & \({0.16}_{-0.03}^{+0.04}\) & \({0.17}_{-0.04}^{+0.02}\) \\
\(\log {U_2}\)                  & \({-0.68}_{-0.24}^{+0.32}\) & \({-0.79}_{-0.19}^{+0.39}\) & \({-0.76}_{-0.17}^{+0.31}\) \\
\(\log {N_{H\,2}}\)             & \({20.10}_{-0.37}^{+0.48}\) & \({20.77}_{-0.43}^{+0.52}\) & \({21.06}\pm{0.86}\) \\
\(F_{2\,(0.3-2)}\)\tnote{a} & \({1.61}\pm{0.21}\) & \({0.77}_{-0.14}^{+0.18}\) & \({0.80}_{-0.15}^{+0.13}\) \\
\(F_{2\,(2-10)}\)\tnote{a} & \({0.12}_{-0.03}^{+0.05}\) & \({0.09}_{-0.03}^{+0.04}\) & \({0.10}_{-0.03}^{+0.05}\) \\
Refl\tnote{c}                                          & \({0.56}_{-0.19}^{+0.22}\) & \({0.40}_{-0.15}^{+0.13}\) & \({0.40}_{-0.07}^{+0.12}\) \\
\(E_{{Si K\alpha}}\)\tnote{d}    & \(1.839^*\)                 & \(1.839^*\)      & \(1.839^*\)   \\
\(F_{{Si K\alpha}}\)\tnote{e}    & \({0.11}\pm{0.05}\) & \({0.10}\pm{0.04}\) & \({0.07}\pm{0.03}\) \\
\(E_{{S K\alpha}}\)\tnote{d}     & \(2.430^*\)                 & \(2.430^*\)   & \(2.430^*\)   \\
\(F_{{S K\alpha}}\)\tnote{e}     & \({0.08}\pm{0.07}\) & \({0.06}\pm{0.04}\) & \({0.06}\pm{0.04}\) \\
\(E_{{Fe K\alpha}}\)\tnote{d}    & \({6.37}_{-0.02}^{+0.03}\) & \({6.58}_{-0.07}^{+0.08}\) & \({6.58}\pm{0.07}\) \\
\(F_{{Fe K\alpha}}\)\tnote{e}    & \({0.60}_{-0.18}^{+016}\) & \({0.78}_{-0.16}^{+0.17}\) & \({0.80}_{-0.16}^{+0.17}\) \\
\hline
\(\chi^2\)(dof)                  & 0.98(47)     & 1.26(57)               & 1.19(95)    \\
\(F_{(0.3-2)}\)\tnote{a} & \({3.36}_{-0.12}^{+0.17}\) & \({2.43}\pm{0.09}\) & \({2.34}_{-0.09}^{+0.07}\) \\
\(F_{(2-10)}\)\tnote{a}  & \({2.21}\pm{0.25}\)           & \({2.38}_{-0.19}^{+0.15}\) & \({2.31}_{-0.10}^{+0.18}\) \\
\hline
\hline
\end{tabular}
       \begin{tablenotes}[para]
                 \item {Notes:}\\
                 \item[a] Unabsorbed flux in units of \({10}^{-13}\mbox{ erg}\mbox{ cm}^{-2}\mbox{ s}^{-1}\).\\
                 \item[b] Plasma temperature in keV.\\
                 \item[c] Normalization of the reflection component in units of \({10}^{-3}\mbox{ photons}\mbox{ keV}^{-1}\mbox{ cm}^{-2}\mbox{ s}^{-1}\)\\
                 \item[d] Line rest-frame energy in keV.\\
                 \item[e] Line fluxes in units of \({10}^{-5}\mbox{ photons} \mbox{ cm}^{-2} \mbox{ s}^{-1}\).\\                
       \end{tablenotes}
\end{threeparttable}
\end{table}

\end{document}